\documentclass[preprint]{aastex}
\usepackage{amsmath,esint}
\usepackage{url,aasmacros}
\usepackage{natbib}

\usepackage{bm}
\usepackage{xspace}
\usepackage{color}
\usepackage{enumitem}

\newcommand{\snia}{{\rm SN~Ia}}
\newcommand{\sneia}{{\rm SNe~Ia}}

\newcommand{\kms}{\ensuremath{\mathrm{km~s}^{-1}}}

\newcommand{\sne}{{\rm SNe}}

\newcommand{\about}{\ensuremath{\sim}}

\newcommand{\hst}{\textit{HST}}
\newcommand{\vd}{\ensuremath{\sigma}}
\newcommand{\glsne}{gLSNe\xspace}
\newcommand{\h}{H\ensuremath{_0}\xspace}
\newcommand{\glsn}{gLSN\xspace}

\newcommand{\km}{\mathrm{km}\xspace}
\newcommand{\sersic}{S\'ersic}
\newcommand{\minion}{\texttt{minion\_1016}}
\newcommand{\altsched}{\texttt{altsched}}
\newcommand{\galsim}{\texttt{GalSim}}
\newcommand{\glafic}{\texttt{glafic}}
\newcommand{\reswidth}{0.95\textwidth}

\shorttitle{Strongly Lensed Supernovae}
\shortauthors{Goldstein, Nugent, and Goobar}

\begin{document}

\title{	
	Rates and Properties of Strongly Gravitationally Lensed Supernovae\\and their Host Galaxies in Time-Domain Imaging Surveys
}

\author[0000-0003-3461-8661]{Daniel~A.~Goldstein}
\altaffiliation{Hubble Fellow}
\affiliation{California Institute of Technology, 1200 East California Blvd, MC 249-17, Pasadena, CA 91125, USA}
\affiliation{Lawrence Berkeley National Laboratory, 1 Cyclotron Road MS 50B-4206, Berkeley, CA, 94720, USA}
\affiliation{Department of Astronomy, University of California, Berkeley, 501 Campbell Hall, Berkeley, CA 94720, USA}
\author[0000-0002-3389-0586]{Peter~E.~Nugent}
\affiliation{Lawrence Berkeley National Laboratory, 1 Cyclotron Road MS 50B-4206, Berkeley, CA, 94720, USA}
\affiliation{Department of Astronomy, University of California, Berkeley, 501 Campbell Hall, Berkeley, CA 94720, USA}
\author[0000-0002-4163-4996]{Ariel Goobar}
\affiliation{The Oskar Klein Centre, Department of Physics, Stockholm University, AlbaNova, SE-106 91 Stockholm, Sweden}

\begin{abstract}
Supernovae that are strongly gravitationally lensed (gLSNe) by galaxies are powerful probes of astrophysics and cosmology that will be discovered systematically by wide-field, high-cadence imaging surveys such as the Zwicky Transient Facility (ZTF) and the Large Synoptic Survey Telescope (LSST). Here we use pixel-level simulations that include  observing strategy, target selection, supernova properties, and dust to forecast the rates and properties of gLSNe that ZTF and LSST will find.  Applying the resolution-insensitive discovery strategy of \cite{gnkc18}, we forecast that ZTF (LSST) can discover 0.02 (0.79) 91bg-like, 0.17 (5.92) 91T-like, 1.22 (47.84) Type Ia, 2.76 (88.51) Type IIP, 0.31 (12.78) Type IIL, and 0.36 (15.43) Type Ib/c \glsne\ per year. We also forecast that the surveys can discover at least 3.75 (209.32) Type IIn \glsne\ per year, for a total of at least 8.60 (380.60) gLSNe per year under fiducial observing strategies. ZTF \glsne\ have a median $z_s=0.9$, $z_l=0.35$, $\mu_\mathrm{tot}=30$, $\Delta t_\mathrm{max}= 10$ days, $\min(\theta)= 0.25^{\prime\prime}$, and $N_\mathrm{img} = 4$.  LSST \glsne\ are less compact and less magnified, with a median $z_s=1.0$, $z_l=0.4$, $\mu_\mathrm{tot}\approx6$, $\Delta t_\mathrm{max} = 25$ days, $\min(\theta)=0.6^{\prime\prime}$, and $N_\mathrm{img} = 2$.  We develop a model of the supernova--host galaxy connection and  find that the vast majority of gLSN host galaxies will be multiply imaged, enabling detailed constraints on lens models with sufficiently deep high-resolution imaging taken after the supernova has faded.  We release the results of our simulations as catalogs at \url{http://portal.nersc.gov/project/astro250/glsne/}.
\end{abstract}

\section{Introduction}
\label{sec:intro}
When a supernova explodes far behind a foreground galaxy, the galaxy's strong gravitational field can create multiple images of the supernova in different places on the sky \citep{1936Sci....84..506E,1937PhRv...51..290Z}.
Because these images travel along different geometric paths and through different gravitational potentials to reach us, they arrive at different times, and in general they can be highly magnified \citep{1964MNRAS.128..295R}. 
Time delays between the multiple images of these ``strongly gravitationally lensed supernovae'' (\glsne) can be used to  measure the Hubble constant \h \citep{refsdal64}, which is currently in tension at the $3.7\sigma$ level \citep{2018arXiv180101120R}, independently of the local distance ladder and the assumed cosmological model \citep[e.g.,][]{2018arXiv180901274B}. 
If a \glsn is discovered before all of its images arrive, early moments of the supernova can be observed by anticipating the appearance of the remaining images \citep[e.g.,][]{gnkc18,2018MNRAS.474.2612S}.
These remarkable attributes make \glsne valuable probes of astrophysics and cosmology. 

To date, only two \glsne with resolved images have been discovered \citep{refsdal_discovery,goobar16}.
Neither has yielded competitive constraints on \h \citep[but see][]{suyu17,bonvin17,2018arXiv180201584G,2018ApJ...853L..31V},  nor  observations of the earliest moments of the supernova light curve. 
However, a new generation of high-cadence, wide-field imaging surveys, exemplified by the Zwicky Transient Facility (ZTF; 2018--2021; Graham et al., in preparation), the Large Synoptic Survey Telescope (LSST; 2021--2033; \citealt{lsst}), and the \textit{Wide-Field Infrared Survey Telescope} (\textit{WFIRST}; 2025--2031; \citealt{wfirst}) is expected to  yield thousands of \glsne\ over the next decade (\citealt{gn17,gnkc18}, see also \citealt{om10}). 
These surveys will cover enough of the sky, to sufficient depth, at a high enough cadence to produce the first statistical samples of \glsne. 
They will also be the first to employ novel detection techniques that will eliminate the need to resolve multiple images for \glsn discovery, furthering the yield. 
Finally, they will implement highly tuned \glsn filters that will lead to early discovery and minimization of false positives. 

To calibrate scientific expectations for the \glsn\ era, reliable forecasts of \glsn\ yields and properties are needed.
\cite{1987ApJ...314..154S} and \cite{1988ApJ...324..786L} carried out the first \glsn\ property forecasts, and  
\cite{1998MNRAS.296..763K},
\cite{2000MNRAS.319..549S},
\cite{2001ApJ...556L..71H},
\cite{dk06}, \cite{om10},
\cite{gn17}, \cite{gnkc18}, and \cite{2018arXiv180307569S} presented  refined calculations.
Each of these studies neglected to account for at least one of the following important effects: observing strategy and conditions, dust, discovery strategy, multiple supernova subtypes and rates, and the supernova-host galaxy connection.
In anticipation of the \glsn\ era, we present the first pixel-level Monte Carlo, ray-tracing, and image simulations of the \glsn population to include a detailed treatment of  these important effects and use them to forecast \glsn\ rates and properties.
In Section \ref{sec:methods}, we describe our models of the supernova, host galaxy,  deflector, and lens galaxy populations. 
In Section \ref{sec:results}, we present the results of our simulations, including \glsn yields and time delay, brightness, and image separation distributions. 
We discuss the implications of our results in Section \ref{sec:discussion} and conclude in Section \ref{sec:conclusion}.
In our calculations we assume a \cite{planck15} cosmology.

\section{Population Models}
\label{sec:methods}

In this section, we describe the  models of the deflector, lens galaxy, supernova, and host galaxy populations that we use to  forecast the rates and properties of \glsne from upcoming surveys.

\subsection{Deflectors}
\label{sec:lenspop}
Although galaxy clusters and late-type galaxies can act as gravitational lenses for background supernovae, we consider only elliptical galaxies as lenses in this analysis.
We model the projected mass distribution of elliptical galaxies as a Singular Isothermal Ellipsoid \cite[SIE;][]{kormann94}, which has shown excellent agreement with observations \citep[e.g.,][]{koopmans09}.
The SIE convergence $\kappa$ is given by:
\begin{equation}
\label{eq:sie}
\kappa(x,y) = \frac{\theta_{E}}{2}
	\frac{\lambda(e)}{\sqrt{(1-e)^{-1}x^2+(1-e)y^2}},
\end{equation}
where
\begin{equation}
\label{eq:einrad}
\theta_{E} = 4 \pi \left(\frac{\vd}{c}\right)^2\frac{D_{ls}}{D_s}.
\end{equation} 
In the above equations, $\vd$ is the velocity dispersion of the lens galaxy, $e$ is its ellipticity, and $\lambda(e)$ is its so-called ``dynamical normalization,'' a parameter related to three-dimensional shape, and $D_{ls}$ and $D_s$ are the angular diameter distances between the lens and the source and the observer and the source, respectively.
Here we make the simplifying assumption that there are an equal number of oblate and prolate galaxies, which \cite{chae03} showed implies $\lambda(e) \simeq 1$. 
We model the velocity distribution of elliptical galaxies as a modified Schechter function \citep{sheth03}:
\begin{equation}
\label{eq:schechter}
dn = \phi(\vd) d\vd = \phi_*\left(\frac{\vd}{\vd_*}\right)^\alpha \exp\left[-\left(\frac{\vd}{\vd_*}\right)^\beta\right]\frac{\beta}{\Gamma(\alpha/\beta)}\frac{d\vd}{\vd},
\end{equation}
where $\Gamma$ is the gamma function, and $dn$ is the differential number of galaxies per unit velocity dispersion per unit comoving volume.
Thus for the lens velocity dispersion distribution, we have:
\begin{equation}
    \vd \sim \phi(\vd).
\end{equation}
We adopt the parameter values \cite{choi07} derived  from SDSS:
$(\phi_*, \vd_*, \alpha, \beta) = (8 \times 10^{-3}~h^3~\mathrm{Mpc}^{-3}, 161~\kms,2.32, 2.67)$.
We assume the mass distribution and velocity function do not evolve with redshift, consistent with the results of \cite{chae07}, \cite{oguri08}, and \cite{bezanson11}. 
Following \cite{collett15}, we draw the lens ellipticity from a velocity dispersion-dependent Rayleigh density:
\begin{equation}
\label{eq:lensellip}
e|\vd \sim \frac{e}{s^2}\exp\left(-\frac{e^2}{s^2}\right),
\end{equation}
where the scale parameter $s=A+ B\vd$, and the fit values are $A=0.38$ and $B=5.7\times10^{-4}\:(\kms)^{-1}$.  
To exclude highly flattened mass profiles, we truncate the  distribution at $e = 0.8$. 
We assume the deflectors have a random orientation, thus for the position angle $\theta_e$ distribution, we have
\begin{equation}
    \theta_e \sim U[0, 2\pi].
\end{equation}

We simulate the effect of lensing by line of sight structures as an external shear term in the deflection potential  \citep[e.g.,][]{kochanek91, keeton97, witt97}.
The deflection potential $\psi$ of the external shear is given by
\begin{equation}
\label{eq:extpot}
\psi(x,y) = \frac{\gamma}{2}(x^2 - y^2)\cos 2 \theta_\gamma + \gamma x y \sin 2 \theta_\gamma,
\end{equation}
where $\gamma$ is the magnitude of the shear, and $\theta_\gamma$ describes its orientation in the image plane.
We assume the shear has a random orientation and a Rayleigh distribution in magnitude with scale parameter $s=0.05$ \citep{wong11}.
Thus the $\gamma$ distribution is
\begin{equation}
    \gamma \sim \frac{\gamma}{s^2}\exp\left(-\frac{\gamma^2}{s^2}\right),
\end{equation}
with $s=0.05$. As the shear orientation is assumed to be random, the $\theta_\gamma$ distribution is
\begin{equation}
    \theta_\gamma \sim U[0, 2\pi].
\end{equation}

The lens redshift distribution can be derived from Equation \ref{eq:schechter}, which gives  the differential number of galaxies per unit velocity dispersion per unit comoving volume.  
We begin with the definition of the comoving volume element,
\begin{equation}
	dV_C = D_H \frac{(1+z_l)^2 D_l^2}{E(z_l)}~dz_ld\Omega,
	\label{eq:comov}
\end{equation}
where $D_H = c / H_0$ is the Hubble distance, $E(z_l) = \sqrt{\Omega_M(1+z_l)^3 + \Omega_\Lambda}$ in our assumed cosmology, and $D_l$ is the angular diameter distance to the lens.
Since $dn = dN/dV_C$, we can combine Equation \ref{eq:schechter} with Equation \ref{eq:comov} to derive the unnormalized, all-sky $(d\Omega=4\pi)$ redshift and velocity dispersion function,
\begin{equation}
\label{eq:galaxydist}
\frac{dN}{d\vd dz_l} =  4\pi D_H \frac{(1+z_l)^2 D_l^2}{E(z_l)}\phi(\vd).
\end{equation}
As $\phi(\sigma)$ has no dependence on $z_l$ we can margialize \vd\ out of Equation \ref{eq:galaxydist} and drop constants to obtain an unnormalized density for $z_l$,
\begin{equation}
    \frac{dN}{dz_l} \propto \frac{(1 + z_l)^2 D_l^2}{E(z_l)}.
    \label{eq:propdist}
\end{equation}
We normalize Equation \ref{eq:propdist} by a constant,
\begin{equation}
    K = \int_{z_{l,\mathrm{min}}}^{z_{l,\mathrm{max}}} \frac{(1 + z_l)^2 D_l^2}{E(z_l)} \, dz_l,
    \label{eq:normfac}
\end{equation}
where $z_{l,\mathrm{min}}$ and $z_{l,\mathrm{max}}$ are the minimum and maximum lens redshifts considered in the simulation, respectively.
We  combine Equations \ref{eq:propdist} and \ref{eq:normfac} to obtain the probability density function for $z_l$, 
\begin{equation}
    \label{eq:galaxypdf}
    p(z_l) = \frac{1}{K} \frac{(1 + z_l)^2 D_l^2}{E(z_l)}.
\end{equation}
%Integrating Equation \ref{eq:schechter} over $z_l$ and $\sigma$ gives the joint probability density  function for $\vd$ and $z_l$:
%\begin{equation}
%\label{eq:galaxypdf}
%p(\vd, z_l) = \frac{1}{N_\mathrm{gal}} \frac{dN}{d\vd dz_l}.
%\end{equation}
%As Equations \ref{eq:galaxydist} and \ref{eq:galaxypdf} show, $z_l$ and $\vd$ are independent random variables, a result that allows us to draw them separately in our Monte Carlo simulations. 
Finally, as a matter of convention, we always take the SIE mass profile centroid coordinates $x_l$ and $y_l$ to be
\begin{align}
    x_l &= 0,\\
    y_l &= 0.
\end{align}
With sampling prescriptions for $e, \gamma, \theta_\gamma, \sigma, z_l, x_l, y_l,$ and $\theta_e$,  we can realize deflectors at random.

\subsection{Lens Galaxies}
\label{sec:lensgal}

We use the Fundamental Plane \citep{1987ApJ...313...59D}, a canonical relation between the mass, size, and brightness of elliptical galaxies, to assign light profiles to  lens galaxies.
Throughout this section, we assume the variables $e,\gamma,\theta_\gamma,\vd,z_l,x_l,y_l,$ and $\theta_e$ have already been sampled as described in Section \ref{sec:lenspop}.
As an ansatz, we model the lens galaxy light profiles as \sersic\ functions with $n=4$ \citep{sersic}.
Such profiles have shown excellent agreement with observations of ellipticals \citep{lackner12}. 
Section \ref{sec:hostpop} includes a more detailed discussion of \sersic\ functions, but for now it is only important that they are specified by seven parameters: an amplitude $I_e$, a size parameter $R_e$, a shape parameter $n$, a centroid position (here $x_l^\prime$ and $y_l^\prime$), an ellipticity (here $e^\prime$), and a position angle (here $\theta_e^\prime$).
The spectra of elliptical galaxies are remarkably uniform, with the primary feature being the break at 4000\AA\ (rest-frame).
Therefore, we model the SEDs of the lens light profiles using the one-component \texttt{Elliptical}\ template of \cite{kinney96}.
We assume that the ellipticities and position angles of the lens light profiles are the same as those of their corresponding mass profiles, i.e., that the light roughly traces mass.
Therefore, for the lens galaxy light profile position angle $\theta_e^\prime$, the lens galaxy light profile ellipticity $e^\prime$, and the lens galaxy light profile centroid coordinates $x_l^\prime$ and $y_l^\prime$, we have
\begin{align}
    \theta_e^\prime &= \theta_e,\\
    e^\prime &= e,\\
    x_l^\prime &= x_l,\\
    y_l^\prime &= y_l.
\end{align}
\cite{bernardi03} express the Fundamental Plane as a multivariate normal relationship between the velocity dispersion \vd, the surface brightness $\mu$, and the effective radius $R_e$, 
\begin{equation}
\label{eq:fp}
\begin{bmatrix}\mu\\ R\\ V\end{bmatrix} \sim \mathcal{N}\left(\begin{bmatrix}\mu_{*,c}\\ R_*\\ V_*\end{bmatrix}, \begin{bmatrix}
\sigma_{\mu}^2 & \sigma_R \sigma_\mu \rho_{R\mu} & \sigma_V \sigma_\mu \rho_{V\mu}\\
\sigma_R \sigma_\mu \rho_{R\mu} & \sigma_R^2 & \sigma_R \sigma_V \rho_{RV} \\
\sigma_V \sigma_\mu \rho_{V\mu} & \sigma_R \sigma_V \rho_{RV} & \sigma_V^2
\end{bmatrix}\right),
\end{equation}
where $V \equiv \log\left(\sigma/ [1 \,\kms]\right)$, $R \equiv \log\left(R_e / [1 \, h_{70}\, \km]\right)$, and $\mu_{*,c}$ is a $k$-corrected $\mu_*$ defined by a correction factor $Q$,
\begin{equation}
\mu_{*,c} = \mu_* - Q z_l.
\end{equation} 
Fitting the model to $i^*$-band photometry of a sample of roughly 9,000 early-type galaxies from SDSS, \cite{bernardi03} find $\sigma_\mu = 0.600$, $\mu_* = 19.40$, $R_* = 0.465$, $\sigma_R = 0.241$, $V_* = 2.201$, $\sigma_V = 0.110$, $\rho_{R\mu} = 0.753$, $\rho_{V\mu} = -0.001$, $\rho_{RV} = 0.542$, and $Q = 0.75$. 
We adopt these values in our simulations.

Using a conditioning identity for multivariate Gaussians,\footnote{\url{https://cs.nyu.edu/~roweis/notes/gaussid.pdf}, Equation 5d.} we can rewrite Equation \ref{eq:fp} to obtain the joint distribution of $\mu$ and $R$,
\begin{equation}
\label{eq:jointdist}
    \begin{bmatrix}\mu \\ R\end{bmatrix}|V \sim \mathcal{N}\left(\frac{V-V_*}{\sigma_V}
    \begin{bmatrix}
        \mu_{*,c} + \sigma_\mu \rho_{V\mu} \\
        R_* +  \sigma_R \rho_{RV}
    \end{bmatrix},
    \begin{bmatrix}
    \sigma_\mu^2(1-\rho_{V\mu}^2) & \sigma_R\sigma_\mu(\rho_{R\mu} - \rho_{RV}\rho_{V\mu})\\
     \sigma_R\sigma_\mu(\rho_{R\mu} - \rho_{RV}\rho_{V\mu}) & 
     \sigma_R^2(1-\rho_{RV}^2)
    \end{bmatrix}
    \right).
\end{equation}
Using Equation \ref{eq:jointdist}, we sample $\mu,R$ pairs given the velocity dispersion \vd. 
We then convert $\mu$ into an $i$-band apparent AB magnitude $m_i$ using the following relation from \cite{bernardi03},
\begin{equation} 
m_i = \mu - 5 \log\left(\frac{R_e / D_l}{1\prime\prime}\right)- 2.5 \log(2\pi) + 10 \log (1 + z_l).
\end{equation}
We then linearly rescale the flux of the \texttt{Elliptical} template so that its $i$-band apparent magnitude is $m_i$. 
We assume that the spectrum of the galaxy is spatially constant, so $m_i$ also fixes $I_e$. 
With the results of Section \ref{sec:lenspop} and sampling prescriptions for $m_i,R_e,\theta_e^\prime,e^\prime,x_l^\prime,$ and $y_l^\prime$, we can realize lens galaxy light profiles at random. 

%\begin{equation}
%p(\mu, R | V) = \mathcal{N}\left(\begin{bmatrix} \mu_{*,c} + \sigma_V\sigma_\mu \rho_{V\mu} \\ R_* + \sigma_V \sigma_R \rho_{VR}\end{bmatrix}
%\frac{V-V_*}{\sigma_V^2}, \begin{bmatrix} \sigma_{\mu}^2 & \sigma_R \sigma_\mu \rho_{R\mu}\\
%\sigma_R \sigma_\mu \rho_{R\mu} & \sigma_R^2 \\ \end{bmatrix} \frac{1}{\sigma_V^2}\right)
%\end{equation}

In our model of the lens galaxy population, we neglect microlensing by lens galaxy stars. 
Studies have shown that microlensing can cause significant errors when using \glsne\ to measure time delays \citep{dk06,gnkc18} or constrain mass models \citep{2018arXiv180207738F}.
However, the effect microlensing on  \glsn\ yields has been shown to be small \citep{gnkc18}.

\subsection{Supernovae}
\label{sec:supernova}

\begin{deluxetable}{cccccc}
  \tablecaption{Details of the supernova population model.
    Magnitudes are given in the Vega system.\label{tab:snparams}}
\tablehead{
\colhead{SN Type} & \colhead{$\mu_{M_B}$} & \colhead{$\sigma_{M_B}$} & \colhead{Template} & \colhead{Template Reference} & \colhead{Luminosity \& Rate Reference}}
\startdata
IIP & $-16.9$ & $1.12$ & SN 2005lc & \cite{sako11} & \cite{2011MNRAS.412.1441L}\\
91bg & $-17.15$ & $0.2$ & Nugent-91bg & \cite{nugent02}  & \cite{2006ApJ...648..868S}\\
Ia & $-19.23$ & $0.1$ & Hsiao v3.0 & \cite{hsiao} & \cite{2006ApJ...648..868S}\\
91T & $-19.3$ & $0.2$ & Nugent-91T & \cite{nugent02} &  \cite{2006ApJ...648..868S}\\
IIL & $-17.46$ & $0.38$ & Nugent-IIL & \cite{gilliland99} &  \cite{2011MNRAS.412.1441L}\\
IIn & $-19.05$ & $0.5$ & Nugent-IIn & \cite{gilliland99} &  \cite{2011MNRAS.412.1441L}\\
Ibc & $-17.51$ & $0.74$ & Nugent-Ibc & \cite{levan05} & \cite{2011MNRAS.412.1441L}
\enddata
\end{deluxetable}
We consider seven different supernova subtypes in this analysis: Type Ia, Type IIP, Type IIn, Type IIL, Type Ib/c, SN 1991bg-like, and SN 1991T-like supernovae. 
Type Ia, SN 1991bg-like, and SN 1991T-like supernovae are believed to result from the thermonuclear explosions of white dwarfs \citep{maozreview}, whereas Type IIP, Type IIL, Type Ib/c, and Type IIn supernovae result from core-collapse in massive stars.
Our model of the supernova population is characterized by two global parameters for each supernova subtype: a mean peak rest-frame $B$-band absolute magnitude in the Vega system $\mu_{M_B}$, and the scatter in this magnitude $\sigma_{M_B}$. 
Throughout this section, we assume that deflector and lens galaxy parameters have already been sampled as described in Sections \ref{sec:lenspop} and \ref{sec:lensgal}.
For each supernova in the simulation, we realize a peak rest-frame $B$-band absolute magnitude $M_B$ according to 
\begin{equation}
M_B \sim \mathcal{N}(\mu_{M_B}, \sigma_{M_B}). 
\end{equation}
In our Monte Carlo simulation, we randomly draw the unlensed angular position of each supernova  uniformly over a circular area of angular radius $\theta_l$ centered on the lens galaxy.
Using another Monte Carlo simulation, we found that in more than 99.9\% of cases, multiply imaged point sources had unlensed positions within 0.9$\theta_E$ of the SIE centroid.
Therefore, in the present calculations, we set $\theta_l = 0.9\theta_E$, where $\theta_E$ is the lens's angular Einstein radius, which can be calculated via Equation \ref{eq:einrad}.
To realize random supernova positions uniformly over this area we first draw two random deviates from the uniform distribution,
\begin{align}
    r &\sim U[0,1],\\
    \theta &\sim U[0, 2\pi],
\end{align}
then convert these into lens-centered Euclidean angular coordinates $x_s$ and $y_s$ via
\begin{align}
    x_s &= \theta_l\sqrt{r}\cos{\theta},\\
    y_s &= \theta_l\sqrt{r}\sin{\theta}.
\end{align}
This ensures that supernovae are realized uniformly over each lens's area of influence.
\begin{figure}
    \centering
    \includegraphics[width=1\textwidth]{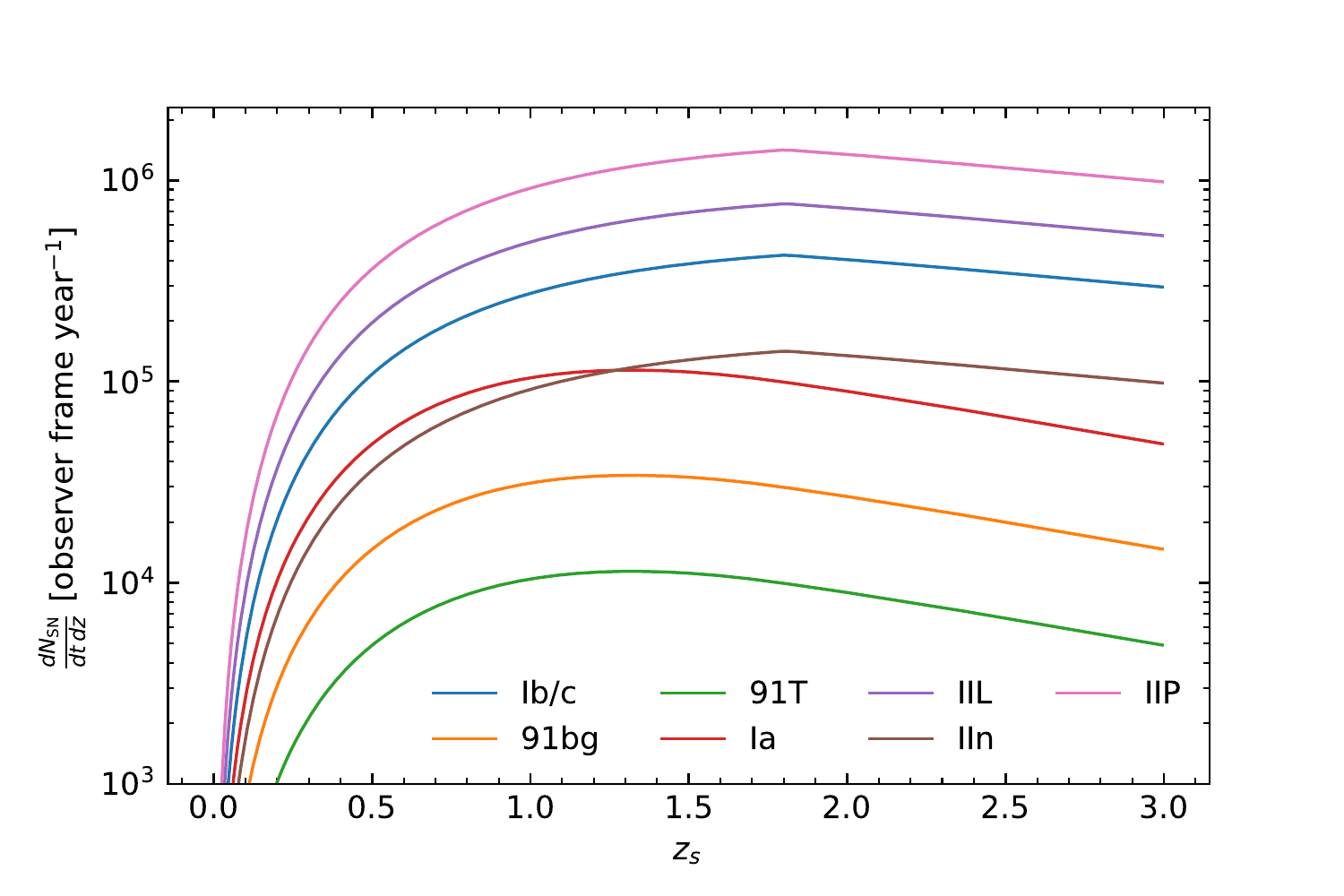}
    \caption{All-sky supernova rates as a function of redshift (observer-frame). 
    In our simulations, supernova redshifts are realized at random from these distributions. 
    The references in Table \ref{tab:snparams} provide the data sources of these curves.}
    \label{fig:snrate}
\end{figure}
We draw a redshift for each supernova from the functions $f_T(z_s)$ shown in Figure \ref{fig:snrate}.
%In our simulations, we sample one supernova per lens.
The normalized Figure \ref{fig:snrate} curves $S_T(z_s)$ give the redshift probability density function for  supernova type $T$, 
\begin{equation}
    \label{eq:snpdf}
    p(z_s) = S_T(z_s),
\end{equation}
where $z_s$ is the source redshift. 
For each supernova subtype, we assume that the spectral evolution is described by a template with one parameter (the overall normalization), and we use the realized $M_B$ to set its value assuming the \cite{planck15} cosmology described in Section \ref{sec:intro}.
With sampling prescriptions for $M_B,z_s,x_s,$ and $y_s$,  we can realize supernovae at random.
Table \ref{tab:snparams} lists the references for our supernova templates, rates, and luminosity functions.

\subsection{Host Galaxies}
\label{sec:hostpop}
The connection between supernovae and their host galaxies is of critical importance to time delay cosmology with \glsne, as lensed host galaxy arcs will provide significant leverage on lens models \citep[e.g.,][]{suyu17}. 
Here we describe an empirical model of the supernova-host galaxy connection that we use to realize hosts for each supernova in our simulation. Throughout this section, we assume that deflector, lens galaxy, and supernova parameters have already been sampled as described in Sections \ref{sec:lenspop}, \ref{sec:lensgal}, and \ref{sec:supernova}.
We consider three types of host galaxies: elliptical galaxies, which have almost no ongoing star formation, S0/a-Sb galaxies, which have a moderate level of ongoing star formation, and late-type/sprial galaxies, which have vigorous ongoing star formation.
As an ansatz, we take the light profiles of the host galaxies in the absence of lensing to be \sersic\ functions with $n=\{1,1,4\}$, respectively.
Only normal \sneia\ and SN1991bg-like events have been observed to be hosted by elliptical or S0/a-Sb galaxies. 
Based on measured rates, we assume these two subclasses of thermonuclear supernovae have a 30\% chance of being hosted by an elliptical, a 35\% chance of being hosted by a S0/a-Sb, and a 35\% chance of being hosted by a late-type/spiral, roughly consistent with the results of \cite{2010ApJ...724..502H}, 
\cite{2011MNRAS.412.1441L},
\cite{2012A&A...544A..81H},  and \cite{2012ApJ...755...61S}. 
In our simulations, Type Ib/c, Type IIP, Type IIL, Type IIn, and SN 1991T-like supernovae can only be hosted by late-type/spiral galaxies.
For simplicity, we assume the spectra of the host galaxies are given by the following \cite{kinney96} templates: \texttt{Elliptical}\ (elliptical), \texttt{Sc}\ (S0/a-Sb), and \texttt{Starburst}\ (late-type/spiral).

\begin{figure}
	\centering
    \includegraphics[width=1\columnwidth]{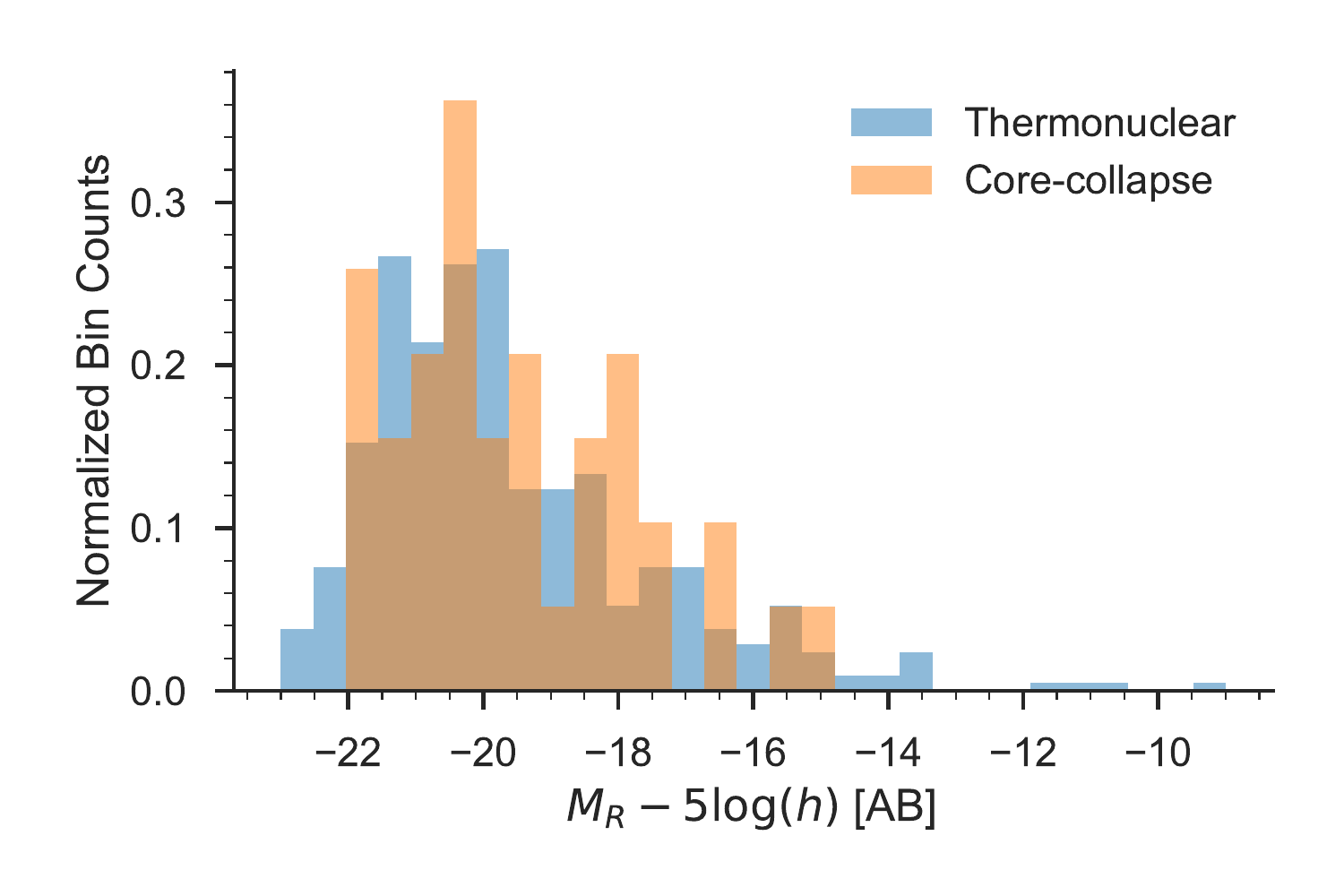}
    \caption{Host galaxy luminosity functions used in our simulations.}
    \label{fig:lumfunc}
\end{figure}

We draw the host galaxy luminosities from two separate luminosity functions: one for the hosts of thermonuclear supernovae (\sneia, SN 1991bg-like, and SN 1991T-like events) and one for the hosts of core-collapse supernovae.
We construct both of our luminosity functions using supernovae discovered by the Palomar Transient Factory \citep[PTF;][]{2009PASP..121.1395L}.
PTF discovered thousands of supernovae to $z\sim0.1$ and obtained spectral confirmation of many of them in a relatively unbiased manner. 
For the core-collapse supernovae, we draw the cosmology-independent host galaxy rest-frame $R$-band absolute magnitude $M_R - 5\log h$ at random from the sample of \cite{2010ApJ...721..777A} confined to $0.01 \leq z \leq 0.05$ to limit the effects of peculiar velocities and to ensure a complete sample. 
For the thermonuclear events, we use a catalog compiled by E. Y. Hsiao and P. E. Nugent (private communication) drawn from the PTF discoveries that overlapped with fields observed by SDSS and BOSS.

%The sample consists of 
%For simplicity, and due to the lack of readily available data, we construct a host galaxy luminosity function for each supernova type but do not further refine by host galaxy type.
%This may be problematic for the thermonuclear events, which can occur in three different types of galaxies. 
Figure \ref{fig:lumfunc} shows the luminosity functions of core-collapse and thermonuclear supernova host galaxies used in the present calculations.
The host galaxy redshift $z_s^\prime$ is fixed to the redshift of the supernova, 
\begin{equation}
    z_s^\prime = z_s.
\end{equation}
The sampled values of $M_R - 5\log h$ and $z_s^\prime$ fix the normalization of the host galaxy spectral template and the host galaxy \sersic\ profile amplitude under the assumption of a \cite{planck15} cosmology. 

Following \cite{2003MNRAS.343..978S}, we take the sizes and intrinsic brightnesses of galaxies to be correlated via the ``size-luminosity relation,''
\begin{equation}
\label{eq:hostreprob}
p(\log R_e^\prime|M_R) = \mathcal{N}(\log\bar{R}_e, \sigma_{\log R_e}),
\end{equation}
where $R_e^\prime$ is the effective radius of the host galaxy \sersic\ profile and $\bar{R}_e$ and $\sigma_{\log{R_e}}$ are global parameters. 
\cite{2003MNRAS.343..978S} find that for elliptical galaxies, $R_e^\prime$ is related to  $M_R$ via
\begin{equation}
\log\left(\frac{\bar{R}_e}{1\,\mathrm{kpc}}\right) = -0.4 a M_{R,c} + b,
\end{equation}
where 
\begin{equation}
M_{R,c} = M_R + 5 \log\left(\frac{0.7}{h}\right).
\end{equation}
Fitting to data from SDSS, \cite{2003MNRAS.343..978S} find $a=0.65$ and $b=-5.06$. 
For S0/a-b and late-type/spiral galaxies, they find
\begin{equation}
\log\left(\frac{\bar{R}_e}{1\,\mathrm{kpc}}\right) = -0.4 \alpha  M_{R,c} + (\beta - \alpha) \log\left[1 + 10^{-0.4(M_{R,c} - M_0)}\right]+ \gamma,
\end{equation}
where fitting the SDSS data give $\alpha=0.26$, $\beta=0.51$, $\gamma = -1.71$, and $M_0 = -20.91$. 
The dispersion in the size-luminosity relation is given by
\begin{equation}
\sigma_{\log R_e} = \sigma_2 + \frac{(\sigma_1 - \sigma_2)}{1 + 10^{-0.8(M_{R,c} - M_0)}},
\end{equation}
for all galaxy types, with $\sigma_1 =0.45$ and $\sigma_2 = 0.27$. 
Having calculated $\sigma_{\log R_e}$ and $\bar{R}_e$ given $M_R$, we can sample a value of $\log R_e^\prime$ using Equation \ref{eq:hostreprob}. 

The next steps are to draw the host galaxy ellipticity $e^{\prime\prime}$ and position angle $\theta_e^{\prime\prime}$. 
We take the  host galaxy orientation to be random, 
\begin{equation}
    \theta_e^{\prime\prime} \sim U[0, 2\pi],
\end{equation}
and to draw ellipticities, we use the results of the Cosmic Evolution Survey \citep[COSMOS;][]{2007ApJS..172....1S}.
COSMOS is a survey designed to probe the correlated evolution of galaxies, star formation, active galactic nuclei, and dark matter with large-scale structure. 
Our access point to COSMOS is the Advanced Camera for Surveys General Catalog \citep[ACS-GC;][]{2012ApJS..200....9G}.
ACS-GC is a photometric and morphological database
containing fits of structural parameters to publicly available data obtained with the Advanced
Camera for Surveys (ACS) instrument aboard \hst.
The catalog was created using the code \texttt{Galapagos} \citep{2007ApJS..172..615H,2011ASPC..442..155H}, which incorporates the source extraction and photometry software SExtractor \citep{1996A&AS..117..393B} and the galaxy light profile fitting algorithm GALFIT \citep{2002AJ....124..266P}.
ACS-GC contains photometry and structural parameters for approximately 305,000 objects (both compact and extended) from COSMOS. 
The COSMOS images were taken with the Wide Field Camera  (WFC) on ACS, through the F814W filter, a broad $i$-band filter spanning the wavelength range 7000 -- 9600\AA, with a scale of 0.05 arcsec pixel$^{-1}$ and a resolution of 0.09$^{\prime\prime}$ FWHM.

We apply the cuts of \cite{2016AJ....152..154G} to create a list of potential supernova host galaxies from the ACS-GC. 
We further subdivide this list into two groups: ``early'' and ``late''-type galaxies, having fitted values of the \sersic\ index in the ACS-GC of $n > 2.5$ and $n \leq 2.5$, respectively.
For elliptical hosts, we draw $e^{\prime\prime}$ at random from the fitted ellipticity values of the ``early'' group, and for S0/a-b and late-type/spiral hosts, we draw $e^{\prime\prime}$ at random from the fitted ellipticity values of the ``late''-type group. 
%Finally, we draw at random a position for the supernova within its host galaxy (in the absence of lensing). 

The last parameters to draw are the unlensed coordinates of the host galaxy centroid $x_h$ and $y_h$. 
Here we take the PDF of supernova positions within the host galaxy to be directly proportional to the light profile, an assumption that has been borne out by observational studies that show supernova positions follow host light \citep{2012ApJ...759..107K}.
Thus we sample offsets $\Delta x$ and $\Delta y$ at random from the host galaxy light profile, then  take 
\begin{align}
    x_h &= x_s - \Delta x,\\
    y_h &= y_s - \Delta y.
\end{align}
The host galaxy light profiles follow a \sersic\ function, defined as,
\begin{equation}
    I(r_c) = I_e \exp\left\{-b_n \left[\left(\frac{r_c}{R_e}\right)^\frac{1}{n} - 1\right]\right\},
    \label{eq:sersic}
\end{equation}
where $r_c$ is an ellipticity-free, host galaxy-centered radial coordinate and $b_n$ is a constant scalar solution to the equation 
\begin{equation}
    \gamma(2n; b_n) = \frac{1}{2}\Gamma(2n),
\end{equation}
in which $\Gamma$ is the Gamma function and $\gamma$ is the incomplete Gamma function.\footnote{An exact, computationally inexpensive method of calculating $b_n$ for a given value of $n$ is to evaluate $\mathtt{gammaincinv(2*n, 0.5)}$ in \texttt{scipy}. \label{fn:bn}}
To sample a position at random from the surface brightness profile we first draw two random deviates $z$ and $\theta^\prime$ uniformly,  
\begin{align}
    z &\sim U[0,1],\\
    \theta^\prime &\sim U[0, 2\pi].
\end{align}
Using the sampled $z$, we solve the following equation\footnote{See Footnote \ref{fn:bn}, but with the substitutions $b_n \rightarrow x$ and $\mathtt{gammaincinv(2*n, 0.5)} \rightarrow \mathtt{gammaincinv(2*n, z)}$.} for $x$:
\begin{equation}
    \gamma(2n; x) = z \Gamma(2n),
\end{equation}
then convert $x$ into the radial coordinate $r_c$ \citep[see, e.g.,][]{2005PASA...22..118G},
\begin{equation}
    r_c = R_e^\prime \left(\frac{x}{b_n}\right)^n.
\end{equation}
We can now write the ellipticity-free host offsets $\Delta x_c$ and $\Delta y_c$ as 
\begin{align}
    \Delta x_c &= r_c \cos\theta,\\
    \Delta y_c &= r_c \sin\theta.
\end{align}
We add ellipticity to obtain $\Delta x_e$ and $\Delta y_e$,
\begin{align}
    \Delta x_e &= \Delta x_c \sqrt{1 - e},\\
    \Delta y_e &= \Delta y_c / \sqrt{1 - e}.
\end{align}
Finally, we account for the position angle of the host galaxy $\theta_e^{\prime\prime}$ by applying a rotation matrix,
\begin{equation}
    \begin{pmatrix}
        \Delta x \\ \Delta y
    \end{pmatrix} = 
    \begin{pmatrix}
        \cos\theta_e^{\prime\prime} & \sin\theta_e^{\prime\prime}\\
        -\sin\theta_e^{\prime\prime} & \cos\theta_e^{\prime\prime}
    \end{pmatrix}^{-1}
    \begin{pmatrix}
    \Delta x_e \\ \Delta y_e
    \end{pmatrix}.
\end{equation}
With  sampling prescriptions for $M_R,R_e^\prime,\theta_e^{\prime\prime},e^{\prime\prime},x_h,$ and $y_h$, we can realize host galaxy light profiles at random. 

\subsection{Sky Distribution}
\label{sec:skydist}

We assign a sky location to each system realized in our simulation, which in turn determines the sampling, signal-to-noise ratio, and filters of its simulated photometry. 
The sky location  also controls the amount of Milky Way dust extinction each system experiences (see Section \ref{sec:extinction}).
To randomly assign a sky position to a \glsn\ system, we  draw two random deviates $u$ and $v$ uniformly,
\begin{align}
    u &\sim U[0,1],\\
    v &\sim U[0,1].
\end{align}
We then convert these to equatorial coordinates $\alpha$ (right ascension) and $\delta$ (declination) via
\begin{equation}
\label{eq:ra}
\delta = \frac{180^\circ \times \arccos(2v -1)}{\pi} - 90^\circ
\end{equation}
and
\begin{equation}
\label{eq:dec}
\alpha = 360^\circ \times u.
\end{equation}
This sampling prescription ensures that systems are distributed uniformly over the celestial sphere.

\subsection{Extinction}
\label{sec:extinction}

\begin{figure}
\centering
\includegraphics[width=1\textwidth]{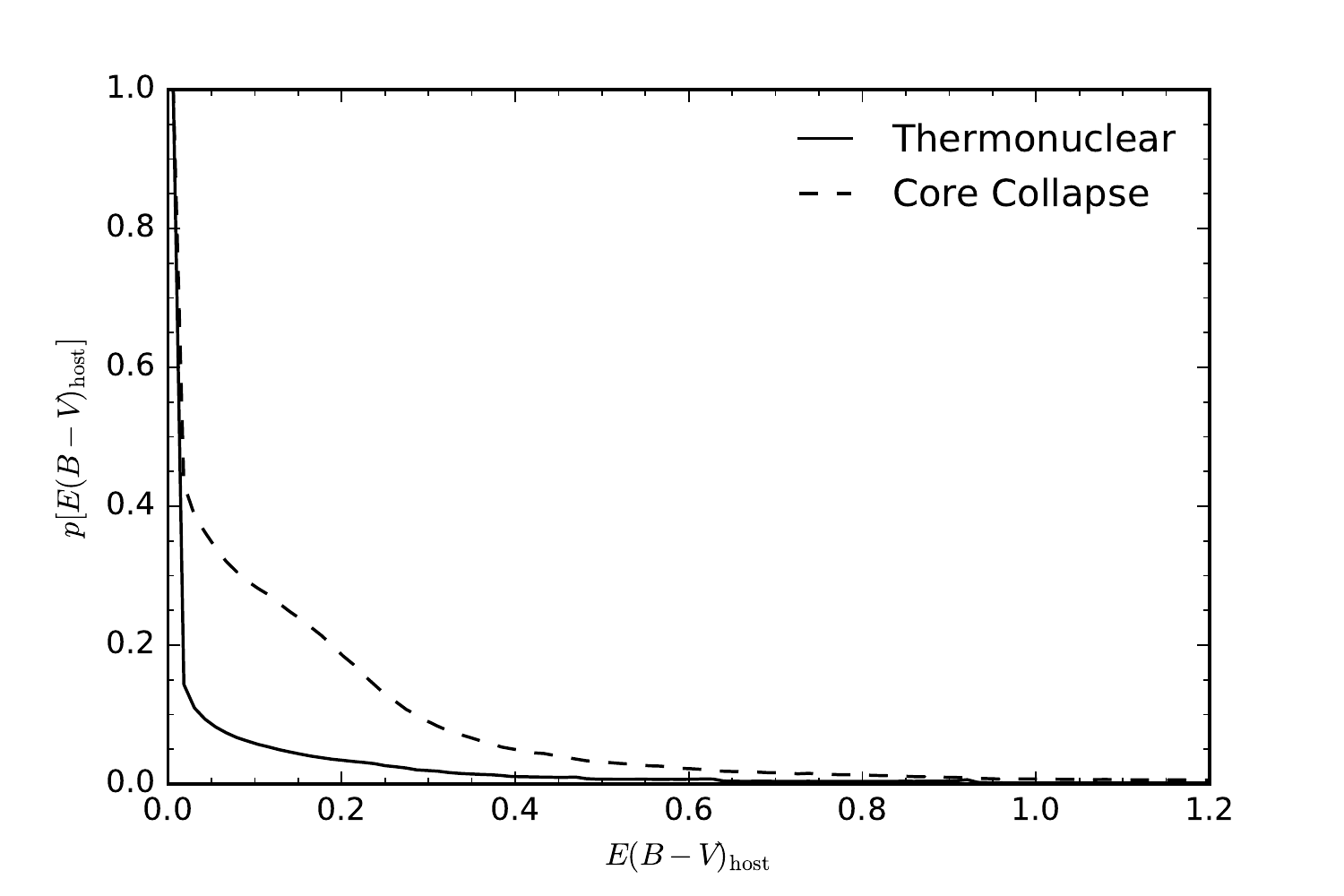}
\caption{$E(B-V)_\mathrm{host}$ distributions for galaxies at random orientations, from \cite{1998ApJ...502..177H}. 
    Host reddenings for Type Ia, SN 1991T-like, and SN 1991bg-like supernovae are drawn from the thermonuclear curve.
    Host reddenings for Type IIP, IIL, IIn, and Ib/c supernovae are drawn from the core-collapse curve.}
\label{fig:ebv}
\end{figure}

After randomly assigning a sky location to each \glsn\ system, we use the extinction maps of \cite{1998ApJ...500..525S} to calculate the associated Milky Way reddening value, $E(B-V)_\mathrm{MW}$. 
We then apply the extinction to the observer-frame spectral time series of the supernova images using a \cite{1989ApJ...345..245C} reddening law with $R_V=3.1$. 
In addition to extinction by dust in the Milky Way, \glsne\ can suffer extinction by dust in their host galaxies. 
Here we assume the host galaxy reddening $E(B-V)_\mathrm{host}$ is distributed according to the thermonuclear and core-collapse extinction distributions of \cite{1998ApJ...502..177H} for galaxies at random orientations, shown in Figure \ref{fig:ebv}.
We apply host extinction to the rest-frame spectral time series of the supernova images using a \cite{1989ApJ...345..245C} reddening law with $R_V=3.1$, the measured Galactic value. \cite{2015MNRAS.453.3300A} showed that there is significant diversity in the value of $R_V$ for the observed host galaxy extinction in Type Ia supernovae and similar conclusions were reached for certain types of core-collapse SNe in \citet{2018A&A...609A.135S}. In particular, lower values of $R_V$ are often found, \citep[see][for a proposed explanation]{2018MNRAS.479.3663B}. By selecting a value of $R_V$ on the upper range observed, we are assuming a relatively large attenuation by dust, $A_V = R_V\cdot E(B-V)$, i.e., a conservative estimate of the SN brightness.
We neglect of extinction by dust in the lens galaxies, which may reduce yields by making lensed images fainter.
SN iPTF16geu showed evidence of extinction due to lens galaxy dust at sub-kpc offsets \citep{goobar16}, but with only one event the frequency and spatial distribution of lens galaxy dust remain unclear.
ZTF and LSST will be able to better constrain lens galaxy dust extinction   by producing large samples of \glsne~Ia.

\subsection{Simulated Surveys}
\label{sec:surveysim}

To simulate realistic light curves and pixel cutouts of our lens systems as they would appear in a  survey, we must account for the survey's unique observing strategy and conditions, instrumental properties, and  visit schedule.
To do this, we use the outputs of software tools that run survey simulations with given science driven desirables; a software model of the telescope and its control system; and models of weather and other environmental variables. 
Such  simulations produce observation histories, which are  records of times, pointings and associated environmental data and telescope activities throughout a simulated survey. 
These histories can be examined to assess whether a simulated survey would be useful for any particular purpose or interest.
We adopt a common format for survey observation histories, consisting of a table with the following columns:
\begin{enumerate}
	\item \texttt{field}: The field ID of the observation. 
    \item \texttt{filter}: The filter in which the observation was taken.
    \item \texttt{time}: The MJD at which the observation began (the shutter-open time).
    \item \texttt{exptime}: The integration time of the exposure.
    \item \texttt{sky\_counts\_per\_pixel}: The sky counts (in electrons) in each pixel. 
	This is not a count rate, but the counts integrated over the entire exposure. 
    This column can optionally also include counts due to other spatially uniform Poisson backgrounds, such as dark current. 
    \item \texttt{psf\_sigma}: The standard deviation (in arcseconds) of the PSF, modeled as a Gaussian.
    \item \texttt{ra}: Right ascension of the center of the pointing.
    \item \texttt{dec}: Declination of the center of the pointing. 
    \item \texttt{night}\ (optional): An integer ID specifying the night of the survey in which the observation was taken, used for grouping and stacking observations. 
\end{enumerate}
In addition to the observation histories, we specify instrumental properties with the following parameters:
\begin{enumerate}
	\item \texttt{pix\_scale}: The plate scale of the camera (arcsec / pixel). 
    \item \texttt{read\_noise}: The read noise of the camera, in electrons. 
    \item \texttt{field\_of\_view}: The field of view of the imager, in deg$^2$. 
    \item \texttt{collecting\_area}: The collecting area of the telescope, in cm$^2$.
\end{enumerate}
In this work, we consider two surveys: ZTF and LSST, the two largest imaging surveys at optical wavelengths during  the periods 2018--2021 and 2021--2032, respectively.
In the following subsections, we describe these surveys and the operations simulations that we use to realize their data. 

\subsubsection{The Zwicky Transient Facility}
\label{sec:ztf}
The Zwicky Transient Facility (ZTF) is an ongoing time-domain imaging survey observing a minimum of 15,000 deg$^2$ in $g$ and $r$-band  ($\delta>-30$ deg) every 3 nights to a depth of at least 20.5 mag, with transient alerts  released in real-time to the public.\footnote{Public alerts can be retrieved from \url{http://ztf.uw.edu}.}
In March 2018, ZTF began science operations, replacing its predecessor, the intermediate Palomar Transient Factory (iPTF), on the 1.2-meter Oschin-Schmidt telescope (P48) at Palomar Observatory near San Diego, California. 
The chief advance of ZTF over iPTF is a new wide-field camera developed for the survey (Smith et al., in preparation).
With its 47 deg$^2$ field of view, the ZTF camera can survey 3,750 deg$^2$ per hour to $g,r\approx20.5$, making it roughly an order of magnitude faster than iPTF.
In addition to the 15,000 deg$^2$ public survey, a subset of 1,600 deg$^2$ is currently monitored six times per night in two filters as a part of the ZTF partnership survey.
Half of the survey area is also monitored in $i$-band every 4 nights.
The remaining  20\% of the survey time is allocated to proposals from collaboration members affiliated with the California Institute of Technology (Caltech) on a competitive basis.
We simulate data from all three ZTF programs in the present work using the simulated ZTF survey of Bellm et al. (in preparation), which uses the same scheduler  as the actual survey.
The scheduler uses Gurobi optimization,\footnote{\url{http://www.gurobi.com/}} a technique for integer programming, to maximize the number of images, weighted by the volume surveyed per image, observed in acceptable cadence windows, while maintaining a balance between the public, Partnership, and Caltech surveys. 
While the observing sequence determined by the scheduler in the simulation is reliable, the observing conditions used  by the simulation are overly optimistic, predicting limiting magnitudes $\sim$21.5 in all filters. 
In reality, ZTF can only reach a limiting magnitude of 20.5 in any filter in a 30-second exposure.
Therefore, in our simulation, we  set the seeing FWHM to 2$^{\prime\prime}$, the survey median, and the limiting magnitude $(5\sigma)$ to 20.5 for all observations. 

\subsubsection{The Large Synoptic Survey Telescope}
\label{sec:lsst}
The Large Synoptic Survey Telescope  is a planned imaging experiment that will  conduct at least two interleaved surveys: a ``wide-fast-deep'' (WFD) survey covering roughly 20,000 deg$^2$ in $ugrizy$ every 2--3 weeks with 30 second exposures $(r_\mathrm{lim} \sim 24)$, and a ``deep drilling'' survey covering a  smaller area at a significantly higher cadence \citep{LSSTObservingStrategyWhitePaper}. 
A new 8m-class telescope and camera with a 9.6 deg$^2$ field-of-view and 0.2$^{\prime\prime}$ pixels, located on the Cerro Pach\'on ridge in northern Chile, are currently under construction to carry out the survey. 
First light and commissioning operations will begin in 2021, followed by science operations in 2022. 
The survey will collect data for 10 years. 

Several detailed candidate observing strategies have been proposed for LSST.
In this analysis we evaluate two of the major ones from the perspective of \glsn\ science: a nominal  observing strategy, known as \texttt{minion\_1016}, and a leading alternative, known as \texttt{altsched}. 
\minion\ divides its time between five interleaved surveys: a ``Universal'' WFD survey (85.1\%), a proposal to monitor the North Ecliptic Spur (6.5\%), a proposal to monitor the Galactic plane (1.7\%), a proposal to monitor the South Celestial Pole (2.2\%), and a proposal to monitor 5 9.6 deg$^2$ ``deep-drilling'' fields (4.5\%).  
The median effective seeing (FWHM) for all proposals in $r$-band is 0.93$^{\prime\prime}$. 
The median single-visit depths for the WFD fields are (23.14, 24.47, 24.16, 23.40, 22.23, 21.57) in the $ugrizy$ bands.

The \minion\ simulation was performed using the software tool \texttt{OpSim} \citep{2014SPIE.9150E..15D}.
\texttt{OpSim}\ uses a greedy algorithm that chooses the best observation at a given time (according to a merit function based on the input science goals), with no look-ahead or long-term strategy. 
\altsched, on the other hand, takes a simpler approach, following a pre-programmed path with no merit function. 
\altsched\ attempts to observe fields at low airmass by observing only on the meridian, optimizing the signal-to-noise ratio (SNR) of the observations. 
Like \minion, \altsched\ retains a dual-visit per night requirement for transient artifact rejection and asteroid orbit linkage, but the two visits are taken in different filters, so colors can be obtained on all objects.
\altsched\ simulations of \snia\ light curves have shown the alternative cadence can lead to significantly better light curve sampling than \minion.
In Section \ref{sec:discussion}, we evaluate both \minion\ and \altsched\ for  the LSST \glsn\ science case.

\subsection{Imaging, Photometry, and Calibration}
\label{sec:phot}
To realize images and photometry of our simulated \glsn\ systems as they would appear in the mock surveys described in Section \ref{sec:surveysim}, we have developed an image-simulation pipeline based on the open-source astronomical image simulation code \galsim\  \citep{2015A&C....10..121R} and the gravitational lensing code \glafic\ \citep{glafic}.
For a given arrangement of supernova, host galaxy, and lens, we first solve the lens equation using \glafic\ to determine the magnifications, time delays, multiplicities, and locations of the lensed supernova images. 
We then use \glafic\ to solve the lens equation again for the magnification and surface brightness profile of the lensed host galaxy.
With this information, we use \galsim\ to model the entire system.
In \galsim\ parlance, we model each lensed supernova image as a \texttt{DeltaFunction}, the lens galaxy as a \texttt{Sersic}, and the lensed host galaxy surface brightness profile as an \texttt{InterpolatedImage}.
We convolve the model with a \texttt{Gaussian} model of the PSF, the width of which is provided by the survey simulation (see Section \ref{sec:surveysim}).
We refer to the noiseless convolved model as $I(x,y)$ and the pixel values of the corresponding  model image as $I_{xy}$. 
To generate an image for viewing, we add \texttt{CCDNoise} to the model consisting of Gaussian read noise, Poisson sky background, and Poisson source noise.

\begin{figure}
	\centering
    \vspace{5mm}
    \includegraphics[width=1\textwidth]{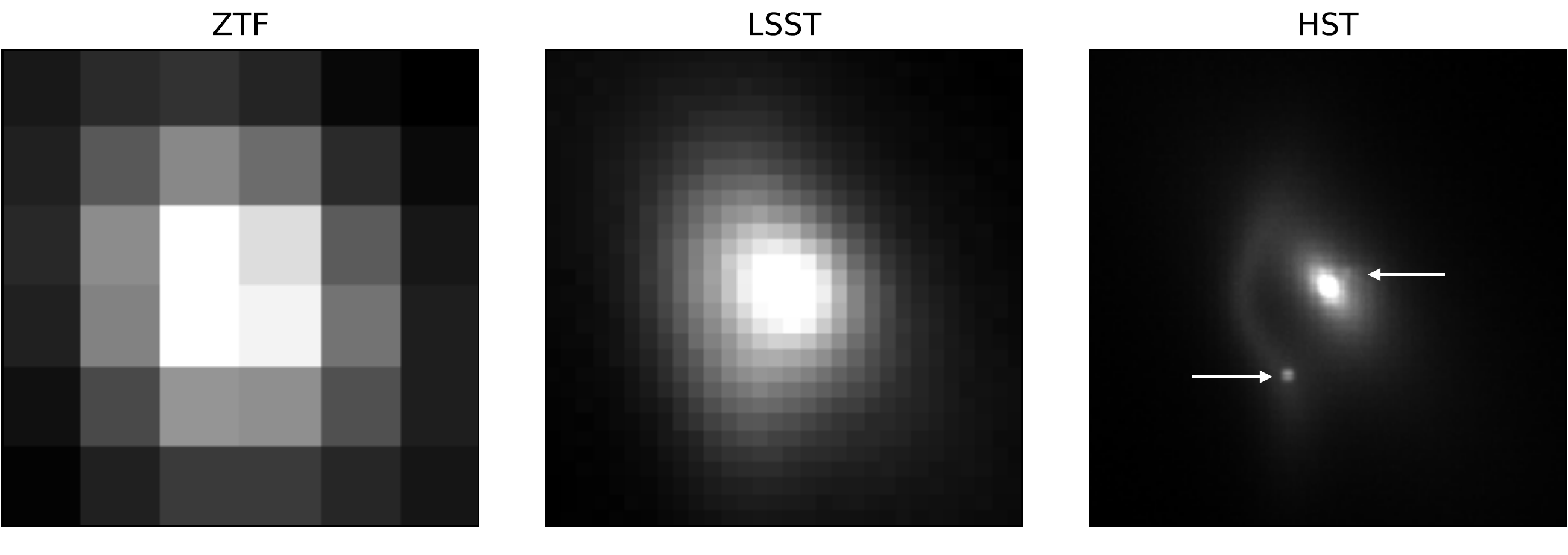}
    \caption{Simulated $r$-band images of the same \glsn, taken at the same epoch, with three different instruments: ZTF (30 second integration), LSST (30 second integration), and \hst\ (1 orbit integration through F625W on WFC3).
    Each panel is  $6^{\prime\prime}\times6^{\prime\prime}$.
    Only in the \hst\ data can the resolved images of the transient be clearly seen; they are marked with arrows.
    ZTF and LSST will be unable to resolve the multiple images of most \glsne, meaning high-resolution follow-up observations will be critical for lens modeling and time delay extraction.}
	\label{fig:imaging}
\end{figure}

We perform photometry using a matched filter, following \cite{2009AnApS...3....6B}.
We assume we have a filter $w_{xy}$ that perfectly matches the shape of the source and is normalized to 1, i.e., $w_{xy} = I_{xy} / \sum I_{xy}$.
We calculate the measured signal as a weighted sum of the image and the filter, via
\begin{equation}
S = \sum_{x,y} w_{xy} I_{xy},
\end{equation}
and we define the noise as the square-root of the signal variance,
\begin{equation}
N = \sqrt{\mathrm{Var}(S)} = \left[\sum_{x,y} w^2_{xy} \sigma^2_{xy}\right]^{1/2},
\end{equation}
where
\begin{equation}
\label{eq:noise}
\sigma^2_{xy} = \mathrm{RN}^2 + I_{xy} + B_{xy}.
\end{equation}
In Equation \ref{eq:noise}, RN is the read noise per pixel in e$^-$ and $B_{xy}$ is the flux in e$^-$ from the background (i.e., the sky, dark current, etc.) at pixel $(x,y)$. 
Finally, we determine the  image zeropoint ZP  via
\begin{equation}
\mathrm{ZP} = 2.5\log S + m,
\end{equation}
where $m$ is the apparent magnitude of the source through some filter in the AB system.
Figure \ref{fig:imaging} shows three example simulated images of the same \glsn\ system generated using our pipeline, taken with three different instruments under representative observing conditions.

To increase our sensitivity to faint transients, we stack observations taken in the same filter in a single night. 
For \minion, this has the effect of combining the two exposures taken in the same filter in a $\sim$30-minute window to reject moving objects into a single observation with a signal-to-noise ratio roughly a factor of $\sqrt{2}$ larger. 
For \altsched, the stacking has no effect, as the strategy performs revisits to reject moving objects in \textit{different} filters to obtain colors.
For ZTF, stacking has no effect on the public MSIP data, which has a typical revisit time of 3-4 days in each filter. 
However, the stacking significantly boosts survey depth in the high-cadence Partnership fields and the Caltech survey. 
In some regions of these proprietary surveys, a single field may be observed as many as six times per night in a single filter, leading to a potential improvement in depth of $2.5\log(\sqrt{6}) \approx 1$ mag over the nominal limiting magnitude of 20.5 in all filters. 
We apply the discovery technique discussed in the next section (Section \ref{sec:discovery}) to the \textit{stacked}, not raw, data.

\begin{figure}
    \centering
    \includegraphics[width=1\textwidth]{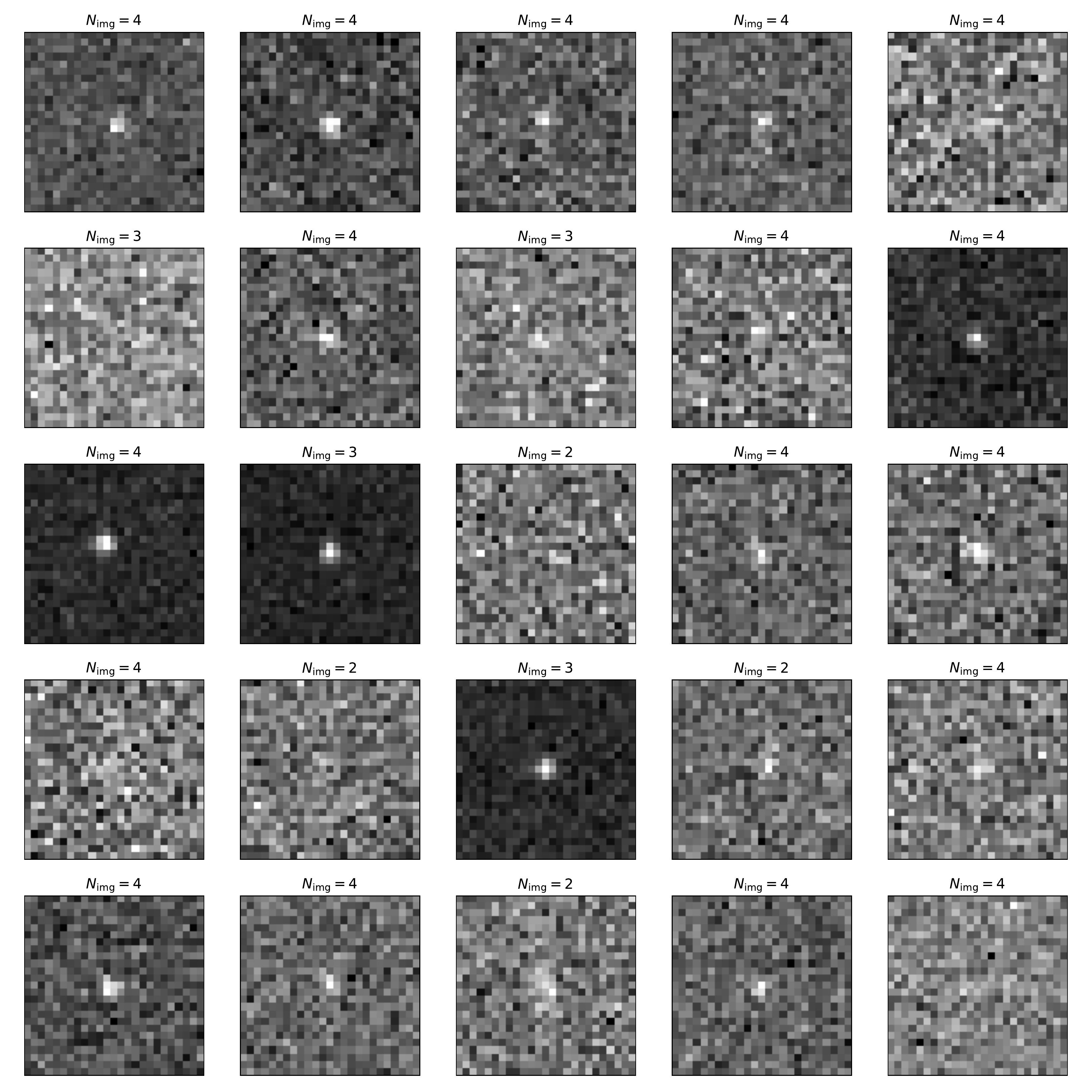}
    \caption{Lens-centered difference image cutouts of 25 randomly selected ZTF \glsne~Ia. 
    Each cutout is $25^{\prime\prime} \times 25^{\prime\prime}$.
    The sources visible in the cutouts contain flux from the \glsn\ images \emph{only}. 
    Lens light and host galaxy light are removed in the subtraction.
    The low spatial resolution of ZTF (1.01$^{\prime\prime}$ pixels) combined with the $2^{\prime\prime}$ FWHM seeing at Palomar Observatory render the survey unable to resolve multiply imaged supernovae, a feature we exploit in Section \ref{sec:discovery}.
    }
    \label{fig:ztf-diff-noisy}
\end{figure}

\begin{figure}
    \centering
    \includegraphics[width=1\textwidth]{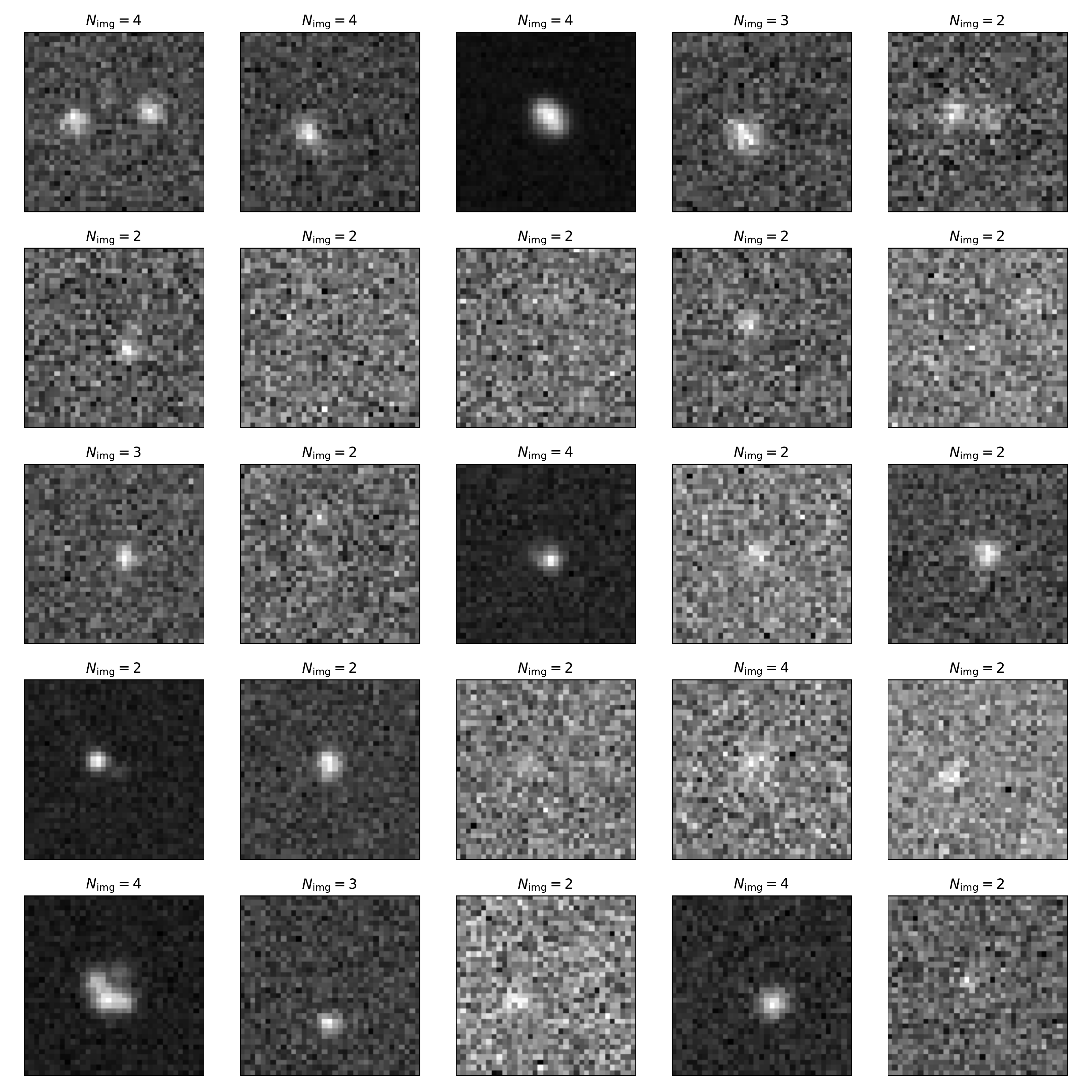}
    \caption{Lens-centered difference image cutouts of 25 randomly selected LSST (\minion) \glsne~Ia.
    Each cutout is $7^{\prime\prime} \times 7^{\prime\prime}$.
    As in Figure \ref{fig:ztf-diff-noisy}, the sources visible in the cutouts contain flux from \glsn\ images only.
    Lens light and host galaxy light are removed in the subtraction. 
    The improved spatial resolution $(0.2^{\prime\prime}$ pixels) of LSST compared to ZTF enables some \glsne\ to be totally or  marginally resolved, but the majority of systems  remain unresolved. 
    LSST must take special care to ensure that its  machine learning algorithm for difference image artifact rejection \citep[e.g.,][]{2015AJ....150...82G} does not reject marginally resolved \glsne, such as the ones in row 5, column 1; row 1, column 4; and row 1, column 5.
    }
    \label{fig:lsst-noisy}
\end{figure}

An important simplification in our simulations is that we treat \glsn\ images as a single object when performing photometry. 
The effect of this assumption is that we can realize a single light curve for each \glsn\ system, the flux of which is  the summed flux of the individual images. 
For ZTF, this is a reasonable assumption, as the large pixels of the detector and the 2$^{\prime\prime}$ seeing at Palomar Observatory ensure \glsne\ cannot be resolved (see Figure \ref{fig:ztf-diff-noisy}).
For LSST, as Figure \ref{fig:lsst-noisy} shows, this assumption should in most cases.
For the cases where the assumption does not hold, and the multiple images of a \glsn\ are resolved, the transient can be detected as two or more bright, nearby transients, as proposed by \cite{om10}.
For simplicity, we also assume perfect image subtractions. 
The main implication of this assumption is that photometric accuracy and source detection are unaffected by proximity to the cores of bright lens galaxies. 

\subsection{Discovery Technique}
\label{sec:discovery}
We simulate the detection and photometric classification of \glsne\ using the technique described in \cite{gnkc18}.
The strategy rests on three observational facts.
First, normal Type Ia supernovae (\sneia) are the brightest type of supernovae that have ever been observed to occur in elliptical galaxies \citep{maozreview}.
Second, the absolute magnitudes of normal \sneia\ in elliptical galaxies are remarkably homogenous, even without correcting for their colors or lightcurve shapes $(\sigma_M \about 0.4\ \mathrm{mag})$, with a component of the population being underluminous \citep{2011MNRAS.412.1441L}.
Finally, due to the sharp 4000\AA\ break in their spectra, elliptical galaxies tend to provide accurate photometric redshifts from large-scale multi-color galaxy surveys such as SDSS.

A high-cadence, wide-field imaging survey can leverage these facts to systematically search for strongly lensed supernovae in the following way.
First, by spatially cross-matching its list of supernova candidates with a catalog of elliptical galaxies for which secure photometric redshifts have been obtained,  supernovae that appear to be hosted by elliptical galaxies can be identified.
The hypothesis that one of these supernovae actually resides in its apparent host can be tested by fitting its broadband light curves with an  \snia\ spectral template (as \sneia\ are the only types of supernovae that occur in ellipticals) fixed to the photometric redshift of the galaxy and constrained to obey $-18.5 > M_B > -20$, a liberal absolute magnitude range for \sneia, assuming a fiducial cosmology. 
If the transient is a lensed supernova at higher redshift, then the spectral template fit will fail catastrophically, as the supernova light curves will be strongly inconsistent with the redshift and brightness implied by the lens galaxy.

We use SALT2 \citep{salt2}, a parametrized \snia\ spectral template that is the standard tool for placing \sneia\ on the Hubble diagram, to perform this technique. 
The template possesses four parameters: $t_0$, $x_0$, $x_1$, and $c$, encoding a reference time, an overall SED normalization, a supernova ``stretch,'' and a color-law coefficient, respectively.
The flux of the template is given by 
\begin{equation}
F_\lambda(\lambda, t) = x_0 [M_0(\lambda, t-t_0) + x_1 M_1(\lambda, t-t_0)] \exp[c\,CL(\lambda)],
\end{equation}
where $M_0$ and $M_1$ are eigenspectra derived from a training sample of measured \snia\ spectra and $CL(\lambda)$ is the average color-correction law of the sample \citep[see][for details]{salt2}. 
The template aims to model the mean evolution of the SED sequence of \sneia\ and its variation with a few dominant components, including a time independent variation with color, whether it is intrinsic or due to extinction by dust in the host galaxy (or both).
Finally, we draw a random reference time $t_r$ for the system uniformly over the duration of the survey,
\begin{equation}
    t_r \sim U[t_\mathrm{min}, t_\mathrm{max}],
\end{equation}
where $t_\mathrm{min}$ and $t_\mathrm{max}$ are the times of the survey's first and last observations, respectively.

We realize broadband photometry of each blended \glsn~Ia using the technique described in Section \ref{sec:phot}. 
Starting from the first observation of the \snia, we fit the light curve with SALT2, fixed to the redshift of the lens galaxy (assumed to be known either as a photometric or spectroscopic redshift) and fixed to obey $-18.5 > M_B > -20$ at that redshift (effectively a constraint on $x_0$).
Additionally, we enforce bounds of $[-0.2, 0.2]$ on $c$ and $[-1, 1]$ on $x_1$, values characteristic of normal \sneia\ \citep{scalzo14a}.
We use the CERN minimization routine \texttt{MIGRAD}\ \citep{minuit} to fit the data.
If the light curve has at least one data point that is at least $5\sigma$ discrepant from the best fit and at least 4 data points with S/N $\geq 5$, then the object is marked ``detected."
If not, then the next observation is added and the process is repeated until the object is detected or all observations are added, resulting in a non-detection.

%\subsection{Monte Carlo}
%\label{sec:mc}
%The lens and source populations are realized in a Monte Carlo simulation.
%We generate $10^5$ lens galaxies with parameters drawn at random from their underlying distributions.
%For each lens galaxy, an effective lensing area of influence is estimated as a $[8 \theta_{E, z_s=\infty}]^2$ box centered on the galaxy.\footnote{This box size was chosen to be large enough to accommodate the effects of ellipticity and external shear.}
%We simulate $5 \times 10^4$ years of \sneia, randomly distributed across the box, rejecting systems where $z_s < z_l$.
%For each remaining source, we solve the lens equation using  \q{glafic}\  \citep{glafic} to determine the macrolensing magnification, image multiplicity, and time delays.
%In total we generated 37,100 multiply imaged systems containing a total of 78,184 images.
%Since our simulation only covers $10^5 / N_\mathrm{gal} \approx 0.026\%$ of the sky, this  corresponds to a rate of 2,675 systems, all sky, per year to $z_s=2$. 

\subsection{Importance Sampling, Sample Weighting, and Rate Calculation}
We perform a separate Monte Carlo simulation for each survey and supernova type, running each simulation until $O(10^5)$ \glsn\ systems are discovered.
In each iteration of the simulation we realize one supernova \textit{behind} the sampled lens in the lensing area of influence.
We run each ZTF simulation for $N=10^8$ iterations, and we run each LSST simulation for $N=10^7$ iterations.
The ZTF simulations require more iterations to converge as ZTF is shallower than LSST, so any given system is less likely to be detected.
To reduce shot noise in our results, we use importance sampling to sample lens and source redshifts, the distributions of which contain almost no probability mass in the crucial region $z \lesssim 0.5$. 
Therefore, each system has an associated importance weight factor $\omega$, 
%The ``weight'' $\omega$ of each detected \glsn\ is given by
\begin{equation}
\omega = \frac{f_\Omega}{\mathcal{A}}\frac{p(z_s)p(z_l)}{q(z_s)q(z_l)},
\label{eq:weight}
\end{equation}
where $p(z_s)$ and $p(z_l)$ are the true densities of $z_s$ and $z_l$ (Equations \ref{eq:galaxypdf} and \ref{eq:snpdf}), $q(z_s)$ and $q(z_l)$ are the sampling densities, $f_\Omega$ is the ratio of sky area imaged by the survey to sky area covered in the simulation, and $\mathcal{A}$ is the factor by which the supernova rate must be multiplied to yield one supernova of the given subtype with $z_s > z_l$ per year in the ``lensing area of influence'' of the lens.
We take the sampling densities to be uniform,
\begin{align}
    q(z_s) &= U[z_{s,\mathrm{min}},z_{s,\mathrm{max}}],\\
    q(z_l) &= U[z_{l,\mathrm{min}},z_{l,\mathrm{max}}],
\end{align}
where $z_{s,\mathrm{min}}$ and $z_{s,\mathrm{max}}$ are the minimum and maximum supernova redshifts considered in the simulation, respectively.
We assume the lenses are uniformly distributed across the sky, so the areal correction factor $f_\Omega$ can be calculated by dividing the total number of lenses in the survey area by the number of lenses $N$ realized in the simulation,
\begin{equation}
    \label{eq:arealcorrfac}
    f_\Omega = \frac{\Omega D_H}{N}\int_{\vd_\mathrm{min}}^{\vd_\mathrm{max}} \phi(\vd) \, d\vd \int_{z_{l,\mathrm{min}}}^{z_{l,\mathrm{max}}}  \frac{(1+z_l)^2 D_l^2}{E(z_l)}\, dz_l,
\end{equation}
where we have integrated Equation \ref{eq:galaxydist} to estimate the total number of lenses in the survey area. 
In Equation \ref{eq:arealcorrfac}, $\Omega$ is the area of the survey in steradians and $\vd_\mathrm{min}$ and $\vd_\mathrm{max}$ are the minimum and maximum lens velocity dispersions considered in the simulation, respectively.  

The number of supernovae per year behind the lens's area of influence is determined by integrating the observer-frame supernova redshift function (Figure \ref{fig:snrate}) from $z_l$ or $z_{s,\mathrm{min}}$ (whichever is larger) to $z_{s,\mathrm{max}}$ and multiplying by the ratio of the lens's area of influence to the full-sky area.
Taking $z_1 = {\max(z_l, z_{s,\mathrm{min}})}$ and $z_2 = {z_{s,\mathrm{max}}}$, we have
\begin{equation}
    \mathcal{A} = \left[\frac{\theta_l^2}{4}\int_{z_1}^{z_2} f_T(z_s) \, dz_s\right]^{-1}.
\end{equation}
The weights specify the contribution of a given discovered system to the overall \glsn\ discovery rate, and have units of [year$^{-1}$]. 
The summed weights provide a Monte Carlo estimate of the \glsn\ discovery rate,
\begin{equation}
    \boxed{\sum_{i=0}^N \omega_i \approx R,}
\end{equation}
where $R$ is the total  discovery rate (in year$^{-1}$).
As with any Monte Carlo estimator, the precision of $R$ increases as the square root of the number of samples $N$. 
The above scheme is roughly $10^3$ times more efficient than sampling all of the parameters of the model brute-force.

\section{Results}
\label{sec:results}

Table \ref{tab:yield} shows the \glsn\ discovery rates $R$ of each simulated survey. 
Our calculations suggest that under nominal survey operations,  ZTF should discover at least 8.60 \glsne\ per year, of which at most 4.1\% are Type Ib/c, 2.0\% are SN 1991T-like, 3.7\% are Type IIL, 14.3\% are Type Ia, 32.1\% are Type IIP,  0.2\% are SN 1991bg-like, and at least  43.6\% are Type IIn.
We find that the \minion\ LSST observing strategy  should discover at least 380.60 \glsne\ per year, of which at most 12.6\% are Type Ia, 1.6\% are SN 1991T-like, 23.3\% are Type IIP, 4.1\% are Type Ib/c, 3.4\% are Type IIL,  0.2\% are SN 1991bg-like, and at least   55.0\% are Type IIn.
The \altsched\ observing strategy should discover at least 341.27 \glsne\ per year, of which at most  4.7\% are Type Ib/c, 3.8\% are Type IIL, 13.9\% are Type Ia, 26.7\% are Type IIP, 1.8\% are SN 1991T-like, 0.3\% are SN 1991bg-like, and at least  45.3\% are Type IIn. 
The Type IIn rates are given as lower limits because \glsne~IIn can be detected in both ZTF and LSST beyond $z_s=3$, the maximum redshift in our simulations, but their rate at $z_s > 3$ is highly speculative.

\begin{deluxetable*}{r||ccc}
\tablecaption{\glsn\ discovery rates (in units of year$^{-1}$) of ZTF and LSST.\label{tab:yield}}
\tablehead{\colhead{SN Type} & \colhead{ZTF} & \colhead{LSST (\texttt{minion\_1016})} & \colhead{LSST (\texttt{altsched})}}
\startdata
%LSST (\texttt{altsched})
Type Ia & 1.23 &  47.84 &  47.42\\
Type IIP & 2.76 &  88.51 & 91.06 \\
Type IIn\tablenotemark{a} & 3.75 & 209.31 & 166.54\\
Type IIL & 0.31 & 11.69 &  13.10\\
Type Ib/c & 0.36 & 14.00 &  16.15 \\
SN 1991bg-like & 0.02 & 0.79 & 0.89\\
SN 1991T-like & 0.17 & 5.41 & 6.09\\
\hline \textbf{Total}\tablenotemark{a} & 8.60 & 380.60 & 341.27
	\enddata
\tablenotetext{a}{Lower limit.}
\end{deluxetable*}

\begin{figure*}
	\centering
    \includegraphics[width=1\textwidth]{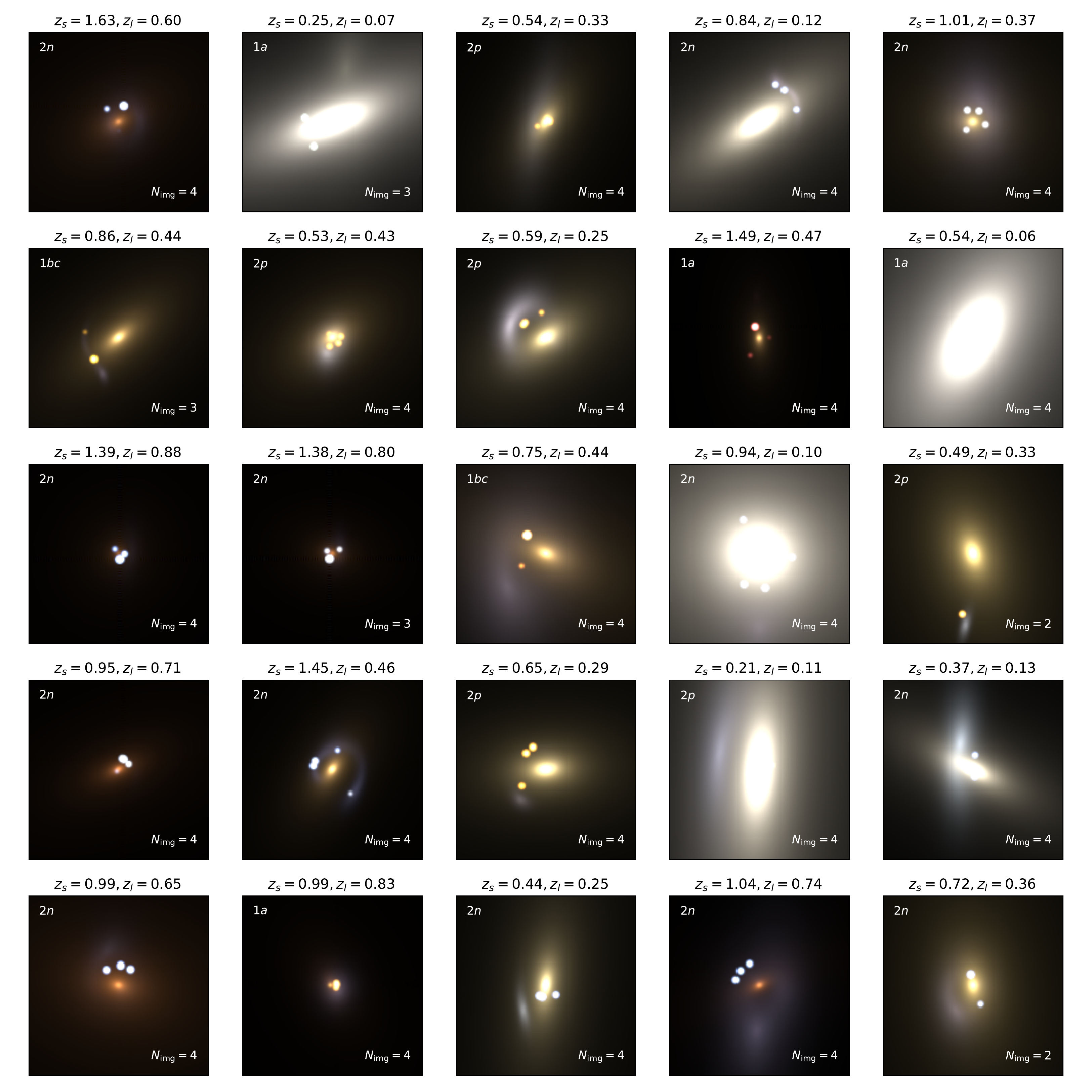}
 \caption{Model (i.e., noiseless)   $6^{\prime\prime}\times6^{\prime\prime}$ composite $gri$ images of 25 randomly-chosen, simulated \glsne, their lens galaxies, and their lensed host galaxies, ``detected'' by ZTF.
 Each image is ``taken'' exactly one night after the transient is detected as a \glsn\ candidate based on a light curve fit to the simulated ZTF data (see Section \ref{sec:discovery}).
 The FWHM of the seeing on the images is $0.1^{\prime\prime}$, and the pixel scale is $0.04^{\prime\prime}$, identical to that of the UVIS channel of the Wide Field Camera 3 (WFC3) on \hst. 
 }
 \label{fig:ztfsne}
\end{figure*}

\begin{figure*}
	\centering
    \includegraphics[width=1\textwidth]{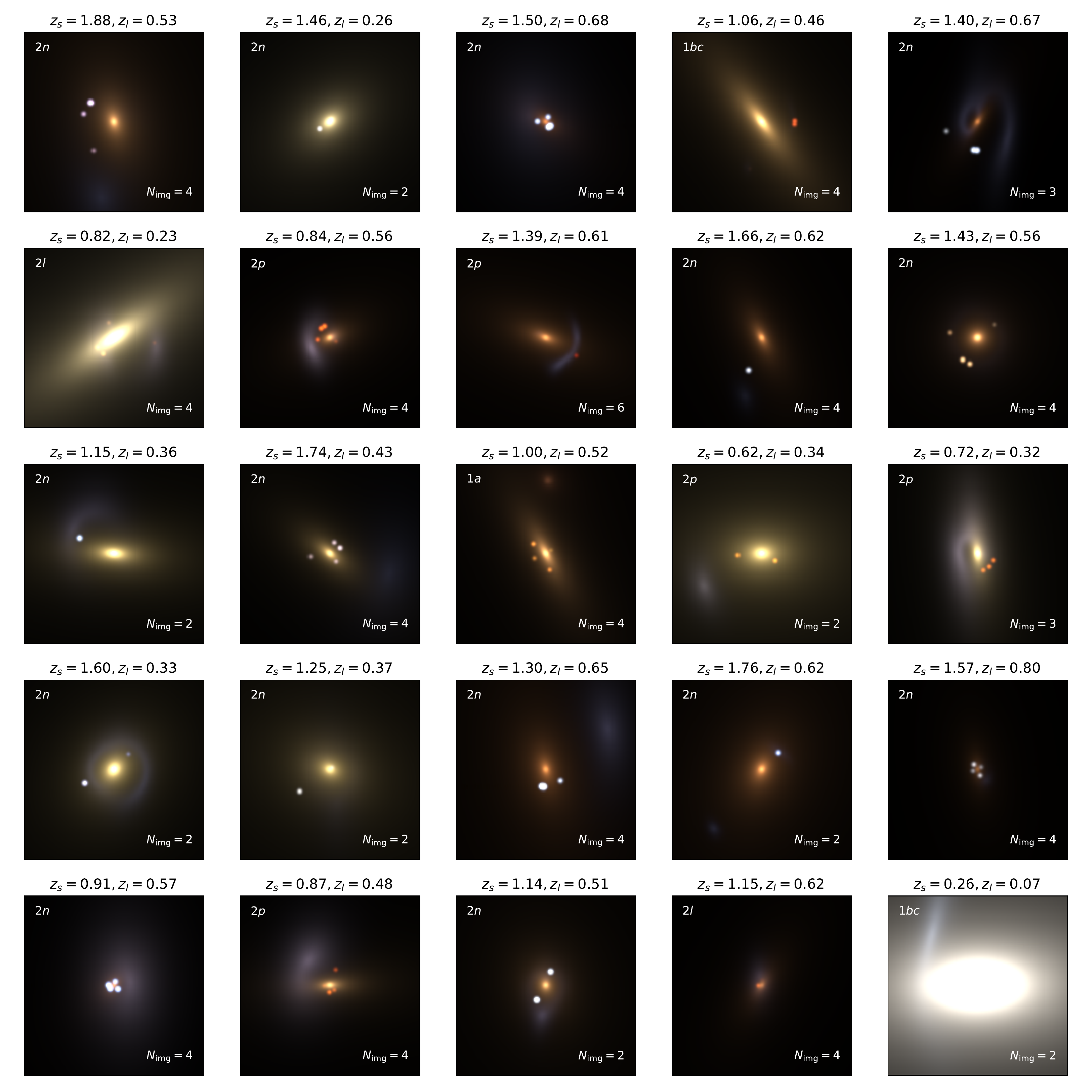}
 \caption{Model (i.e., noiseless)   $6^{\prime\prime}\times6^{\prime\prime}$  composite $gri$ images of 25 randomly-chosen, simulated \glsne, their lens galaxies, and their lensed host galaxies, ``detected'' by LSST under the \minion\ observing strategy.
 Each image is ``taken'' exactly one night after the transient is detected as a \glsn\ candidate based on a light curve fit to the simulated LSST data (see Section \ref{sec:discovery}).
 The FWHM of the seeing on the images is $0.1^{\prime\prime}$, and the pixel scale is $0.04^{\prime\prime}$, identical to that of the UVIS channel of WFC3. 
 The systems in this mosaic are generally  less compact and less magnified than those in Figure \ref{fig:ztfsne}, reflecting the increased depth and red-sensitivity of LSST compared to ZTF.}
 \label{fig:lsstsne}
\end{figure*}

\begin{figure*}
    \gridline{\fig{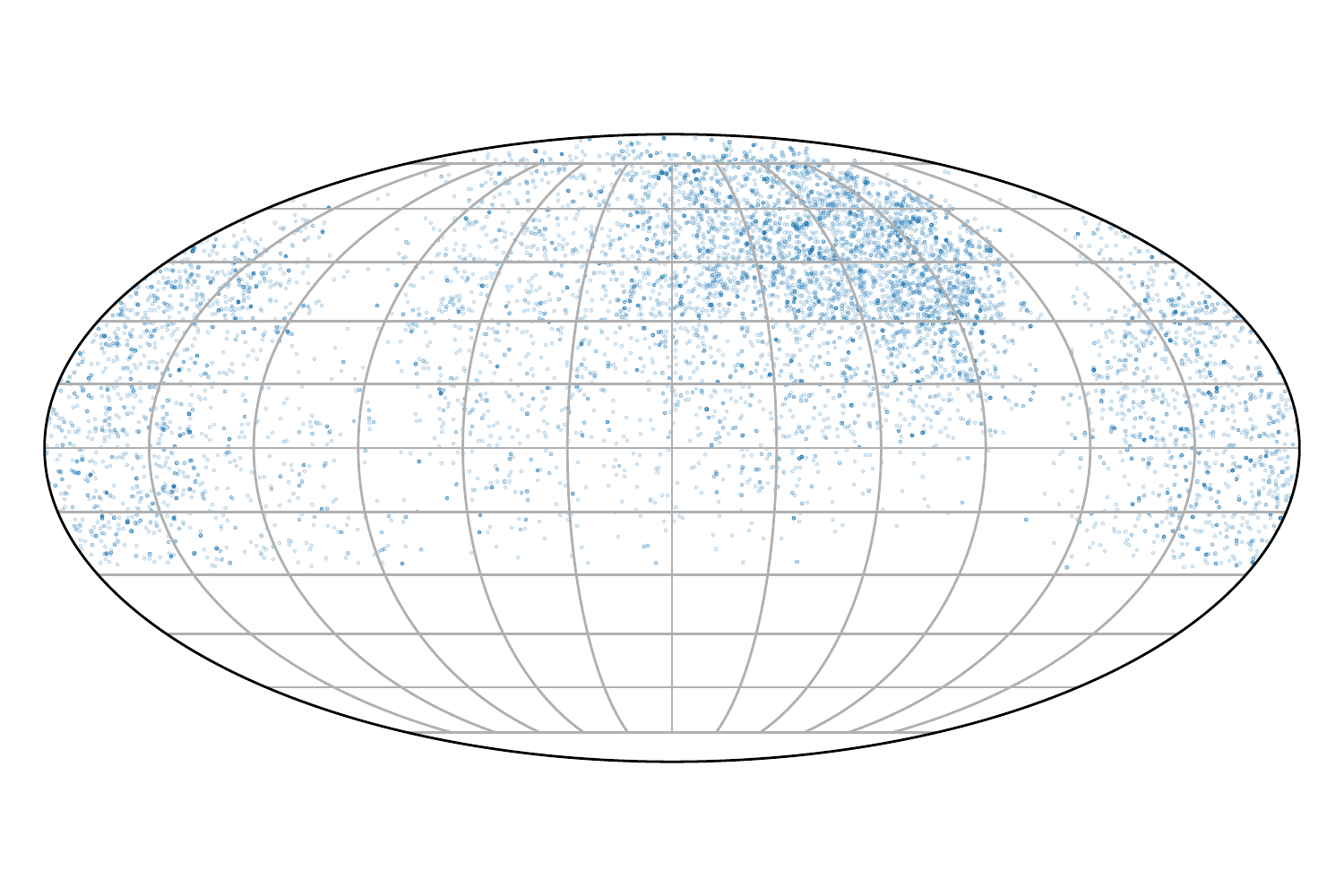}{0.8\textwidth}{(a)
Sky distribution of \glsne\ (all types) detected by ZTF in the simulation.
The discovered \glsne\ are concentrated in the high-cadence and $i$-band fields.}}
% 12109
    \gridline{\fig{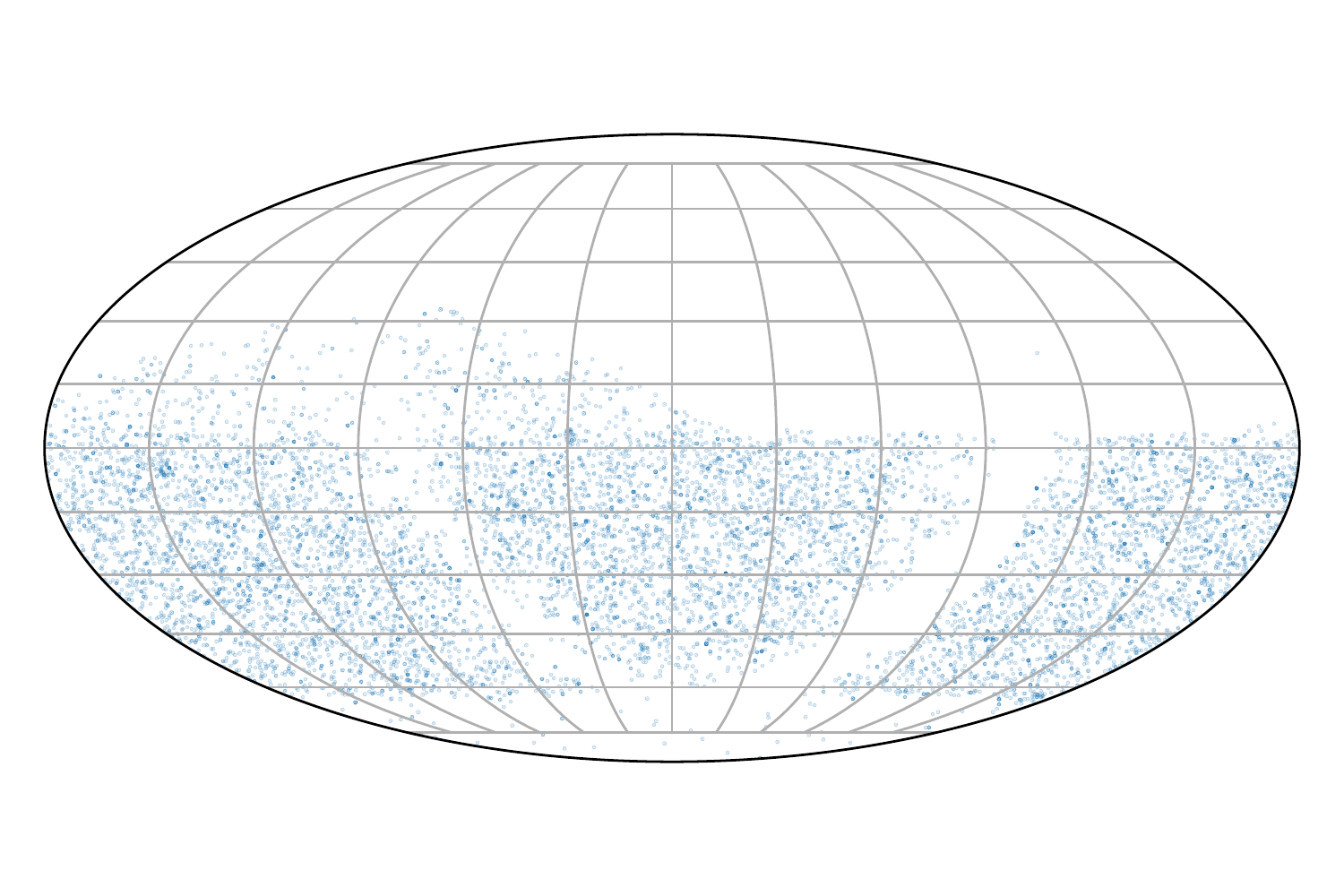}{0.8\textwidth}{(b)
Sky distribution of \glsne\ (all types) detected by \minion\ in the simulation.
The discovered \glsne\ are relatively uniformly distributed across the survey footprint, except for the Galactic plane, which has high extinction and yields almost no \glsne, and the north ecliptic spur, which has high airmass and yields \glsne at a reduced rate.}}
    \caption{Sky distributions of \glsne\ discovered in the simulations. \label{fig:skydist1}}
\end{figure*}

\begin{figure*}
\centering
\includegraphics[width=0.8\textwidth]{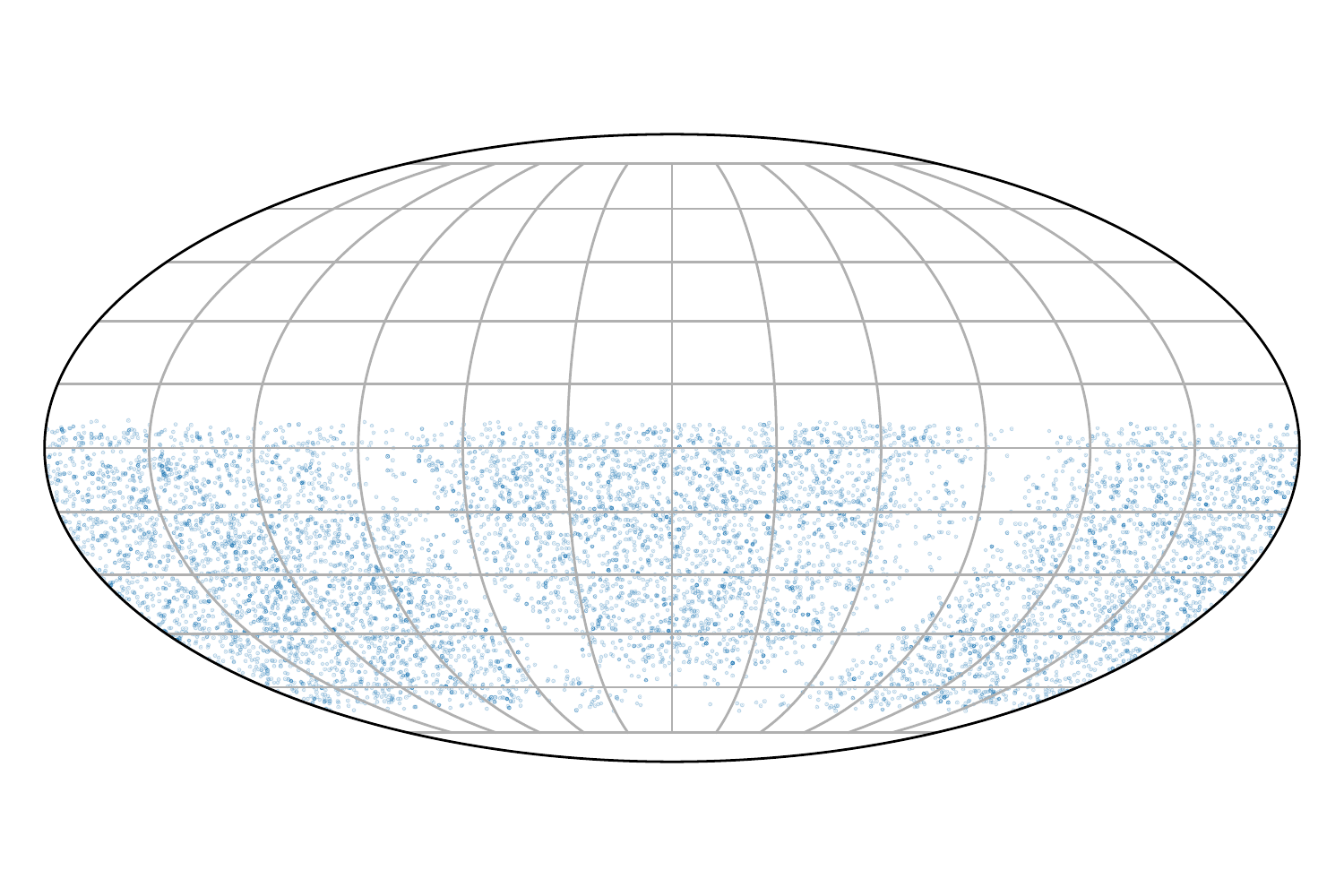}
\caption{
Sky distribution of \glsne\ (all types) detected by \altsched\ in the simulation.
The discovered \glsne\ are  uniformly distributed across the survey footprint, except for the Galactic plane, which has high extinction.}
\label{fig:skydist2}
\end{figure*}

\begin{figure*}
    \centering
    \includegraphics[width=\reswidth]{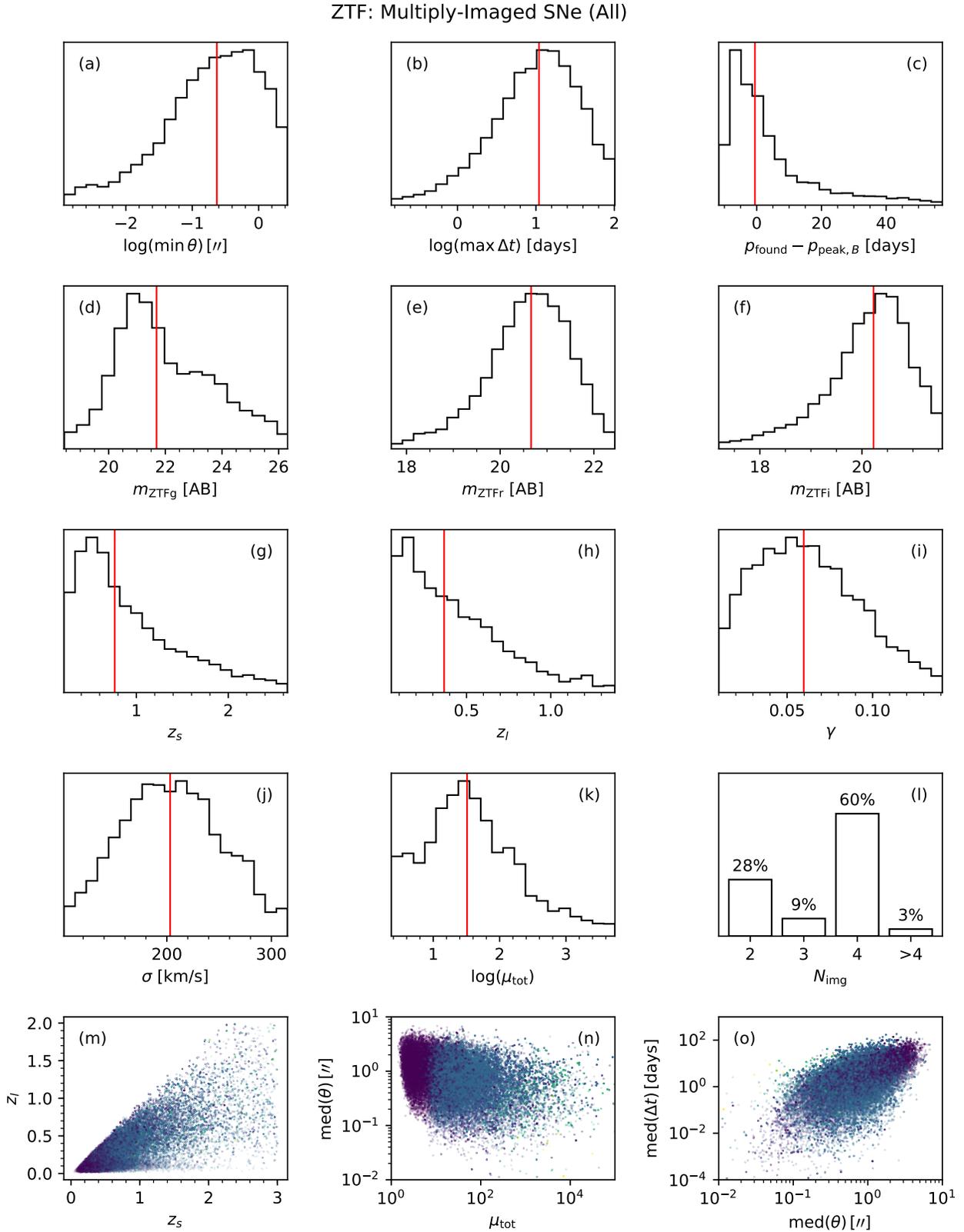}
    \caption{Monte Carlo results for ZTF  supernovae (all types).
    See Table \ref{tab:mcfigs} for a description of each subpanel.}
    \label{fig:ztf-all}
\end{figure*}

\begin{figure*}
    \centering
    \includegraphics[width=\reswidth]{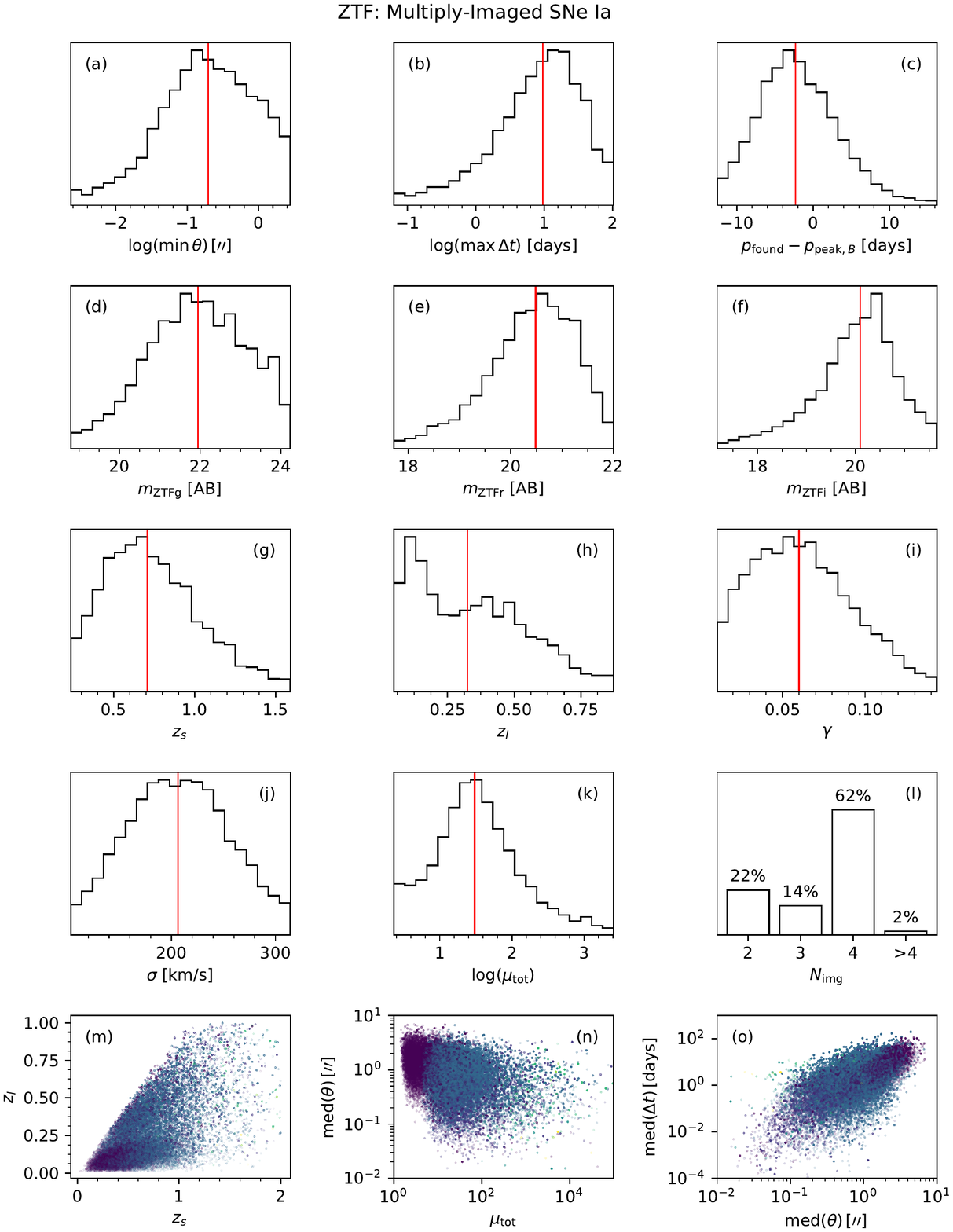}
    \caption{Monte Carlo results for ZTF Type Ia supernovae. See Table \ref{tab:mcfigs} for a description of each subpanel.}
    \label{fig:ztf-1a}
\end{figure*}

\begin{figure*}
    \centering
    \includegraphics[width=\reswidth]{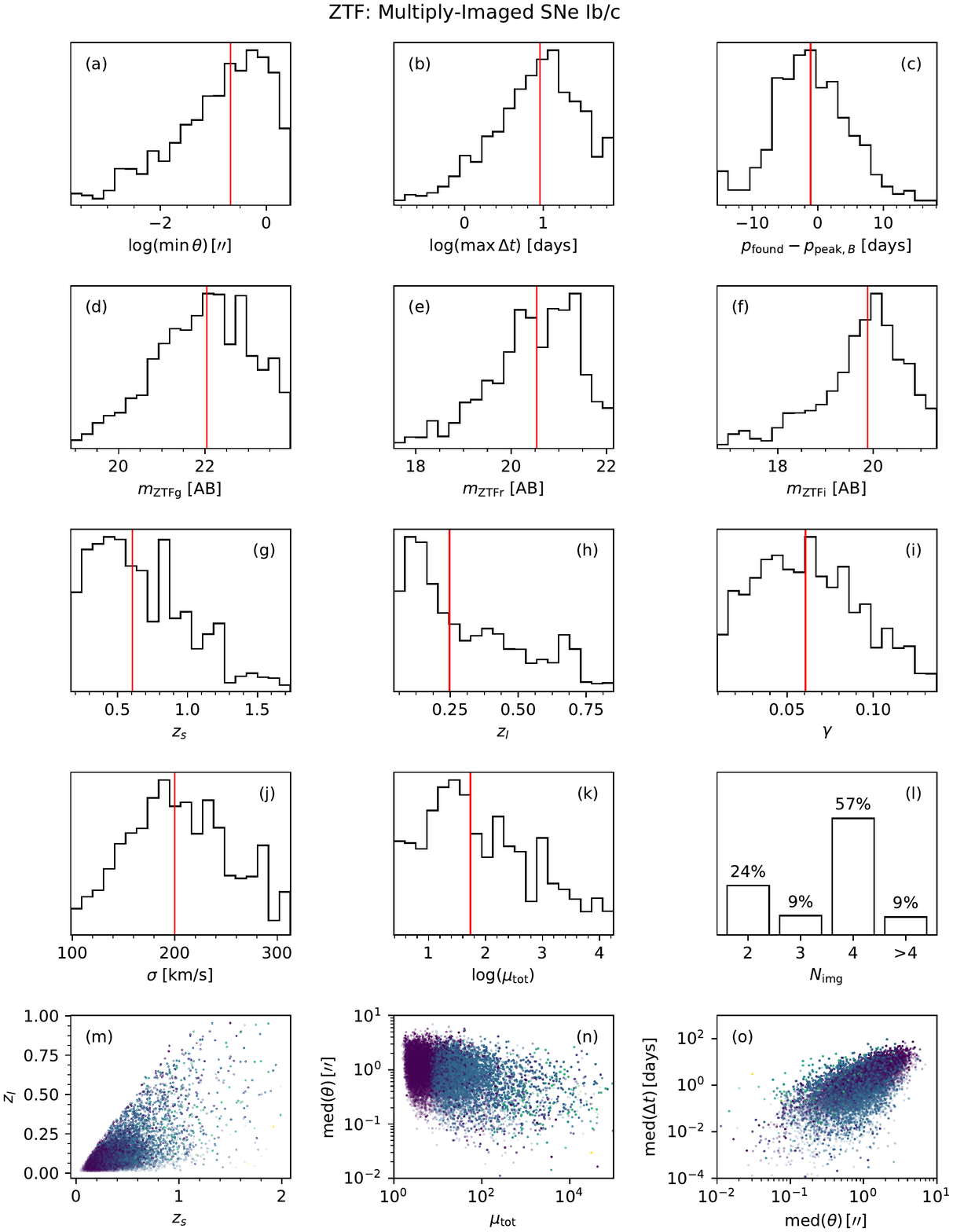}
    \caption{Monte Carlo results for ZTF Type Ib/c supernovae. See Table \ref{tab:mcfigs} for a description of each subpanel.}
    \label{fig:ztf-1bc}
\end{figure*}

\begin{figure*}
    \centering
    \includegraphics[width=\reswidth]{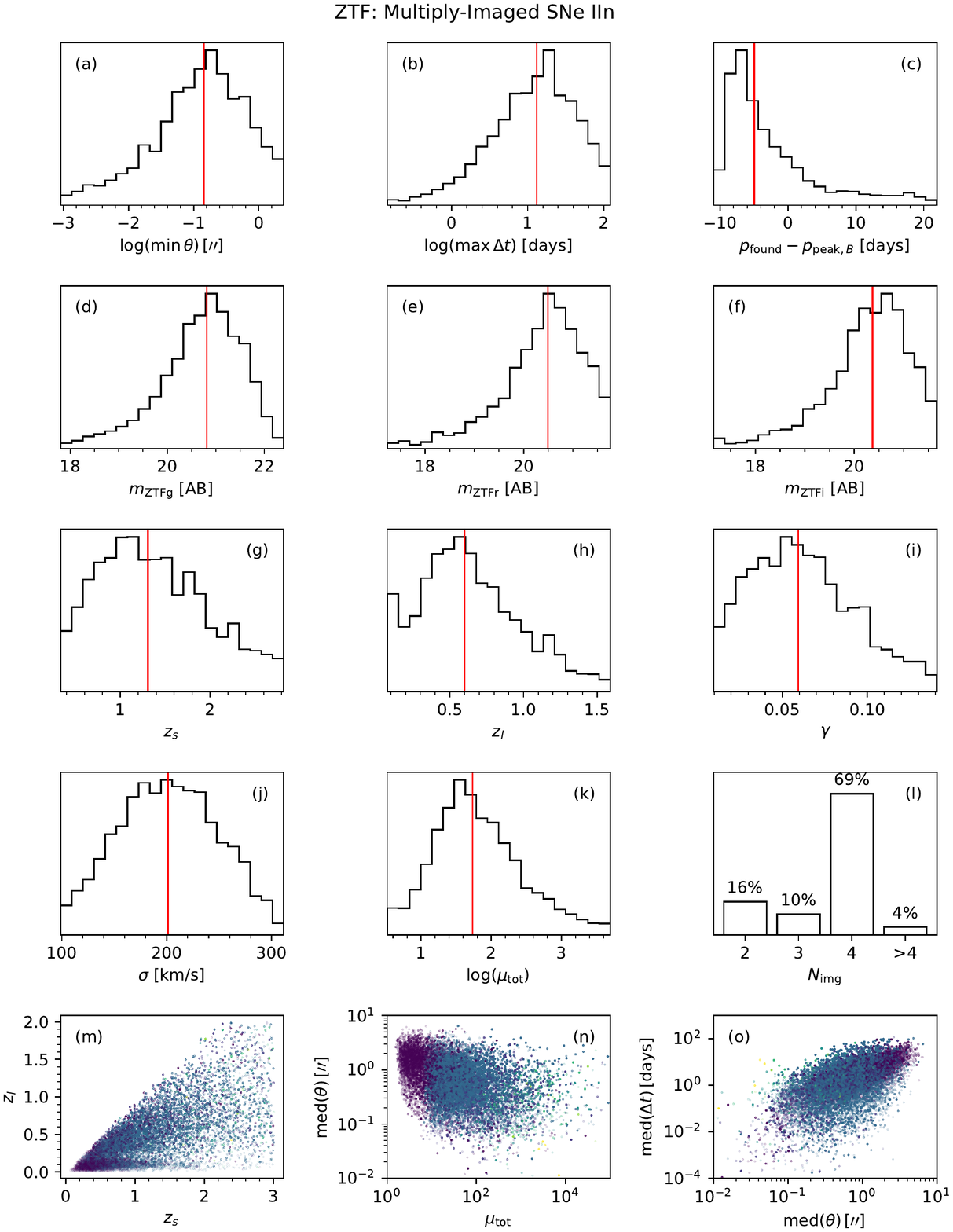}
    \caption{Monte Carlo results for ZTF Type IIn supernovae. See Table \ref{tab:mcfigs} for a description of each subpanel.}
    \label{fig:ztf-2n}
\end{figure*}

\begin{figure*}
    \centering
    \includegraphics[width=\reswidth]{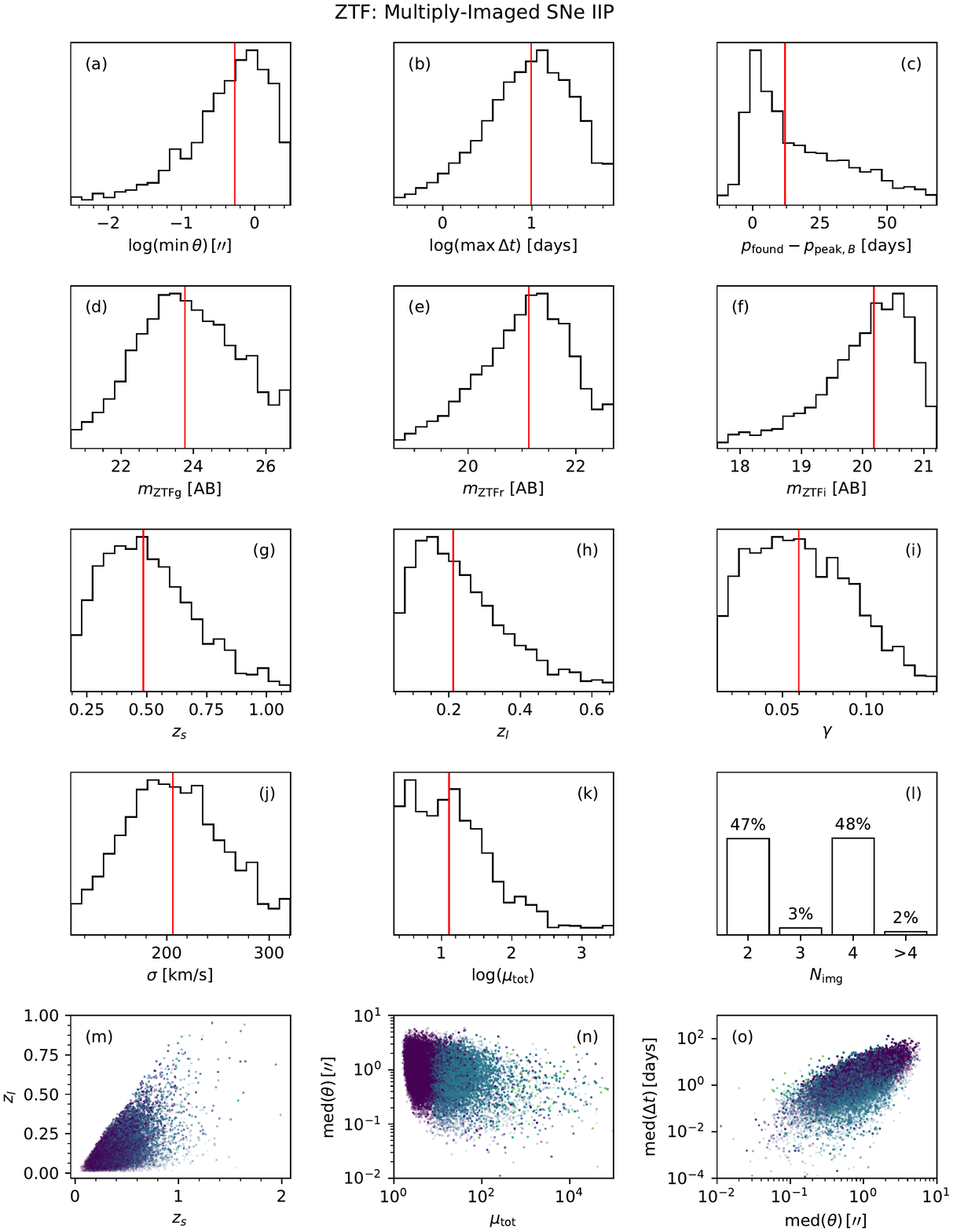}
    \caption{Monte Carlo results for ZTF Type IIP supernovae. See Table \ref{tab:mcfigs} for a description of each subpanel.}
    \label{fig:ztf-2p}
\end{figure*}

\begin{figure*}
    \centering
    \includegraphics[width=\reswidth]{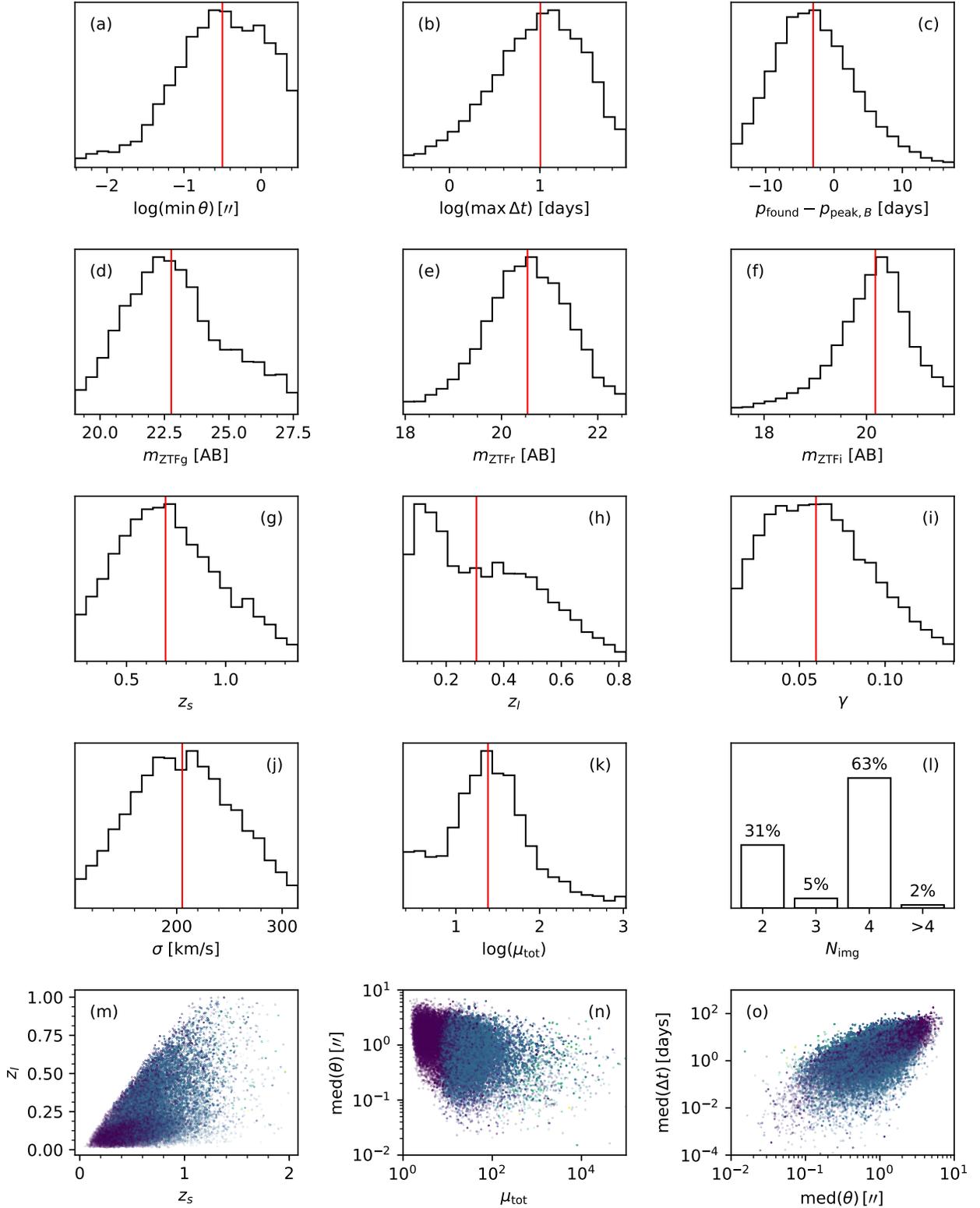}
    \caption{Monte Carlo results for ZTF SN 1991T-like supernovae. See Table \ref{tab:mcfigs} for a description of each subpanel.}
    \label{fig:ztf-91t}
\end{figure*}

\begin{figure*}
    \centering
    \includegraphics[width=\reswidth]{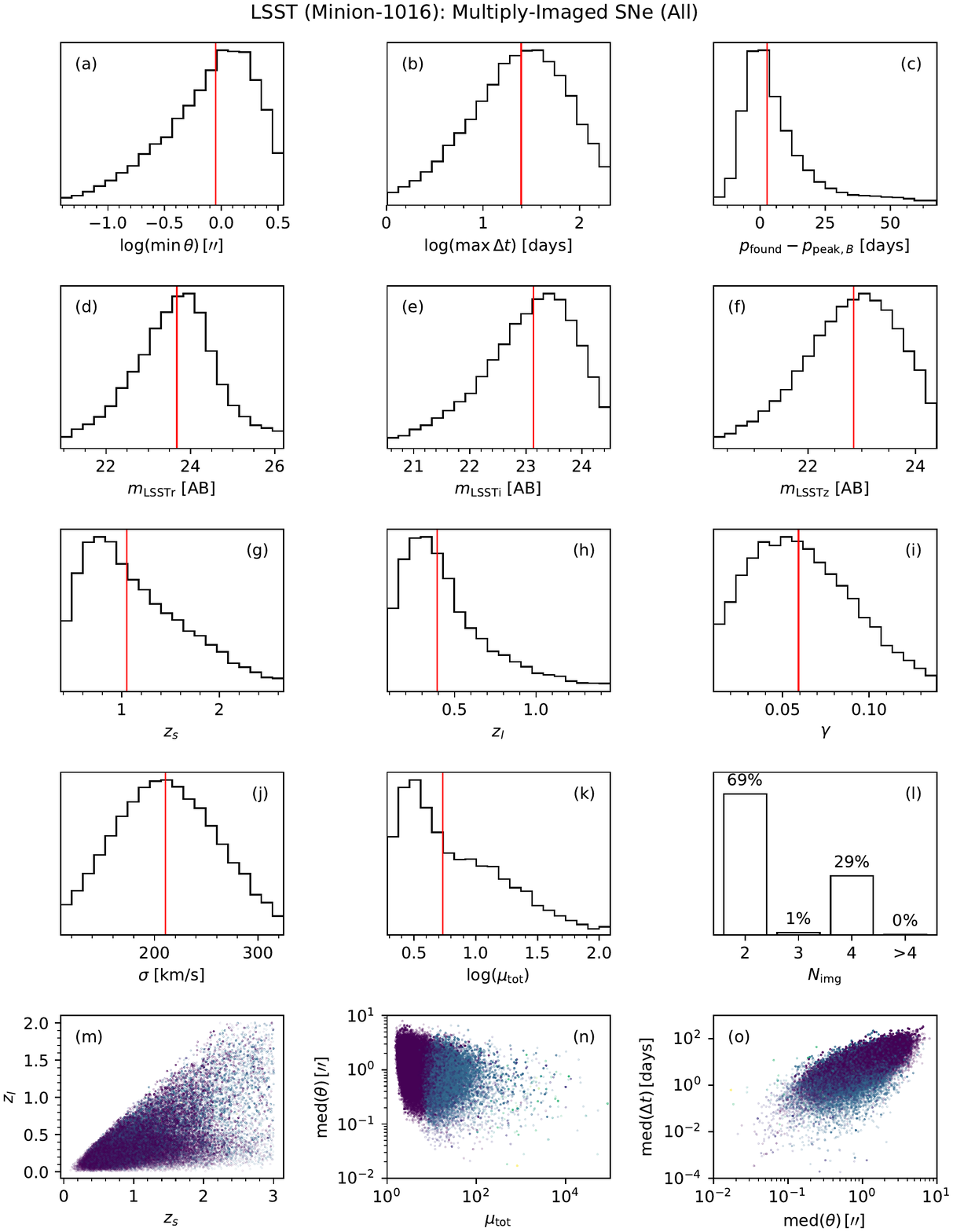}
    \caption{Monte Carlo results for LSST (\minion) supernovae (all subtypes). See Table \ref{tab:mcfigs} for a description of each subpanel.}
    \label{fig:minion-all}
\end{figure*}

\begin{figure*}
    \centering
    \includegraphics[width=\reswidth]{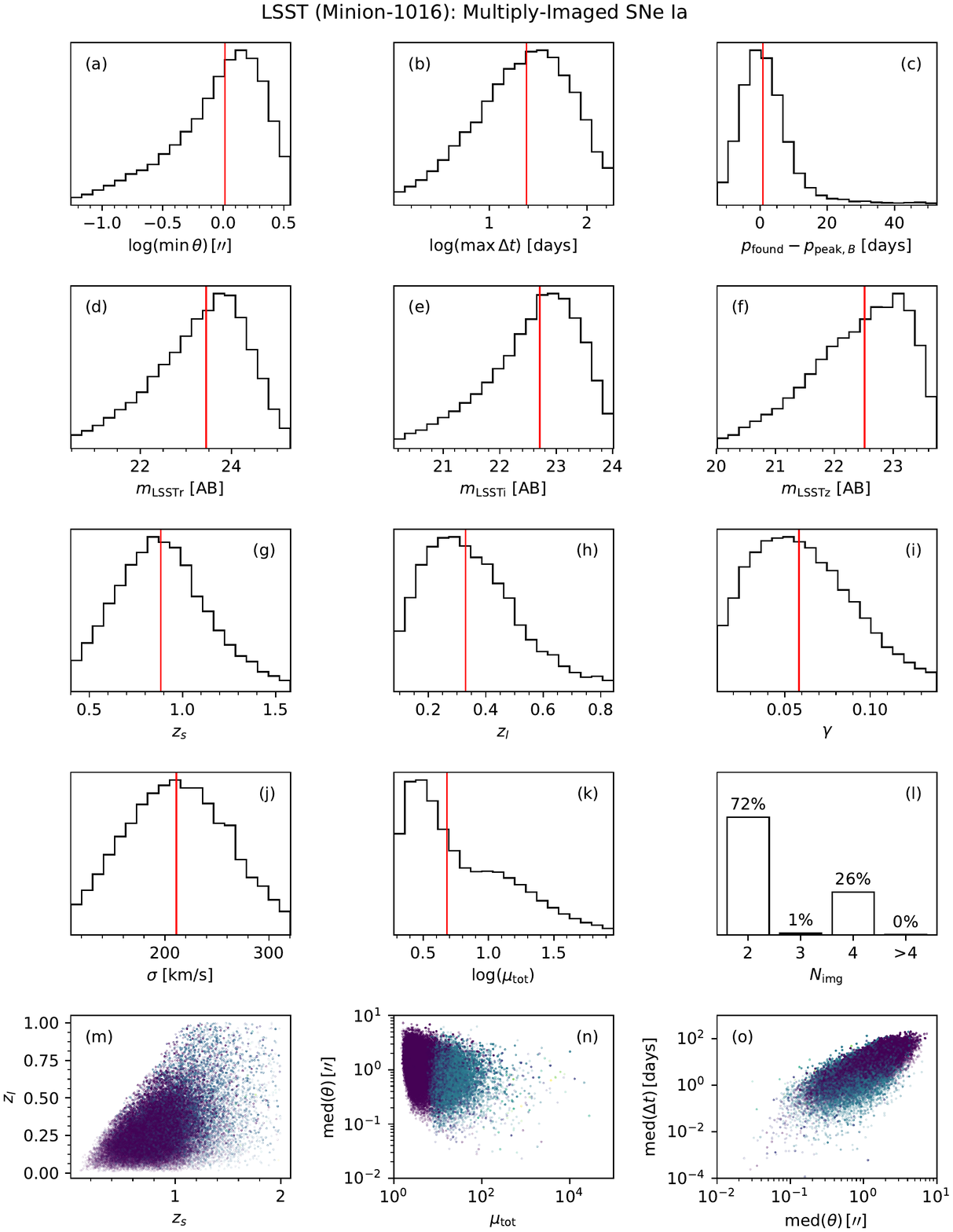}
    \caption{Monte Carlo results for LSST (\minion) Type Ia supernovae. See Table \ref{tab:mcfigs} for a description of each subpanel.}
    \label{fig:minion-1a}
\end{figure*}

\begin{figure}
    \centering
    \includegraphics[width=\reswidth]{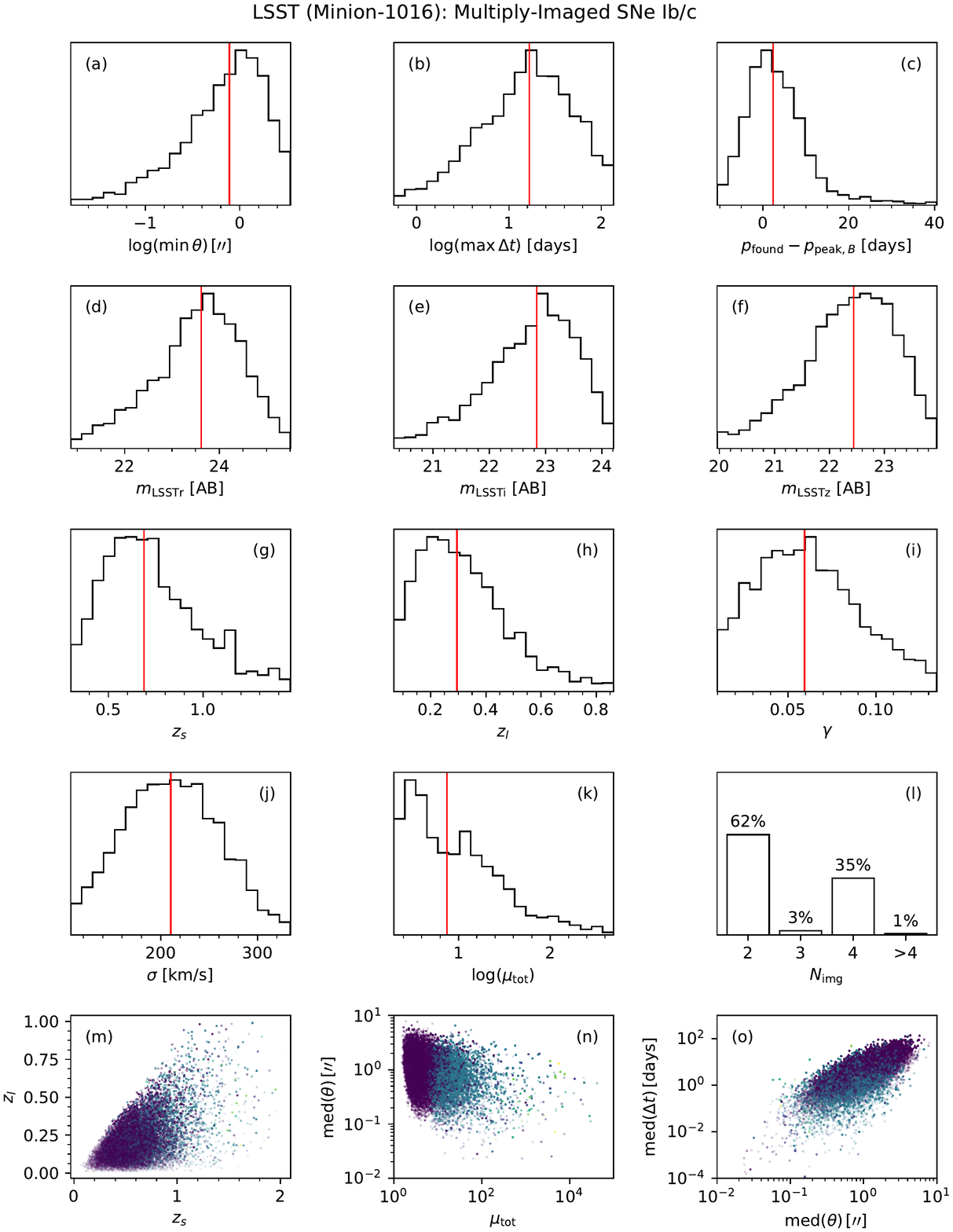}
    \caption{Monte Carlo results for LSST (\minion) Type Ib/c supernovae.  See Table \ref{tab:mcfigs} for a description of each subpanel.}
    \label{fig:minion-1bc}
\end{figure}

\begin{figure*}
    \centering
    \includegraphics[width=\reswidth]{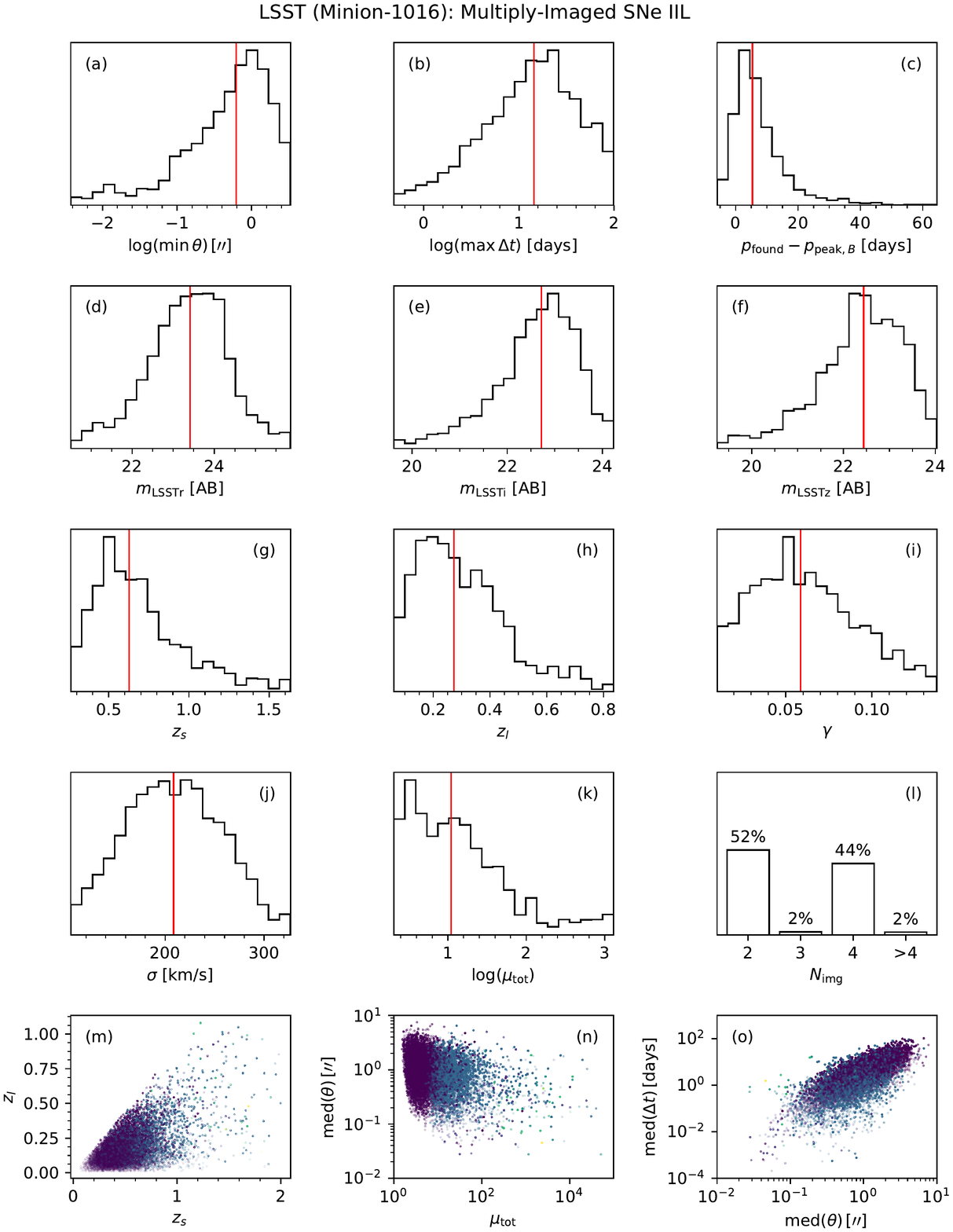}
    \caption{Monte Carlo results for LSST (\minion) Type IIL supernovae.  See Table \ref{tab:mcfigs} for a description of each subpanel.}
    \label{fig:minion-2l}
\end{figure*}

\begin{figure*}
    \centering
    \includegraphics[width=\reswidth]{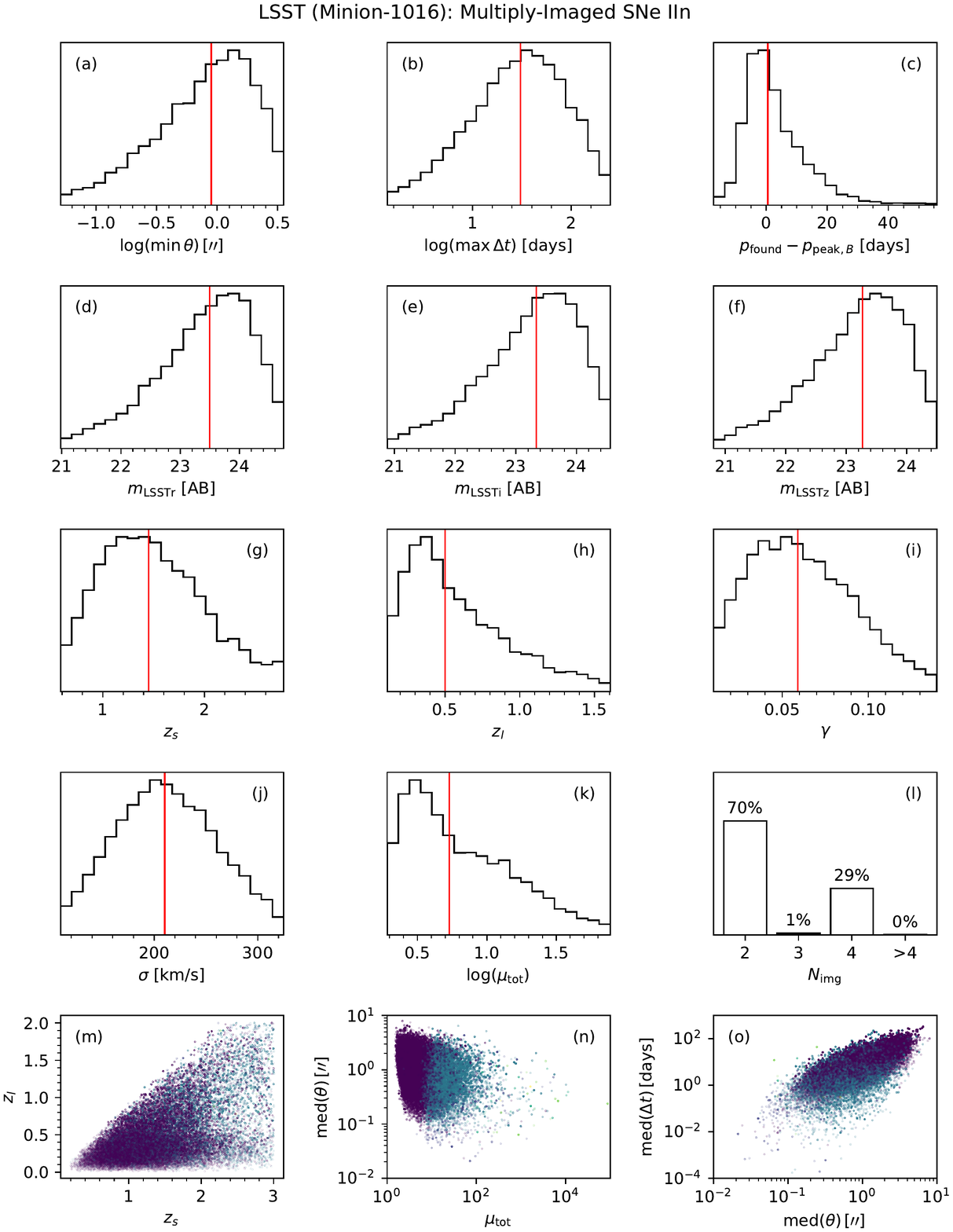}
    \caption{Monte Carlo results for LSST (\minion) Type IIn supernovae.  See Table \ref{tab:mcfigs} for a description of each subpanel.}
    \label{fig:minion-2n}
\end{figure*}

\begin{figure*}
    \centering
    \includegraphics[width=\reswidth]{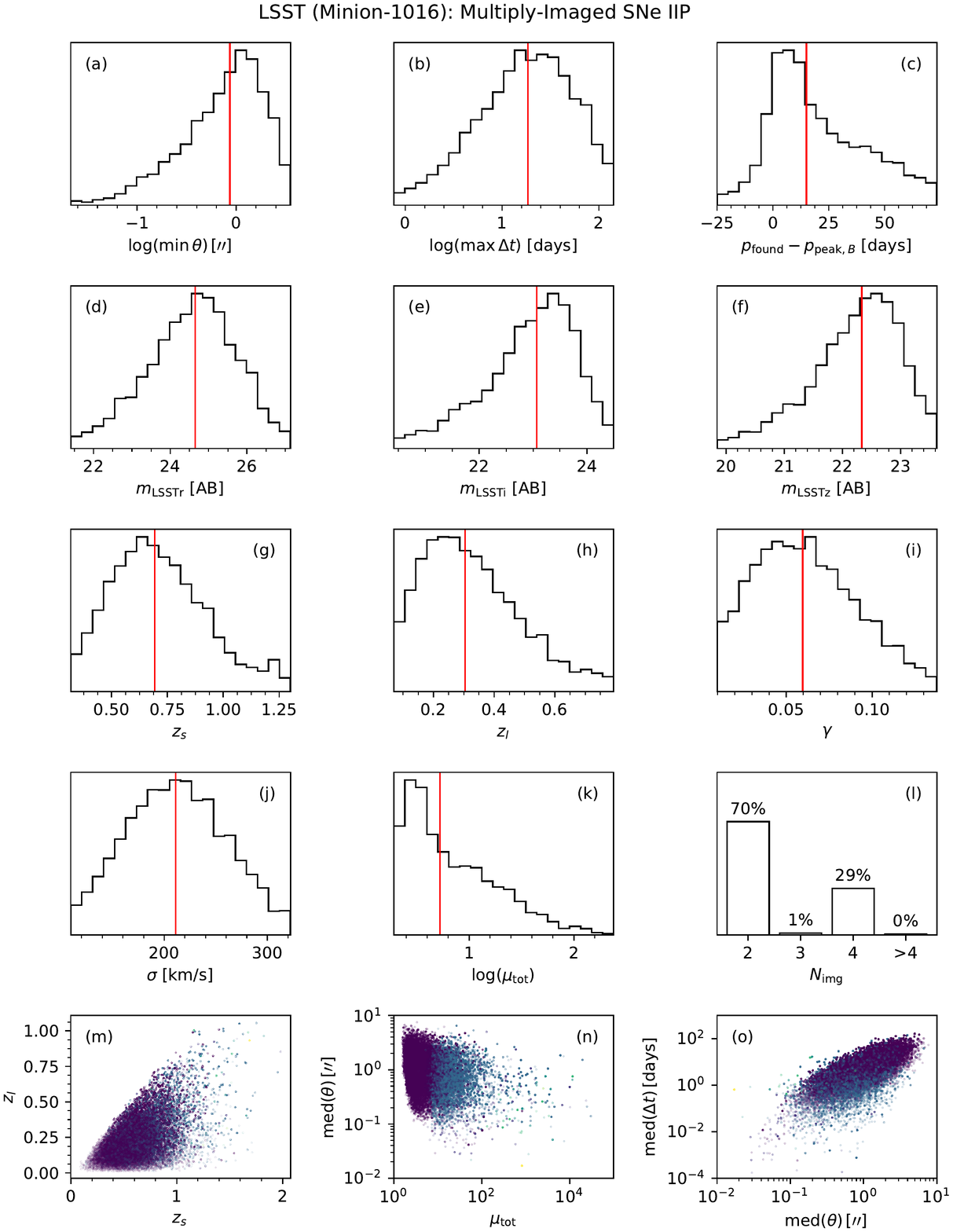}
    \caption{Monte Carlo results for LSST (\minion) Type IIP supernovae.  See Table \ref{tab:mcfigs} for a description of each subpanel.}
    \label{fig:minion-2p}
\end{figure*}

\begin{figure*}
    \centering
    \includegraphics[width=\reswidth]{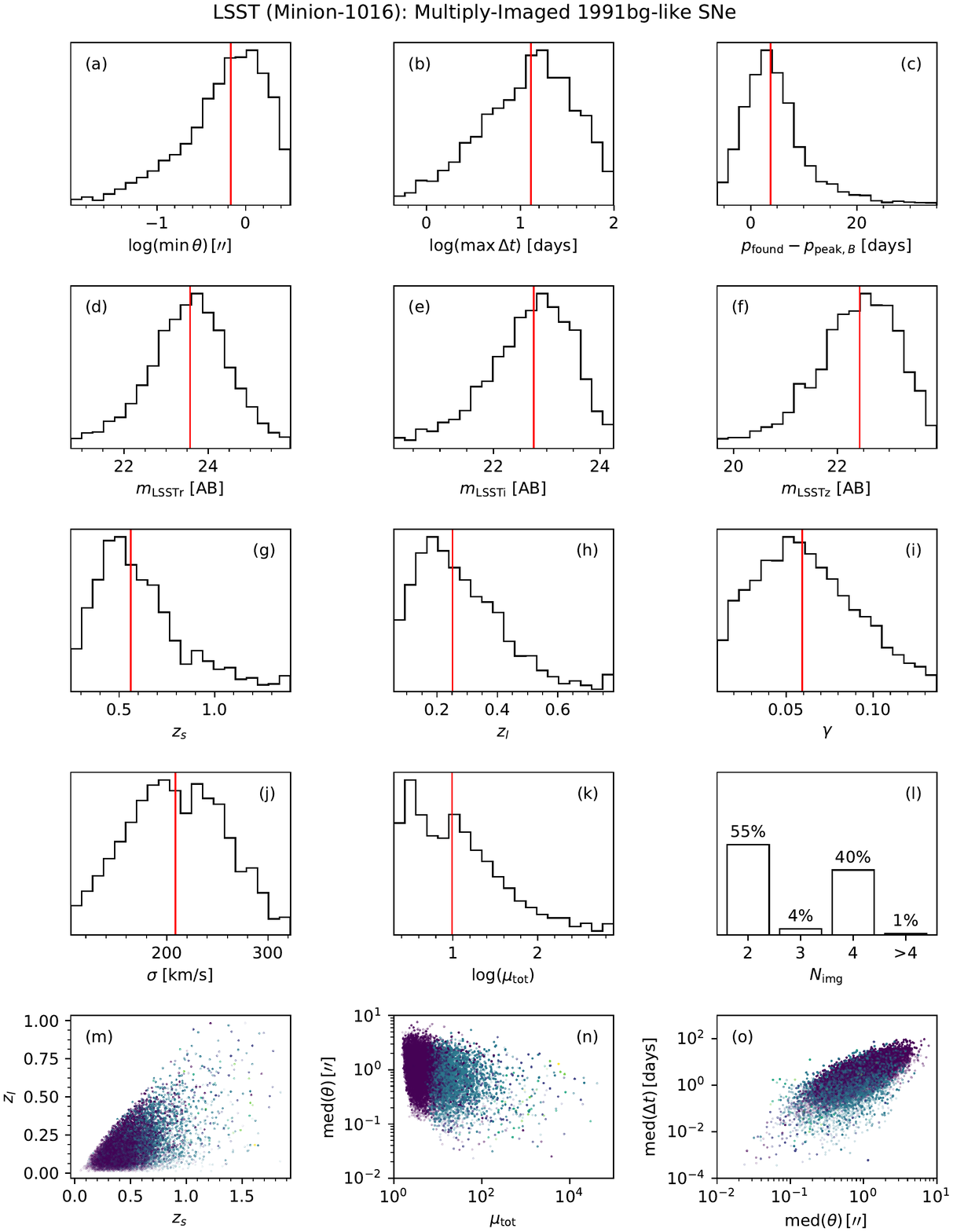}
    \caption{Monte Carlo results for LSST (\minion) SN 1991bg-like supernovae.  See Table \ref{tab:mcfigs} for a description of each subpanel.}
    \label{fig:minion-91bg}
\end{figure*}

\begin{figure*}
    \centering
    \includegraphics[width=\reswidth]{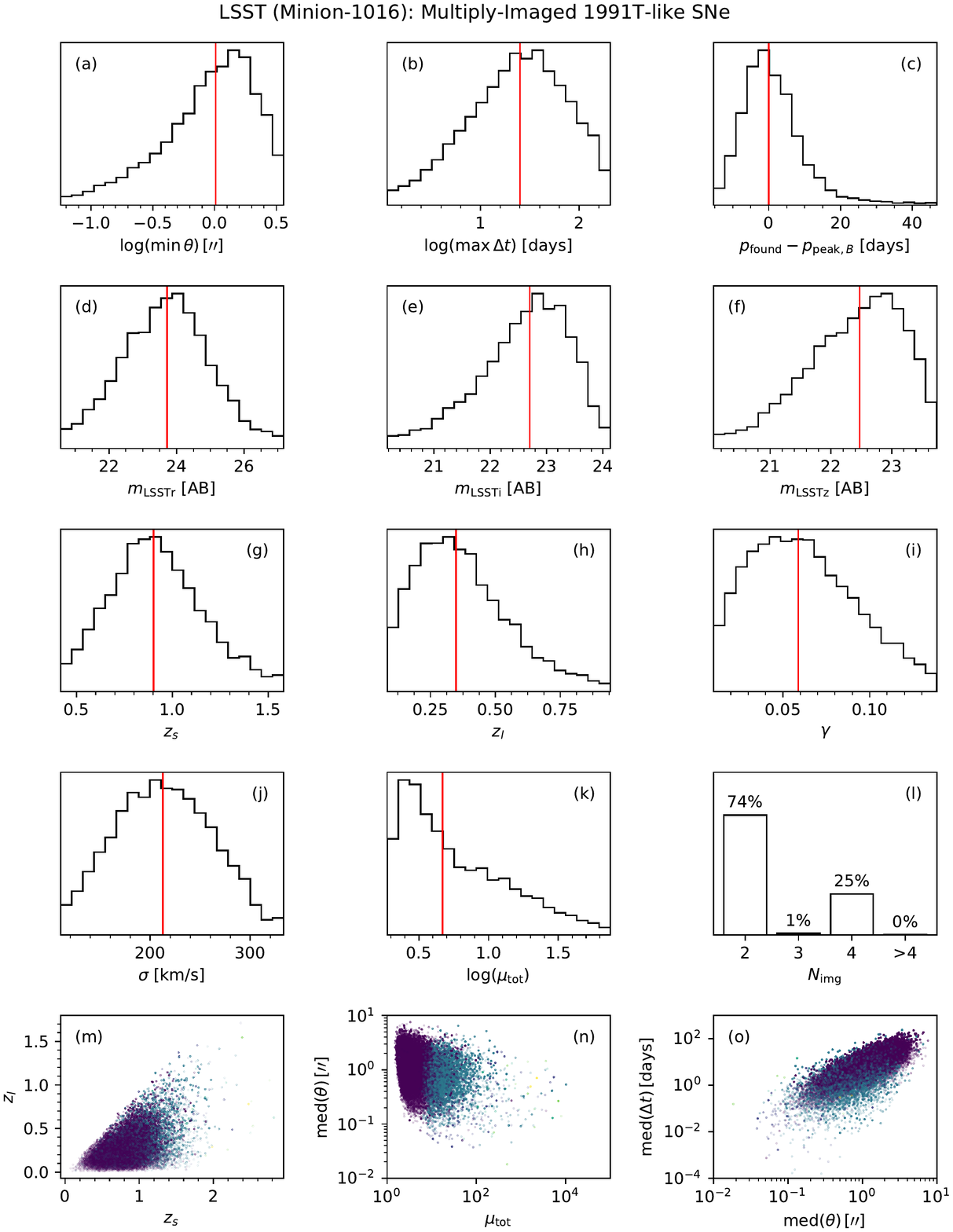}
    \caption{Monte Carlo results for LSST (\minion) SN 1991T-like supernovae.  See Table \ref{tab:mcfigs} for a description of each subpanel.}
    \label{fig:minion-91T}
\end{figure*}

\begin{figure*}
    \centering
    \includegraphics[width=\reswidth]{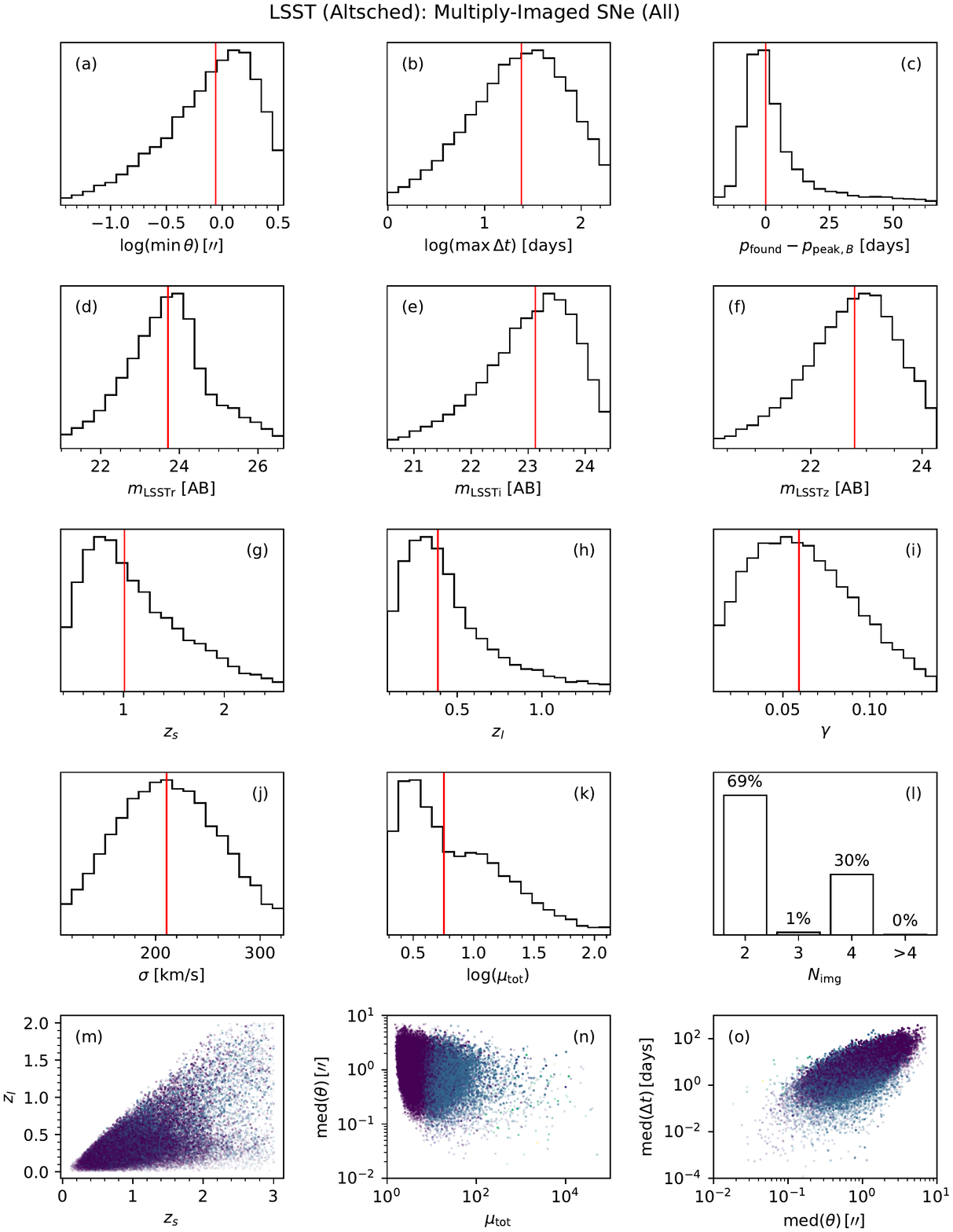}
    \caption{Monte Carlo results for LSST (\altsched) supernovae (all types).  See Table \ref{tab:mcfigs} for a description of each subpanel.}
    \label{fig:altsched-all}
\end{figure*}

\begin{figure*}
	\centering
    \gridline{\fig{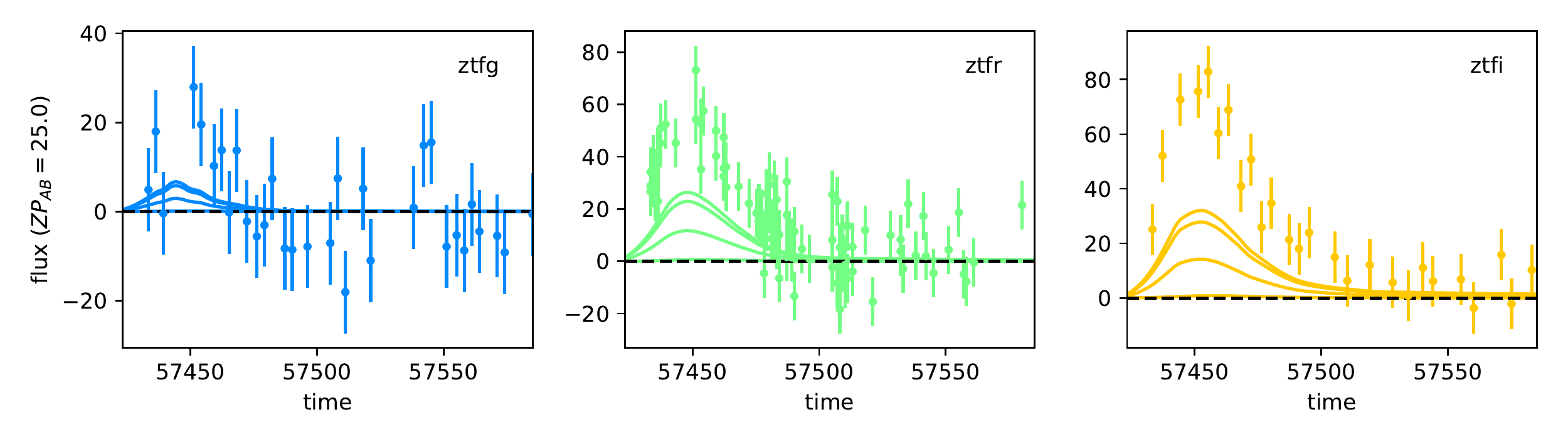}{1\textwidth}{(a)
    Light curves of a quadruply-imaged \glsn~Ia with $z_s=0.74$, $z_l=0.38$. 
    The images have time delays (relative to the earliest image) of 0.08, 0.10, and 4.46 days, and lensing amplifications of 8.9, 4.6, 10.3, and 0.3. 
    }}
% 56030
    \gridline{\fig{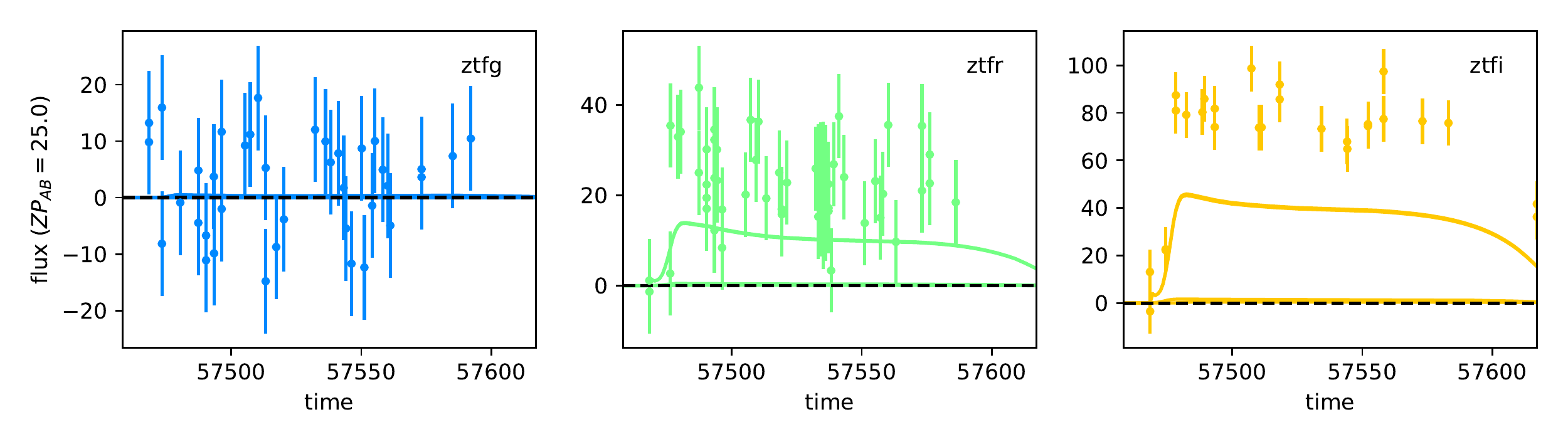}{1\textwidth}{(b)
    Light curves of a quadruply-imaged \glsn~IIP with $z_s=0.69$, $z_l=0.42$. 
    The images have time delays (relative to the earliest image) of 7.6, 2.1, and 2.1 days, and lensing  amplifications of 2.8, 0.8, 76.6, and 77.7. 
    }}
        \caption{Simulated ZTF light curves of \glsne. 
        The solid lines show the model light curves of the individual images.
        The photometric data are realized from the sum of the model light curves. \label{fig:ztflc}}
\end{figure*}

\begin{figure*}
% 23359
    \gridline{\fig{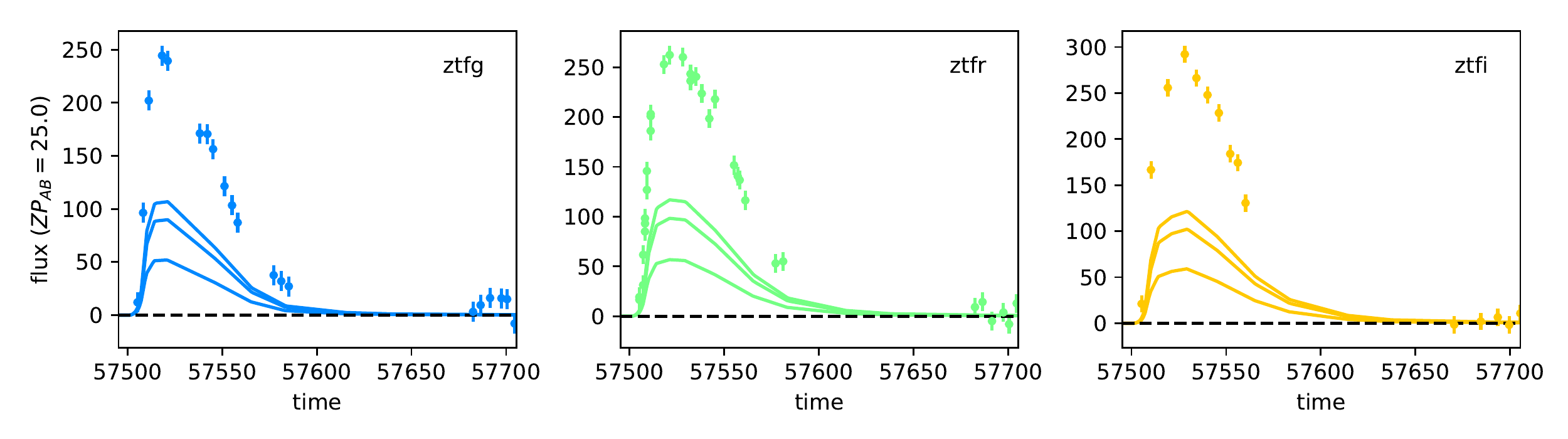}{1\textwidth}{(a)
    Light curves of a triply-imaged \glsn~IIn with $z_s=0.38$, $z_l=0.15$. 
    The images have time delays (relative to the earliest image) of 0.34 and  0.30 days, and lensing amplifications of 4.2, 8.7, and 7.3. 
    }}
% 12109
    \gridline{\fig{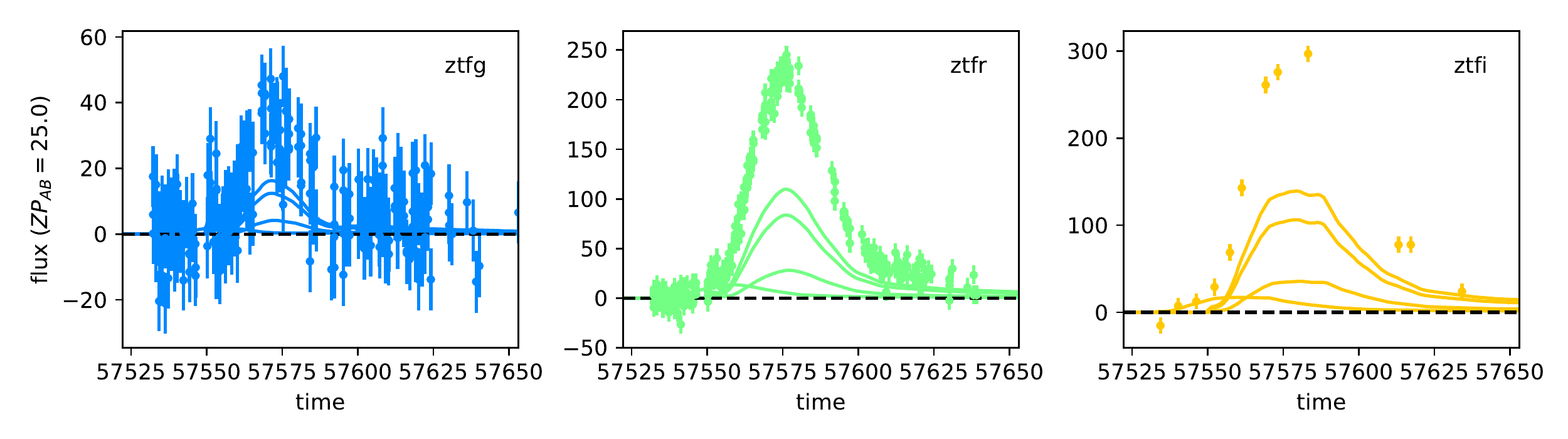}{1\textwidth}{(b)
    Light curves of a quadruply-imaged \glsn~Ib/c with $z_s=0.62$, $z_l=0.40$. 
    The images have time delays (relative to the earliest image) of 18.74, 17.75, and 17.72 days, and lensing amplifications of 6.1, 3.0, 18.1, and 23.6. 
    }}
    \caption{Simulated ZTF light curves of \glsne. 
        The solid lines show the model light curves of the individual images.
        The photometric data are realized from the sum of the model light curves. \label{fig:ztflc2}}

\end{figure*}

\begin{figure*}
	\centering
    \includegraphics[width=1\textwidth]{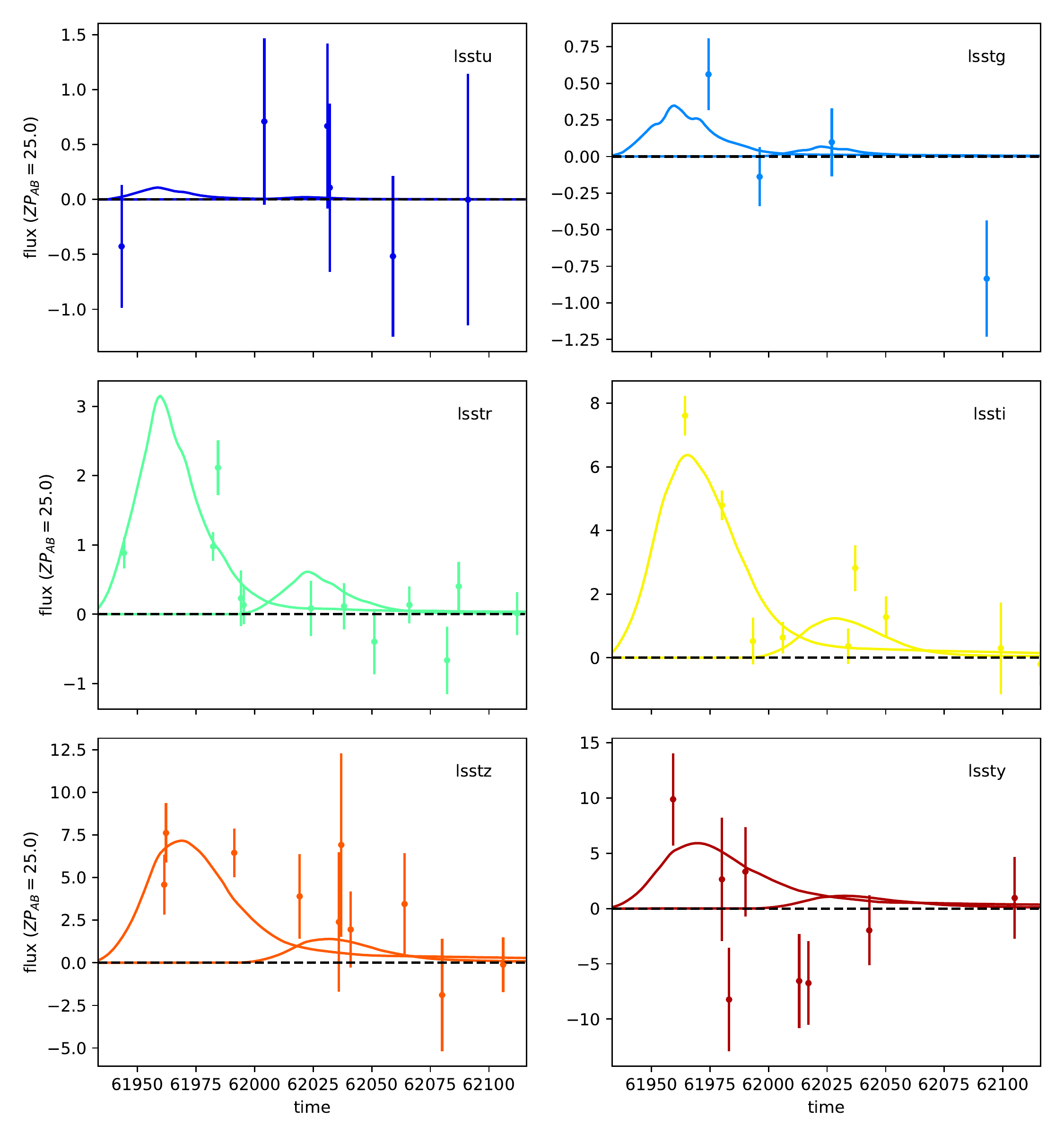}
    \caption{Simulated LSST \minion\ light curves of a \glsn~Ia with two images. 
    The system has $z_s=0.98$, $z_l=0.36$. 
    The images have a time delay of 62.9 days, and lensing amplifications of 3.3 and 0.6. 
    The  lines show the model light curves of the individual images.
    The photometric data are realized from the sum of the model light curves.
    Single-filter revisits taken within 30 minutes of one another to reject asteroids have been combined via stacking into single light curve points for clarity.
    }
    \label{fig:minion-1a-lc}
\end{figure*}

\begin{figure*}
	\centering
    \includegraphics[width=1\textwidth]{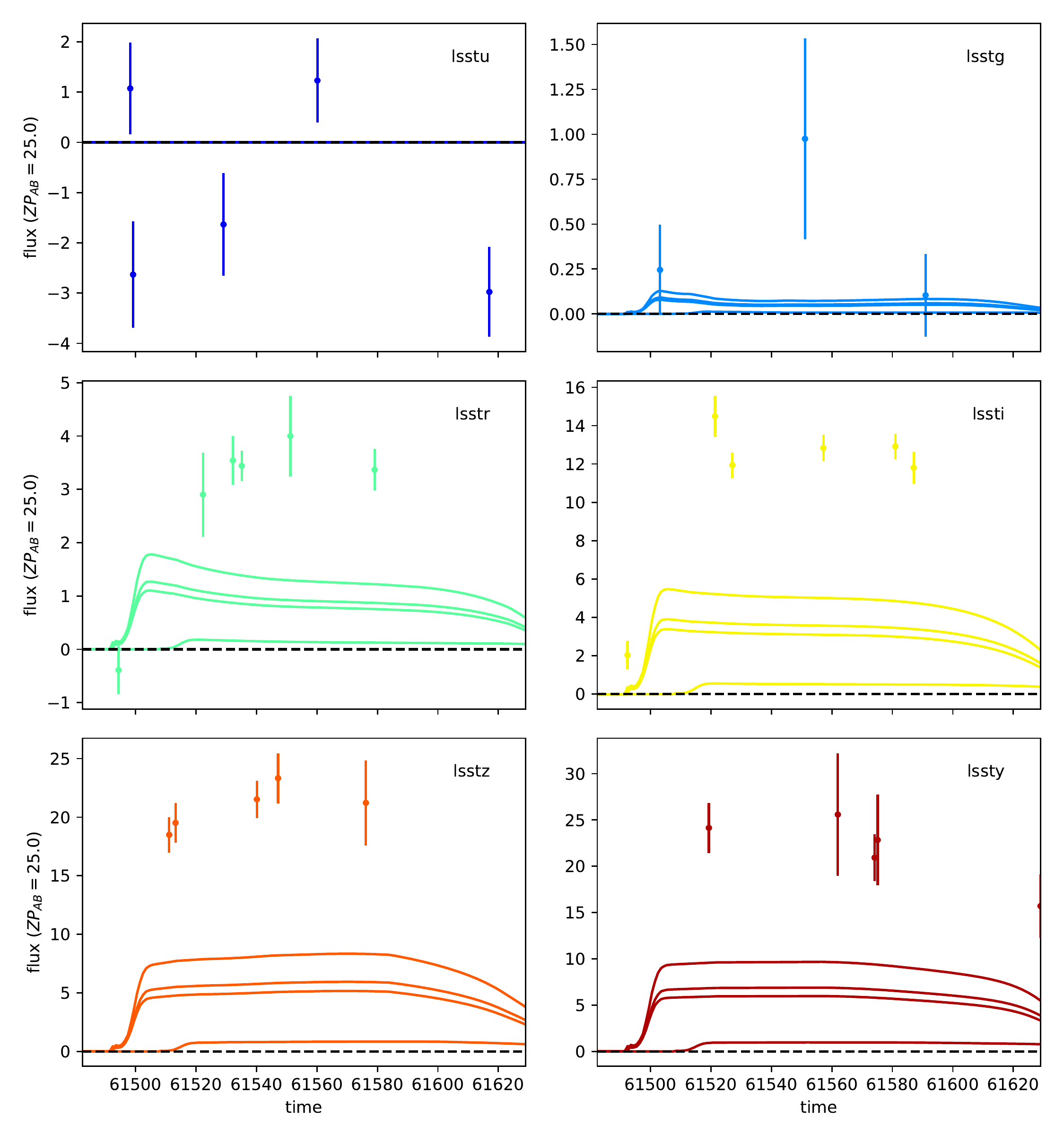}
    \caption{Simulated LSST \minion\ light curves of a \glsn~IIP with four images. 
    The system has $z_s=0.58$, $z_l=0.17$. 
    The images have time delays relative to the earliest image of 0.16, 15.66, and 0.46 days, and lensing amplifications of 8.6, 9.9, 1.4, and 13.8. 
    The  lines show the model light curves of the individual images.
    The photometric data are realized from the sum of the model light curves.
    Single-filter revisits taken within 30 minutes of one another to reject asteroids have been combined via stacking into single light curve points for clarity.}
    \label{fig:minion-2p-lc}
\end{figure*}

\begin{figure*}
	\centering
    \includegraphics[width=1\textwidth]{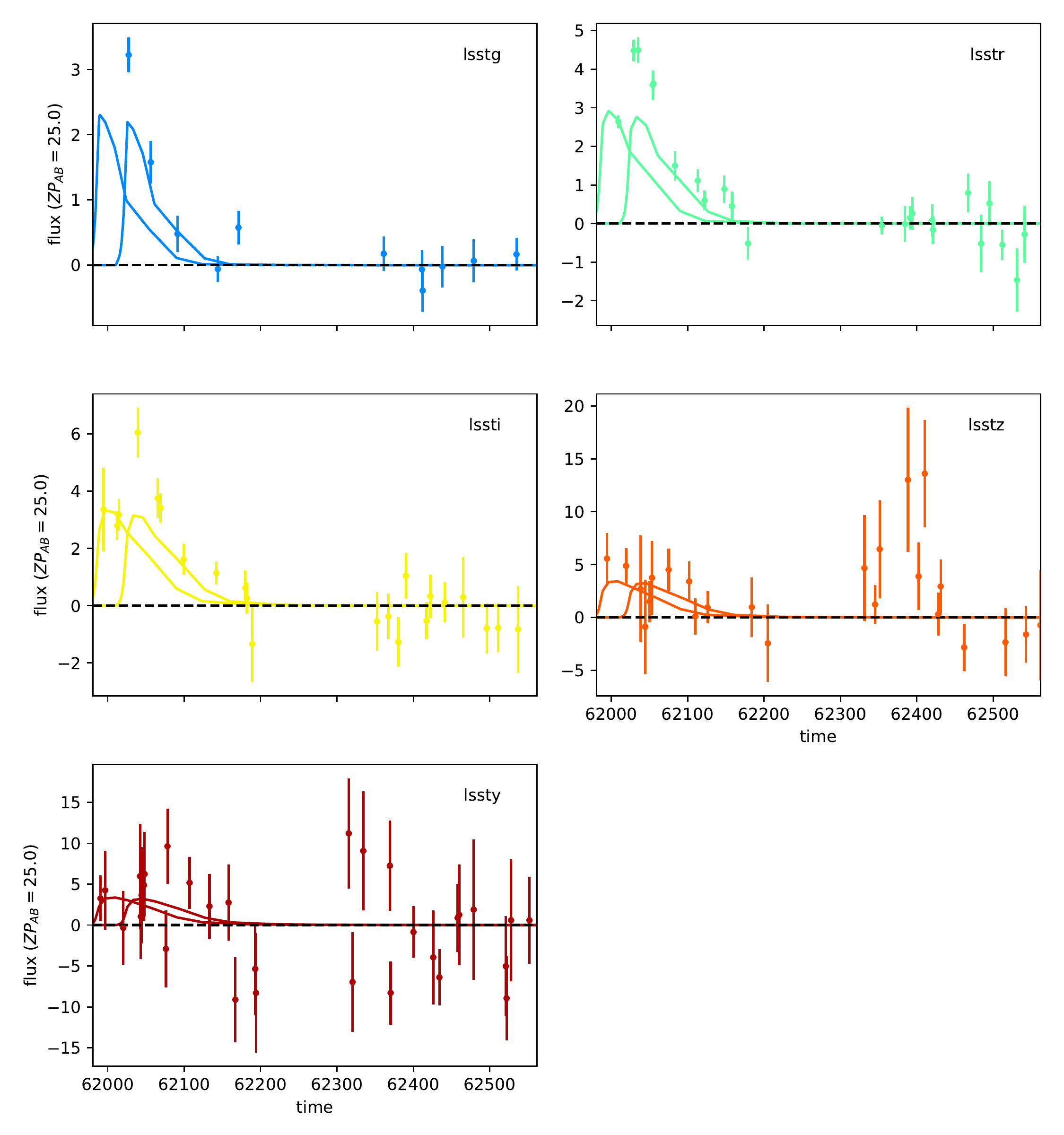}
    \caption{Simulated LSST \minion\ light curves of a \glsn~IIn with two images. 
    The system has $z_s=1.52$, $z_l=0.21$. 
    The images have a time delay of 36.8 days, and lensing amplifications of 1.8 and 1.7. 
    The  lines show the model light curves of the individual images.
    The photometric data are realized from the sum of the model light curves.
    Single-filter revisits taken within 30 minutes of one another to reject asteroids have been combined via stacking into single light curve points for clarity.}
    \label{fig:minion-2n-lc}
\end{figure*}

\begin{figure*}
	\centering
    \includegraphics[width=1\textwidth]{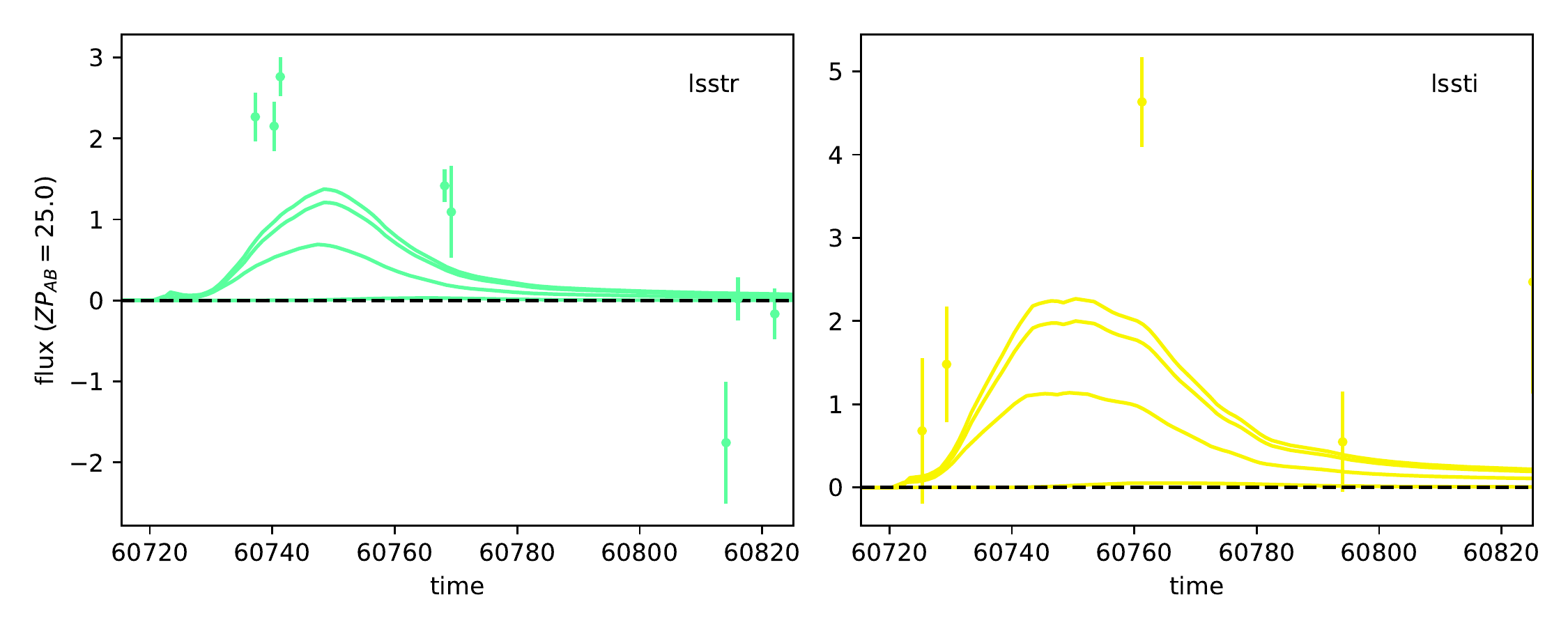}
    \caption{Simulated LSST \minion\ light curves of a \glsn~Ib/c with four images. 
    The system has $z_s=0.68$, $z_l=0.22$. 
    The images have time delays relative to the earliest image of 1.26, 1.17, and 16.73 days, and lensing amplifications of 2.65, 4.64, 5.27, and 0.12. 
    The  lines show the model light curves of the individual images.
    The photometric data are realized from the sum of the model light curves.
    Single-filter revisits taken within 30 minutes of one another to reject asteroids have been combined via stacking into single light curve points for clarity.}
    \label{fig:minion-1bc-lc}
\end{figure*}

\begin{figure*}
	\centering
    \includegraphics[width=1\textwidth]{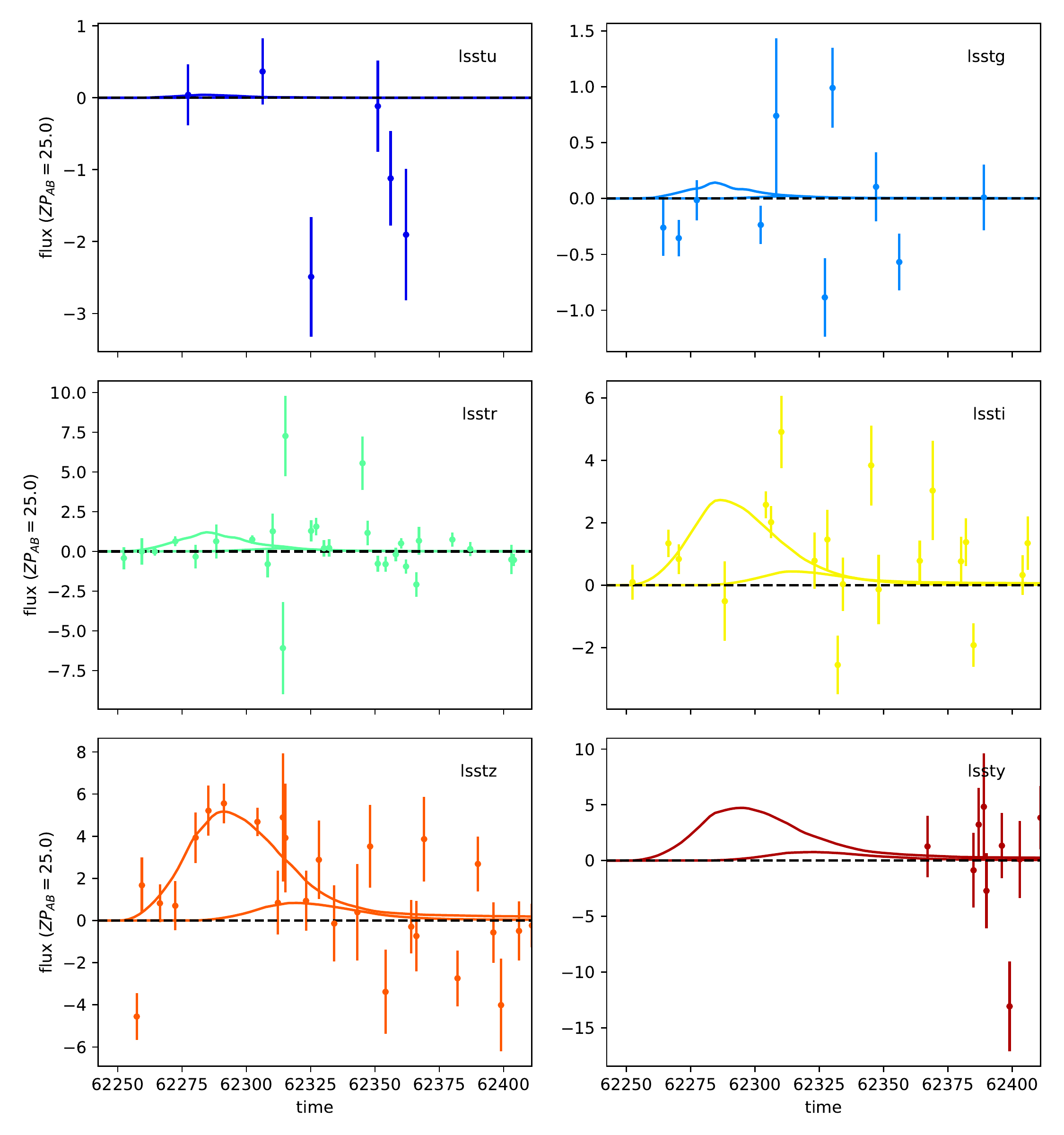}
    \caption{Simulated LSST \altsched\ light curves of a \glsn~Ia with two images. 
    The system has $z_s=1.17$, $z_l=0.19$. 
    The images have time delays relative to the earliest image of 1.26, 1.17, and 16.73 days, and lensing amplifications of 2.65, 4.64, 5.27, and 0.12. 
    The  lines show the model light curves of the individual images.
    The photometric data are realized from the sum of the model light curves.
}
    \label{fig:altsched-1a-lc}
\end{figure*}

\begin{figure*}
	\centering
    \includegraphics[width=1\textwidth]{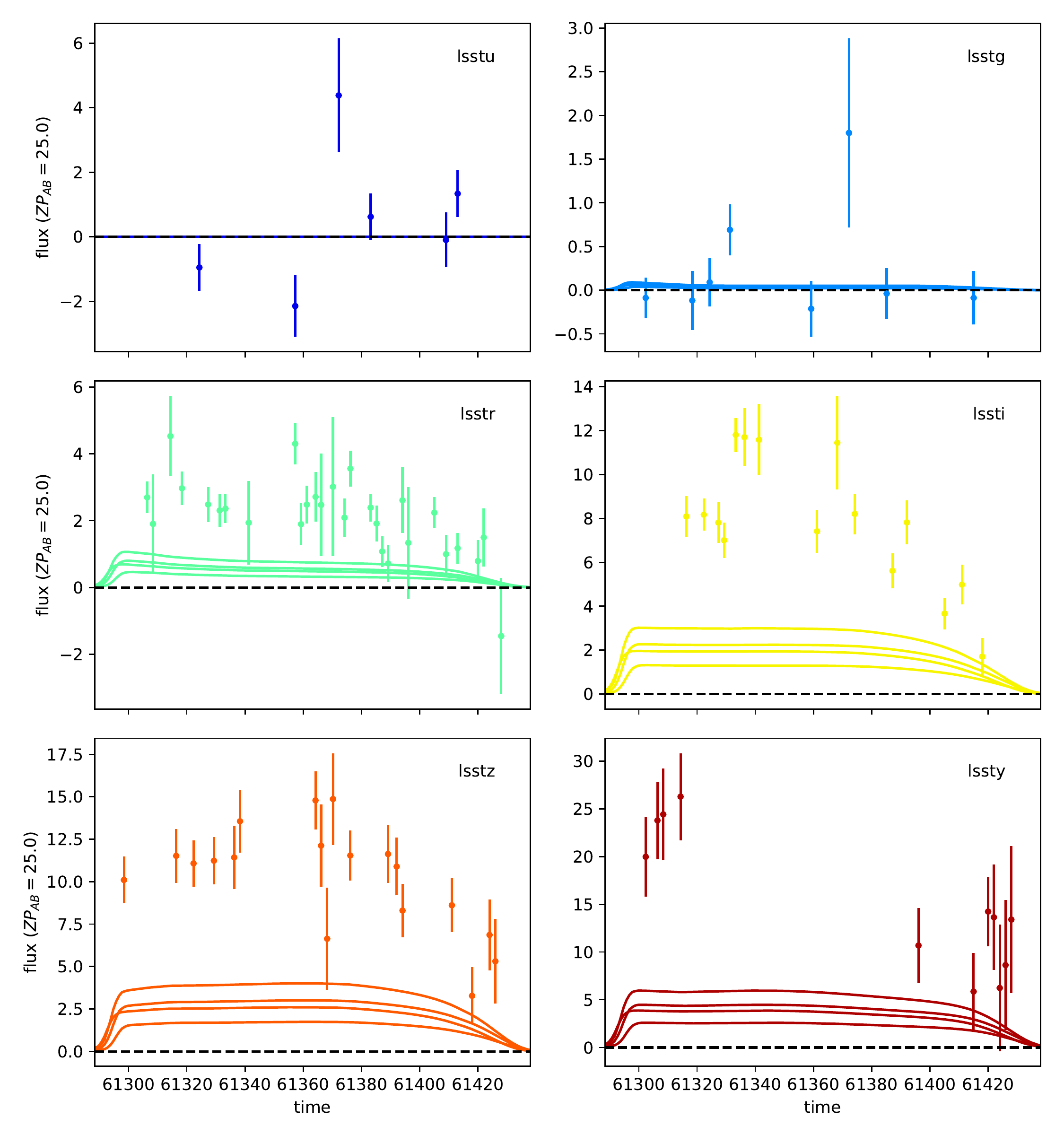}
    \caption{Simulated LSST \altsched\ light curves of a \glsn~IIP with four images. 
    The system has $z_s=0.53$, $z_l=0.14$. 
    The images have time delays relative to the earliest image of 1.32, 1.90, and 3.00 days, and lensing amplifications of 4.05, 6.23, 4.68, and 2.71. 
    The  lines show the model light curves of the individual images.
    The photometric data are realized from the sum of the model light curves.
    }
    \label{fig:altsched-2p-lc}
\end{figure*}

\begin{figure*}
	\centering
    \includegraphics[width=1\textwidth]{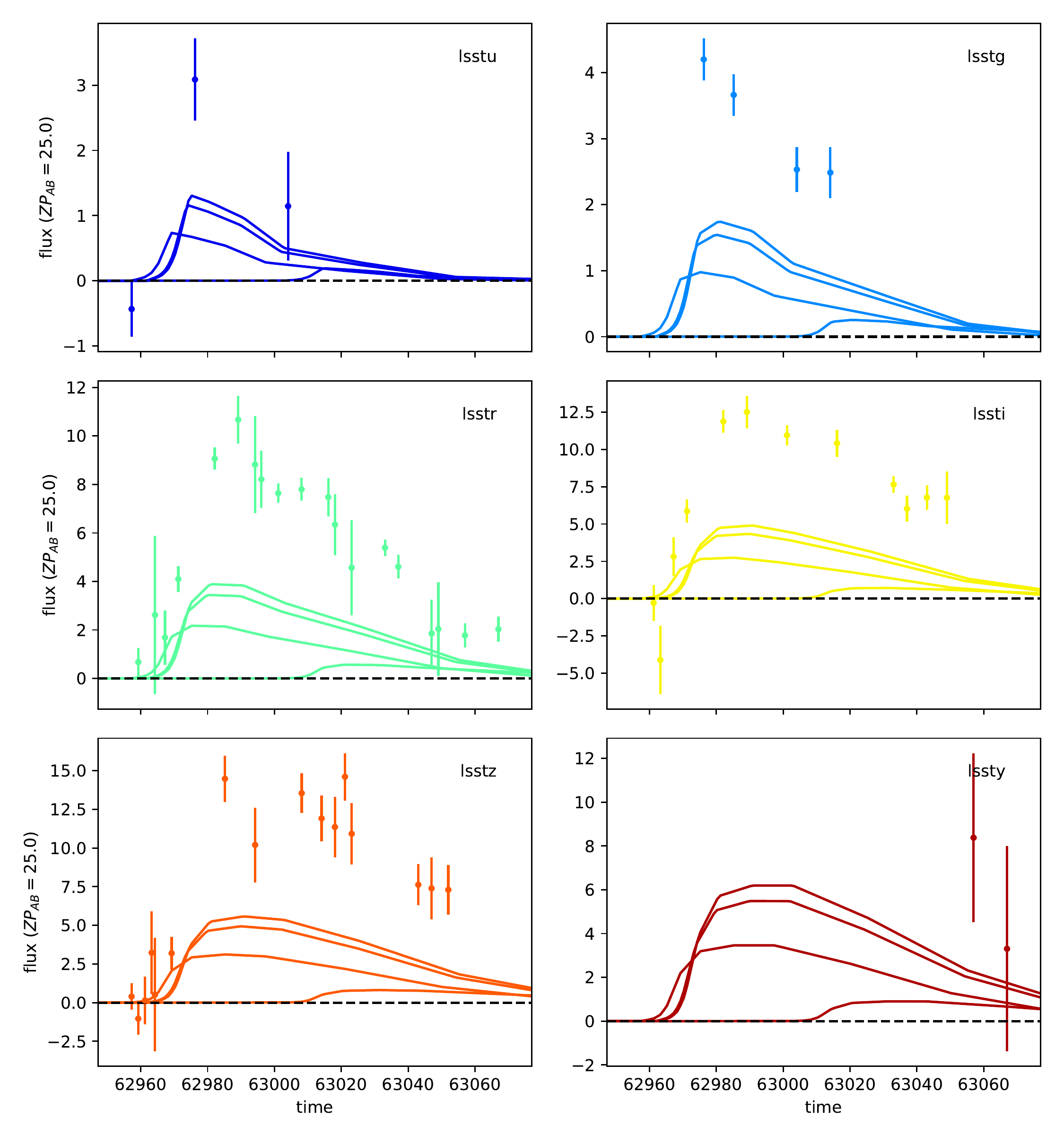}
    \caption{Simulated LSST \altsched\ light curves of a \glsn~IIn with four images. 
    The system has $z_s=1.02$, $z_l=0.46$. 
    The images have time delays relative to the earliest image of 45.37, 5.49, and 4.59 days, and lensing amplifications of 4.7, 1.2, 8.4, and 7.4.
    The  lines show the model light curves of the individual images.
    The photometric data are realized from the sum of the model light curves.
}
    \label{fig:altsched-2n-lc}
\end{figure*}

\begin{figure*}
	\centering
    \includegraphics[width=1\textwidth]{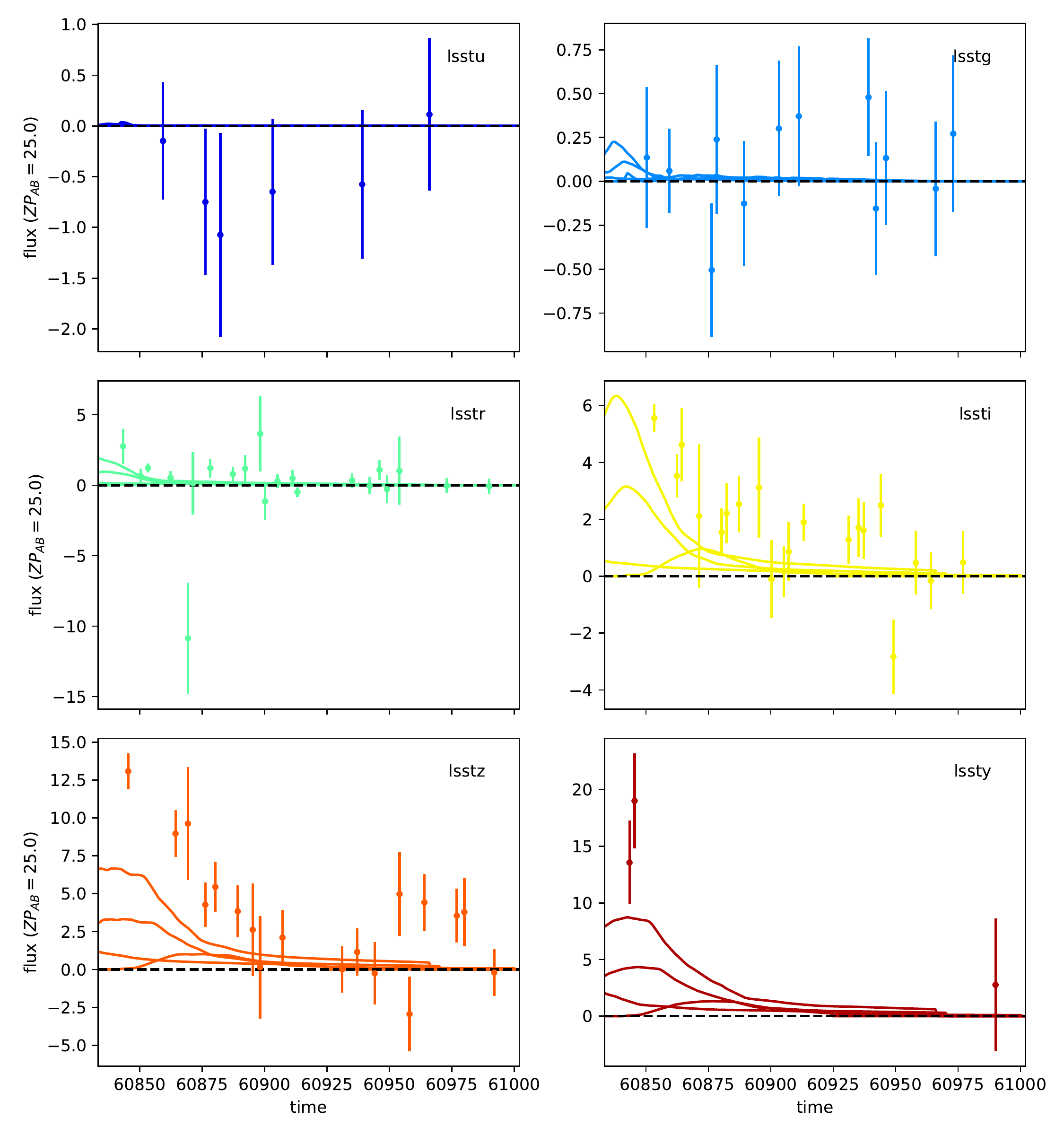}
    \caption{Simulated LSST \altsched\ light curves of a \glsn~Ib/c with four images. 
    The system has $z_s=0.88$, $z_l=0.25$. 
    The images have time delays relative to the earliest image of 46.44, 42.42,  and 76.43 days, and lensing amplifications of 1.8, 1.4, 2.8, and 0.4.
    The  lines show the model light curves of the individual images.
    The photometric data are realized from the sum of the model light curves.
}    
    \label{fig:lsst-lastlc}
\end{figure*}

\begin{figure*}
    \centering
    \fig{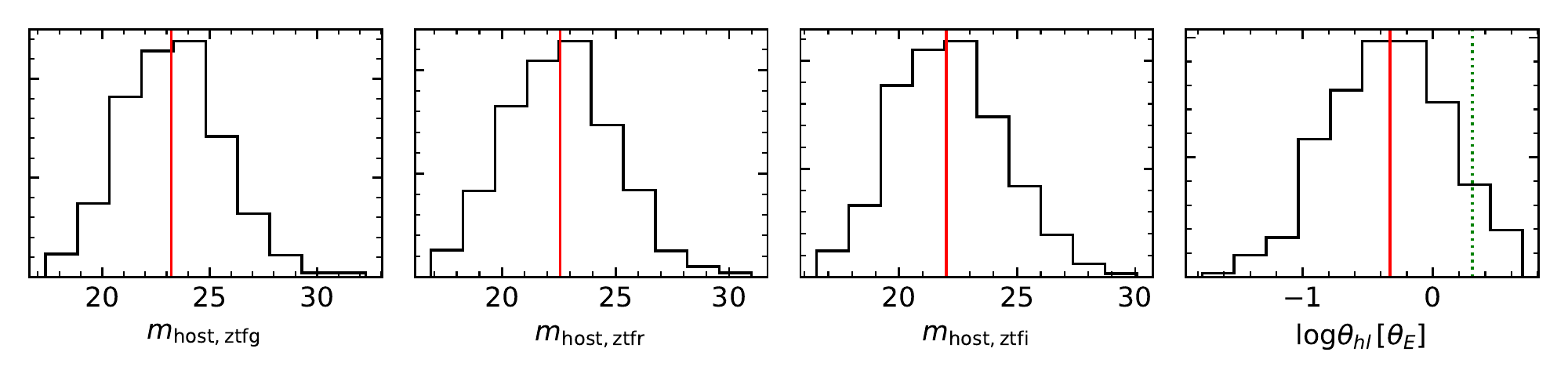}{1\textwidth}{(a) Apparent magnitude [AB] and host-lens separation distributions for ZTF \glsn\ host galaxies (all SN types).}
    \fig{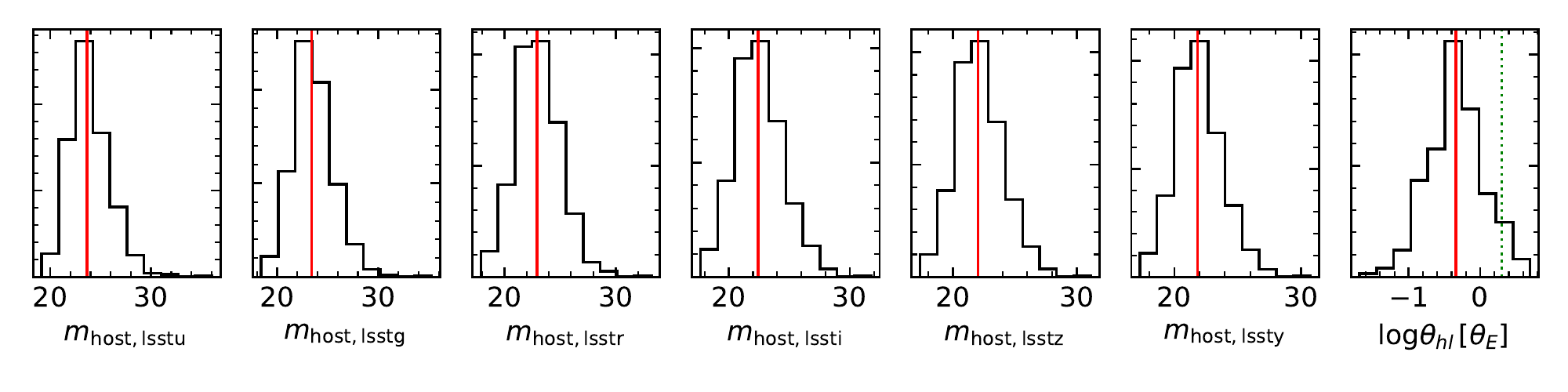}{1\textwidth}{(b) Apparent magnitude [AB] and host-lens separation distributions for LSST (\minion) \glsn\ host galaxies (all SN types).}
    \fig{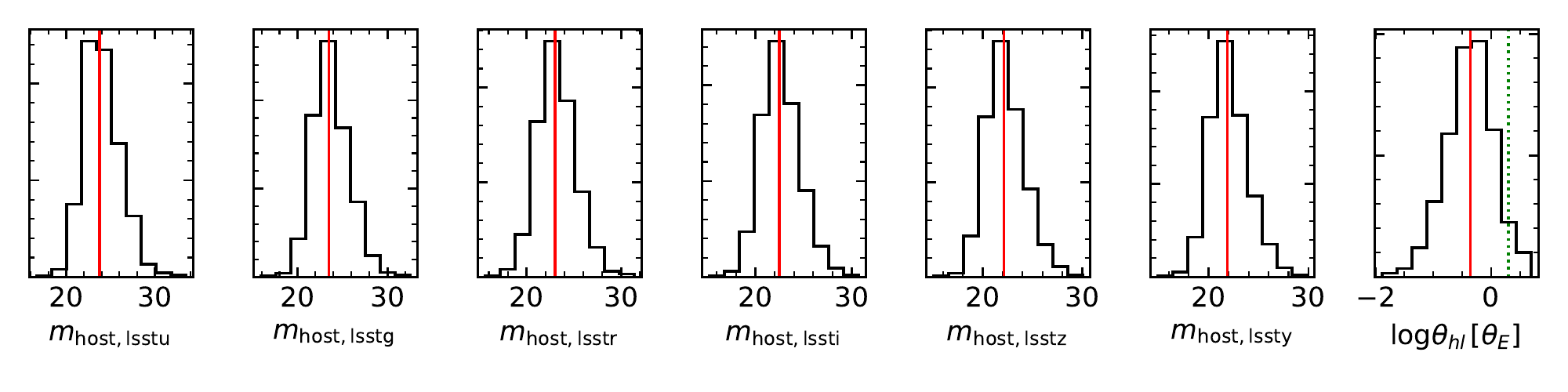}{1\textwidth}{(c) Apparent magnitude [AB] and host-lens separation distributions for LSST (\altsched) \glsn\ host galaxies (all SN types).}
    \caption{Lensed host galaxy property distributions for the three surveys considered in this analysis. 
    Red vertical lines indicate medians. 
    The  quantity $\theta_{hl}$ gives the separation between the unlensed position of the host galaxy centroid and the lens galaxy centroid in units of the Einstein radius $\theta_E$. 
    Hosts with centroids separated from the lens centroid by less than $2\theta_E$ (green dashed line) have a significant likelihood of being multiply imaged and can thus provide significant constraints on the lens model after the supernova has faded. 
    }
    \label{fig:host}
\end{figure*}

Color-composite images of randomly selected \glsne, drawn in proportion to their weights, discovered by ZTF and LSST (\minion) are shown in Figures \ref{fig:ztfsne} and \ref{fig:lsstsne}, respectively.
Figures \ref{fig:skydist1} and \ref{fig:skydist2} show the sky distributions of detected \glsne.
Figures \ref{fig:ztf-all} -- \ref{fig:altsched-all} summarize the results of our Monte Carlo simulations, presenting the distributions of several key observables and  parameters of detected systems.
Table \ref{tab:mcfigs} describes the subpanels in each figure, and red lines in histogram panels indicate medians. 
Figures \ref{fig:ztflc} -- \ref{fig:lsst-lastlc} show multi-band light curves of \glsne\ from ZTF and LSST. 
In those figures, the solid lines reflect the true underlying light curves of each image, while the  photometric  data are realized from the sum of the images.
The ZTF photometry is unstacked, reflecting the survey's high intranight cadence, whereas the LSST photometry is combined nightly into single point per filter for clarity. 
Figure \ref{fig:host} shows distributions of lensed host galaxy  apparent magnitudes and separations (relative to the lens centroid) in units of $\theta_E$.  
If the lens-host centroid distance is less than $2\theta_E$, there is a strong likelihood that the host galaxy is multiply imaged and can thus provide useful constraints on the lens model. 

Figures \ref{fig:ztfsne} -- \ref{fig:lsstsne}, \ref{fig:ztf-all}, \ref{fig:minion-all}, and \ref{fig:altsched-all} show that ZTF and LSST are sensitive to different populations of \glsne. 
ZTF \glsne\ have a median $z_s=0.9$, $z_l=0.35$, $\mu_\mathrm{tot}=30$, $\Delta t_\mathrm{max}= 10$ days, $\min \theta= 0.25^{\prime\prime}$, and $N_\mathrm{img} = 4$. 
LSST \glsne\ have a median $z_s=1.0$, $z_l=0.4$, $\mu_\mathrm{tot}\approx6$, $\Delta t_\mathrm{max} = 25$ days, $\min \theta=0.6^{\prime\prime}$, and $N_\mathrm{img} = 2$. 
Synthesizing this information, the ZTF \glsne\ tend to be  more compact, highly magnified, and have  shorter time-delays  than their LSST counterparts.
Additionally, ZTF \glsne\ are more likely to be quads than \glsne\ from LSST.
The \glsn\ iPTF16geu discovered by ZTF's predecessor iPTF was broadly consistent with this picture: it was a compact $(\mathrm{med}\,\theta \sim 0.3^{\prime\prime})$, highly magnified $(\mu \sim 90)$, quad with short time delays ($\Delta t < 1$ day). 
The \glsne\ from LSST will be better suited to time-delay cosmology. 
Their longer time delays and wider separations will enable more precise constraints on \h\ and better models of the mass profile. 
However, they will be fainter, and thus require larger telescopes and more observing time for follow-up observations. 
Table \ref{tab:ztf} shows that just 10\% of the \glsne\ ZTF will find will come from the public data alone. 
The proprietary data, notably the high-cadence data and the $i$-band survey, will be critical for discovering \glsne.

\begin{deluxetable}{r||p{0.76\textwidth}}[!htbp]
\tablecaption{Description of the subpanels in Figures \ref{fig:ztf-1a} -- \ref{fig:minion-91T}. \label{tab:mcfigs}}
\tablehead{\colhead{Subpanel} & 
		   \colhead{Description} }
\startdata
a &  The smallest angular separation, in arcseconds, between two images in the system (alternatively, the angular resolution required to completely resolve the system).\\
\hline b & The largest time delay between two images in the system.\\
\hline c & The rest-frame phase of the blended light curve on the date of discovery relative to rest-frame $B$-band maximum.\\
\hline d & Peak observer-frame AB magnitude of the \glsn\ in $g$ (ZTF) or $r$ (LSST).\\
\hline e & Peak observer-frame AB magnitude of the \glsn\ in $r$ (ZTF) or $i$ (LSST).\\
\hline f & Peak observer-frame AB magnitude of the \glsn\ in $i$ (ZTF) or $z$ (LSST).\\
\hline g & The source redshift.\\
\hline h & The lens redshift.\\
\hline i &  The magnitude of the external shear.\\
\hline j & The SIE velocity dispersion.\\
\hline k & The total lensing amplification of the \glsn\ images.\\
\hline l & The number of \glsn\ images in the system.\\
\hline m & The correlation between source and lens redshift, color coded by image multiplicity. Purple points correspond to double images, blue to quads, and redder colors to systems with more than four images.\\
\hline n & The correlation between total magnification and image separation, color coded as (m). \\
\hline o & The correlation between median image separation and median time delay, color coded as (m). 
\enddata
\end{deluxetable}

%\begin{figure*}
%	\centering
%\includegraphics[width=0.8\textwidth]{ztf.pdf}
%\caption{Sky distribution of \glsne\ (all types) detected by ZTF in the simulation.}
%\label{fig:sky}
%\end{figure*}


\begin{deluxetable}{rcccc}
\tablecaption{Fraction of \glsne\ discovered in the ZTF simulation that have $i$-band data (Partnership), high-cadence data (Partnership), and exclusively MSIP (public survey) data. \label{tab:ztf}}
\tablehead{\colhead{SN Type} &   \colhead{$i$ [\%]} &  \colhead{High Cadence [\%]} &  \colhead{MSIP Only [\%]}}
\startdata
Type Ia      &             77.7 &                77.3 &      12.4 \\
Type IIP      &           82.0 &                73.9 &      10.5 \\
Type IIn      &             71.3 &                73.1 &      16.2 \\
Type Ib/c     &          80.4 &                76.8 &      10.8 \\
Type IIL      &          81.1 &                75.1 &      10.7 \\
SN 1991bg-like    &          81.7 &                75.6 &       9.9 \\
SN 1991T-like &        77.3 &                75.5 &      13.0 
\enddata
\end{deluxetable}

\section{Discussion}
\label{sec:discussion}

\subsection{Comments on the LSST Observing Strategy}
Broadly speaking, candidate observing strategies for LSST can be arranged on a spectrum in which area and season length are traded for sampling and depth. 
In this analysis we have investigated strategies from both ends of this spectrum. 
\minion\ covers a large area with relatively poor light curve sampling, while \altsched\ covers a smaller area with better sampling and greater depth.
Table \ref{tab:yield} shows that the nominal LSST observing strategy \minion\ discovers roughly the same number of  \glsne\ as the alternative strategy \altsched, and Figures \ref{fig:minion-all} and \ref{fig:altsched-all} show that the greatest difference in the \glsne\ discovered under the two strategies is the discovery phase (see panel (c) of both figures).
\altsched\ discovers \glsne\ earlier than \minion\ due to its higher-quality light curves. 
A key result of this analysis is that for LSST, the  improved light curve sampling and depth of surveys like \altsched\  can compensate for the corresponding loss in area / season length by discovering more \glsne\ per square degree. 
Moreover,  the simulated \altsched\ survey only used 85\%  of the total LSST observing time, so it is possible that the \altsched\ yields presented here are too low by a factor of $\sim$1.17. 
Because the \glsn\ yields of \altsched\ are comparable to those of \minion, which has significantly more area (26,100 deg$^2$ compared to \altsched's 21,460 deg$^2$),\footnote{The yields of \glsne~IIn appear to be higher in \minion\ than \altsched, but this is an artifact of the high redshifts needed to fully simulate the \glsn~IIn population. The lower limits given have the ratio of the areas of the two surveys, indicating that both \minion\ and \altsched\ are fully probing the population to $z_s=3$. With an accurate model  of the supernova rate at extremely high redshifts, it is likely that both \minion\ and \altsched\ would converge to similar \glsn~IIn yields.} but the resulting light curves have significantly better sampling and are discovered earlier, we conclude that \altsched\ is a superior strategy for finding \glsne, enabling faster spectroscopic follow-up and more observations of \glsne\ while they are in the achromatic phase.

\subsection{Host Galaxy Properties and Implications for Lens Modeling}
\label{sec:host}

Figure \ref{fig:host} suggests that in both ZTF and LSST, at least 90\% of lensed host galaxy centroids will be within $2\theta_E$ of their associated lens galaxy centroids, making it extremely likely that they will be multiply imaged.
The median apparent magnitudes of the hosts from both surveys are roughly 22 in the redder filters, placing them well within reach of space-based imaging facilities such as \textit{HST}, \textit{JWST}, and \textit{WFIRST}, and larger ground-based facilities, especially those with adaptive optics systems. 
Combined with the fact that \glsne\ fade away, enabling a more precise reconstruction of the lensed hosts compared to lensed AGNs, this suggests that host galaxy modeling will not be a limiting factor in \glsn\ time delay cosmology. 

\subsection{Triple Images and other Exotic Configurations}
Figures \ref{fig:ztfsne} -- \ref{fig:lsstsne} and  \ref{fig:ztf-all} -- \ref{fig:altsched-all} show that ZTF and LSST will occasionally discover \glsne\ with three or more than four lensed images. 
These exotic configurations are uncommon but legitimate predictions of our population model. 
Triple image systems, such as row three, column five of Figure \ref{fig:lsstsne}, are a consequence of ellipticity in the SIE mass profile. 
When an SIE lens becomes sufficiently elliptical, part of its inner ``diamond'' caustic can extend beyond the outer ``oval caustic'' in a configuration known as a ``naked cusp'' \citep{collettcunnington16}. 
If a source is located in the naked cusp, it will form three adjacent lensed images in a curve around the mass profile. 

\glsne\ with more than four images are even rarer than \glsne\ lensed by  naked cusps, but they may still be discovered occasionally with LSST (it is extremely unlikely that ZTF will find any).
They are a consequence of a nonzero core radius in the SIE lens potential, which itself is a consequence of ellipticity. If a supernova is located sufficiently close to the core of an elliptical SIE, it is possible that more than four images will form -- in our simulations, systems with as many as eight images formed.
These systems are extremely  magnified $\mu \sim 10^4-10^6$, and have vanishingly small time delays and separations.
For this reason, they may be straightforward to detect, but will provide almost no useful information for cosmology. 
They may, however, enable high signal-to-noise-ratio spectroscopy of very high redshift supernovae, for which spectroscopy cannot currently be obtained.
This would be useful for studying the evolution of the supernova population with redshift. 

\subsection{A Bimodal Lens Redshift Distribution for ZTF \glsne~Ia}
As Figures \ref{fig:ztf-1a} (h) and \ref{fig:ztf-91t} (h) show, the lens redshift distributions for Type Ia and SN 1991T-like supernovae in ZTF are bimodal, with a first peak at $z_l \approx 0.1$ and a second at $z_l \approx 0.4$. 
This is due a selection effect introduced by the discovery strategy described in Section \ref{sec:discovery}, which biases the survey against discovering \sneia\ with two images in lenses with $z_l \gtrsim 0.15$. 
In such systems, the flux amplification  from lensing, which is usually on the order of a factor of a few, compensates for the reduction in flux caused by the fact that the supernova is at a higher redshift than the lens galaxy, making the overall flux of the transient compatible with an \snia\ hosted by the lens.
Thus a dearth of \glsne~Ia with two images occurs for $z_l \gtrsim 0.15$, causing the bimodal distribution. 
Other types of \glsne\ in ZTF do not have bimodal lens redshift distributions because of their core-collapse nature. 
The colors of core-collapse supernovae  are so different from those of normal \sneia\ that they are still identified by the discovery  when their overall fluxes are consistent with those of \sneia\ hosted by the lens galaxy.

\subsection{The Prevalence of \glsne~IIn}
\label{sec:2n}
Both ZTF and LSST will discover  \glsne~IIn more frequently than any other \glsn\ subtype.
SN Refsdal at $z_s=1.49$, the first identified \glsn\ with resolved images, was a peculiar type of interacting supernova, similar to a Type IIn \citep{refsdal_classification}.
Relatively speaking, unlensed Type IIn supernovae are uncommon, making up just 8--12\% of the observed core-collapse supernova rate \citep{2011MNRAS.412.1441L}. 
However, Type IIn supernovae are extremely bright (roughly 2 magnitudes brighter than Type IIp supernovae) and blue.
Their colors are so different from those of Type Ia supernovae that they are trivially identified by the discovery strategy detailed in Section \ref{sec:discovery}. 
As their volumetric rate follows the star formation rate (see Figure \ref{fig:snrate}), they are extremely common at high redshift \citep[e.g.,][]{2016A&A...594A..54P}, just beyond the flux limit of most imaging surveys.

Flux amplification from gravitational lensing will allow future synoptic imaging surveys to tap into this high-redshift population. 
This will enable unprecedented spectroscopic studies of the high redshift core-collapse  and interacting supernova populations. 
While in general the evolution of SNe IIn is slow, with the SED dominated by a black body continuum which slowly gets colder, several of these events show abrupt rises shortly after explosion as well as periods in which the interaction increases or decreases abruptly. 
These maybe suitable for time delay measurements, but will be the focus of future research.
Because these \glsne\ will be so numerous, increased focus should be placed on maximizing their scientific return.

\subsection{iPTF16geu: remarkable fluke or evidence of  physics not captured by current lensing models?}

\begin{figure}
	\centering
    \includegraphics[width=1\textwidth]{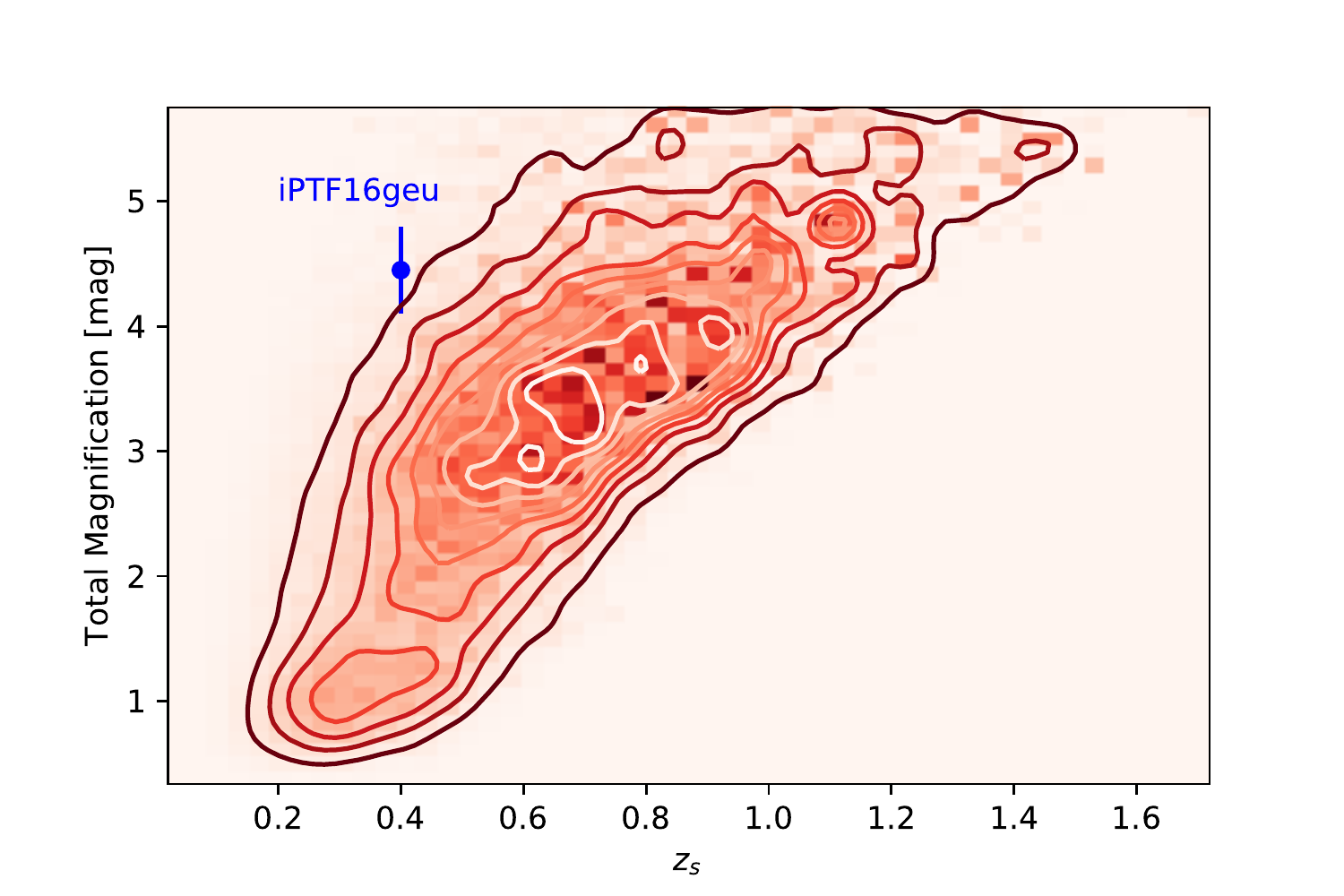}
    \caption{Joint distribution of lensing amplification and source redshift for \glsne~Ia found by the simulated ZTF survey and lensed by smooth galaxy lenses. 
    iPTF16geu, marked with a blue star, was dramatically more magnified than expected at its redshift. 
    Additional events are necessary to address the origin of this discrepancy.}
    \label{fig:geumag}
\end{figure}

iPTF16geu \citep{goobar16}, the only \glsn~Ia with resolved images discovered to date, is notable for its remarkably high magnification.
Accounting for extinction, its four supernova images had a total magnification $40 \lesssim \mu \lesssim 90$, significantly larger than the predicted $\mu \sim 25$ \citep{more16}. 
Amplification by unresolved lens galaxy  stars (microlensing) was proposed as an explanation for this anomaly \citep{more16}.
In a subsequent investigation, \cite{2017arXiv171107919Y} used microlensing ray-tracing simulations to show that microlensing alone could not account for the large observed flux anomalies. 
This may indicate that the anomaly is due to  millilensing, but a systematic study of millilensing  induced by lens-galaxy substructures on \glsne\ has yet to be performed. 

Thus the origin of the large magnification of iPTF16geu  remains a mystery, but the simulations presented in this paper can help place this discrepancy in context.
iPTF, the survey that found iPTF16geu, used the same telescope as ZTF (the P48), to observe the same region of sky to roughly the same depth, but at a lower cadence.
Thus the ZTF results presented here should be quite similar to those for iPTF.
Figure \ref{fig:geumag} shows the joint distribution of $z_s$ and $\mu_\mathrm{tot}$ for \glsne~Ia discovered in ZTF, showing that iPTF16geu is significantly more magnified than expected for its redshift $(>5\sigma)$.
Was iPTF16geu a remarkable fluke, or is there fundamental physics at play that our models for lensing do not capture?
Searches for new strongly lensed \sne\ with ZTF will likely resolve this intriguing question.

\section{Conclusion}
\label{sec:conclusion}
In this article, we have presented detailed simulations of the \glsn\ population and made predictions of the  properties and rates of \glsne\ that  forthcoming synoptic time-domain imaging surveys will find. 
ZTF should discover roughly 20 \glsne\ over the course of a three-year survey, and LSST should find roughly 3,500 over its 10-year lifetime.
Most host galaxies will be multiply imaged, enabling detailed lens modeling if sufficiently deep high-resolution imaging is obtained. 
ZTF and LSST are sensitive to different \glsn\ populations.
ZTF is most sensitive to compact, highly magnified quads with short time delays, whereas LSST is more sensitive to fainter doubles, which in general are less magnified and have longer delays. 
This will give LSST an advantage for time-delay cosmology if it can obtain the follow-up resources needed to extract spectroscopy and time delays from these transients. 
Our inclusion of dust decreases the expected \glsn~Ia rate over the predictions of \cite{gnkc18}, which did not include dust, by a factor of $\sim$2, but the predictions remain largely consistent with those of \cite{gn17}. 
This study has found that \glsne~IIn will be the most frequently discovered by both ZTF and LSST. 
With respect to  LSST observing strategy, we find that strategies that produce  dense light curves at the expense of a larger survey area can yield comparable numbers of \glsne, but the better-sampled surveys discover these \glsne\ earlier and produce higher-quality light curves.

\acknowledgements
D.A.G. gratefully acknowledges Masamune Oguri for sharing the \texttt{glafic}\ source code, which made the Monte Carlo simulations in this paper significantly more efficient,  Ravi Gupta for useful conversations about supernova host galaxies, Tom Collett for improving the deflector mass model, Eric Bellm for sharing his simulated ZTF survey, and Rollin Thomas and Shane Canon for assistance with \texttt{shifter}\ at NERSC.
D.A.G. acknowledges support from Hubble Fellowship grant HST-HF2-51408.001-A.
Support for Program number HST-HF2-51408.001-A is provided by NASA
through a grant from the Space Telescope Science Institute, which is
operated by the Association of Universities for Research in Astronomy,
Incorporated, under NASA contract NAS5-26555.
A.G acknowledges funding from the Swedish Research Council and the Swedish National Space Agency. 
D.A.G. and P.E.N. acknowledge support from the DOE under grant DE-AC02-05CH11231, Analytical Modeling for Extreme-Scale Computing Environments. 
This research used resources of the National Energy Research Scientific Computing Center, a DOE Office of Science User Facility supported by the Office of Science of the U.S. Department of Energy under Contract No. DE-AC02-05CH11231.

\bibliography{ref}

\end{document}